%% file: root.tex
\documentclass[12pt,oneside]{book}

\newcommand{\UCthesistitle}{Short Quantum Games}
\newcommand{\UCauthor}{Gustav Gutoski}
\newcommand{\UCmonth}{September}
\newcommand{\UCyear}{2005}

\usepackage{fullpage}

\input{header}

\begin{document}

\frontmatter


\begin{titlepage}

\begin{center}
THE UNIVERSITY OF CALGARY \\
\vfill
\UCthesistitle
\vfill
by
\vfill
\UCauthor
\vfill
A THESIS \\
SUBMITTED TO THE FACULTY OF GRADUATE STUDIES \\
IN PARTIAL FULFILMENT OF THE REQUIREMENTS FOR THE \\
DEGREE OF MASTER OF SCIENCE
\vfill
DEPARTMENT OF COMPUTER SCIENCE
\vfill
CALGARY, ALBERTA \\
\UCmonth, \UCyear
\vfill
\copyright\ \UCauthor\ \UCyear
\end{center}

\end{titlepage}

%
%

\newlength{\signwidth}
\newenvironment{signing}[1]{\settowidth{\signwidth}{#1}%
\def\signline{ \null\vskip 1em\makebox[\signwidth]{\hrulefill}\\}
\def\newsigncolumn{\end{minipage}\hfill\begin{minipage}[t]{\signwidth}}%
\noindent\begin{minipage}{\textwidth}
\raggedleft
\null
\begin{minipage}[t]{\signwidth}
}{\null\end{minipage}\end{minipage}
\null
\par
\vfill
\noindent \rule{2.5in}{.01in}\null
\\
\noindent Date
}


\setcounter{page}{2}

\addcontentsline{toc}{chapter}{Approval Page}

\begin{center}
THE UNIVERSITY OF CALGARY \\
FACULTY OF GRADUATE STUDIES
\end{center}

\

\noindent
The undersigned certify that they have read, and recommend to the Faculty of
Graduate Studies for acceptance, a thesis entitled ``\UCthesistitle'' submitted
by \UCauthor\ in partial fulfilment of the requirements for the degree of
Master of Science.

\begin{signing}{Department of Computer Science}
\signline
Supervisor, Dr.~John Watrous \\
Department of Computer Science \\
\signline
Dr.~Peter H\o yer \\
Department of Computer Science \\
\signline
Dr.~Michael Lamoureux \\
Department of Mathematics and Statistics \\
\end{signing}


\include{abstract}







{

\addcontentsline{toc}{chapter}{Contents}
\tableofcontents

\pagebreak

\addcontentsline{toc}{chapter}{List of Figures}
\listoffigures

}

\include{glossary}


\mainmatter

\markright{}

\include{intro}

\include{defs}

\include{QIPinSQG}
\include{SQGinEXP}
\include{end}

\backmatter


\addcontentsline{toc}{chapter}{Bibliography}
\bibliographystyle{plain}
\bibliography{references}

\end{document}

%% file: header.tex
\usepackage{amsmath} 
\usepackage{amssymb} 
\usepackage{amsthm} 
\usepackage{mathrsfs} 

\theoremstyle{definition}
\newtheorem{thm}{Theorem}[chapter]
\newtheorem{lm}[thm]{Lemma}
\newtheorem{cor}[thm]{Corollary}
\newtheorem{fact}[thm]{Fact}
\newtheorem{con}[thm]{Conjecture}

\newcommand{\cls}{\mathsf}        
\newcommand{\hilb}{\mathcal}      

\newcommand{\poly}{\textit{poly}}

\newcommand{\tinyspace}{\mspace{1mu}}

\DeclareMathOperator{\tr}{tr}              
\DeclareMathOperator{\rank}{rank}          

\newcommand{\bra}[1]{\langle#1|}

\newcommand{\ket}[1]{|#1\rangle}

\newcommand{\braket}[1]{\langle#1\rangle}
\newcommand{\Braket}[1]{\left\langle#1\right\rangle}
\newcommand{\inner}{\braket}
\newcommand{\Inner}{\Braket}

\newcommand{\paren}[1]{(#1)}
\newcommand{\Paren}[1]{\left(#1\right)}
\newcommand{\set}[1]{\{#1\}}
\newcommand{\Set}[1]{\left\{#1\right\}}
\newcommand{\mapset}[2]{\mathbf{#1}\paren{#2}}
\newcommand{\Mapset}[2]{\mathbf{#1}\Paren{#2}}

\newcommand{\abs}[1]{|\tinyspace#1\tinyspace|}
\newcommand{\Abs}[1]{\left|\tinyspace#1\tinyspace\right|}
\newcommand{\norm}[1]{\lVert\tinyspace#1\tinyspace\rVert}
\newcommand{\Norm}[1]{\left\lVert\tinyspace#1\tinyspace\right\rVert}
\newcommand{\tnorm}[1]{\norm{#1}_{\mathrm{tr}}}
\newcommand{\TNorm}[1]{\Norm{#1}_{\mathrm{tr}}}
\newcommand{\dnorm}[1]{\norm{#1}_{\diamond}}
\newcommand{\DNorm}[1]{\Norm{#1}_{\diamond}}

\newcommand{\ptr}[2]{\tr_{#1}\paren{#2}}   
\newcommand{\Ptr}[2]{\tr_{#1}\Paren{#2}}


%% file: abstract.tex
\chapter{Abstract}

In this thesis we introduce quantum refereed games, which are quantum
interactive proof systems with two competing provers.
We focus on a restriction of this model that we call
\emph{short quantum games} and we prove an upper bound and a lower bound on the
expressive power of these games.

For the lower bound, we prove that every language having an ordinary quantum
interactive proof system also has a short quantum game.
An important part of this proof is the establishment of a quantum measurement
that reliably distinguishes between quantum states chosen from disjoint convex
sets.

For the upper bound, we show that certain types of quantum refereed games,
including short quantum games, are decidable in deterministic exponential time
by supplying a separation oracle for use with the ellipsoid method for convex
feasibility.

%% file: glossary.tex
\chapter{List of Abbreviations}

What follows is a list of the complexity classes discussed in this thesis.
Each entry in the list consists of the name of a complexity class, the page in
which the class is first used or defined, and an informal definition of that
class.

\begin{description}

\item[$\cls{P}$, page \pageref{P}.]

The class of languages decidable by a deterministic polynomial-time Turing
machine.

\item[$\cls{NP}$, page \pageref{NP}.]

The class of languages decidable by a deterministic polynomial-time Turing
machine with the help of polynomial-size ``proof'' strings.

\item[$\cls{coNP}$, page \pageref{coNP}.]

The class of languages whose complements are in $\cls{NP}$.

\item[$\cls{PSPACE}$, page \pageref{PSPACE}.]

The class of languages decidable by a deterministic polynomial-space Turing
machine.
A polynomial-space Turing machine visits at most a polynomial number of
cells on its tape before halting.

\item[$\cls{EXP}$, page \pageref{EXP}.]

The class of languages decidable by a deterministic exponential-time Turing
machine.

\item[$\cls{NEXP}$, page \pageref{NEXP}.]

Same as $\cls{NP}$ except that the ``proof'' strings may have exponential size.

\item[$\cls{coNEXP}$, page \pageref{coNEXP}.]

The class of languages whose complements are in $\cls{NEXP}$.

\item[$\cls{IP}(c,s)$, page \pageref{IP}.]

The class of languages that have interactive proof systems with completeness
error $c$ and soundness error $s$.

\item[$\cls{RG}(c,s)$, page \pageref{RG}.]

The class of languages that have refereed games with completeness error $c$ and
soundness error $s$.

\item[$\cls{RG}_1(c,s)$, page \pageref{RG1}.]

The class of languages that have one-round refereed games with completeness
error $c$ and soundness error $s$.
A round consists of a message from the verifier to each of the provers in
parallel followed by their responses.

\item[$\cls{BQP}$, page \pageref{BQP}.]

The class of languages decidable by a polynomial-time uniform family of quantum
circuits.
Widely considered to be the quantum analogue of $\cls{P}$.

\item[$\cls{QIP}(c,s)$, page \pageref{QIP}.]

The class of languages that have quantum interactive proof systems with
completeness error $c$ and soundness error $s$.

\item[$\cls{QRG}(c,s)$, page \pageref{QRG}.]

The class of languages that have quantum refereed games with completeness error
$c$ and soundness error $s$.

\item[$\cls{SQG}(c,s)$, page \pageref{SQG}.]

The class of languages that have short quantum games with completeness error $c$
and soundness error $s$.
A short quantum game is a one-round quantum refereed game in which the verifier
may process the yes-prover's response before sending a message to the no-prover.

\item[$\cls{SQG}_*(c,s)$, page \pageref{SQGstar}.]

Same as $\cls{SQG}(c,s)$ except that the verifier cannot send a message to the
yes-prover.

\item[$\cls{DQIP}(c,s)$, page \pageref{DQIP}.]

The class of languages that have double quantum interactive proof systems with
completeness error $c$ and soundness error $s$.
A double quantum interactive proof system is a quantum refereed game in which
the verifier exchanges several messages with only the yes-prover and then
several messages with only the no-prover.

\end{description}

%% file: intro.tex
\chapter{Introduction}
\label{ch:intro}

In this thesis we define a new model of computation that incorporates existing
models based upon the notions of competitive interaction and quantum
information.
We focus on a variant of this new model with a restricted protocol for
interaction and we prove lower and upper bounds on the power of this restricted
model.
In the process of proving these bounds, we develop new computational and
information-theoretic tools that may prove useful in other fields in computer
science and physics.

This introductory chapter starts with an informal overview of our results in
Section \ref{sec:intro:overview}.
We offer a survey of relevant topics from complexity theory in Section
\ref{sec:intro:complexity} and a review of quantum information in Section
\ref{sec:intro:quantum}.
Chapter \ref{ch:defs} formalizes the fundamental concepts discussed in this
thesis---the interested reader can find a more precise statement of the
contributions of this thesis in Section \ref{subsec:defs:contributions} on page
\pageref{subsec:defs:contributions}.
Chapters \ref{ch:QIPinSQG} and \ref{ch:SQGinEXP} are devoted to proving the
results stated in that section.
We conclude with Chapter \ref{ch:end}, which mentions some open questions and
possible directions for future research.

\section{Overview}
\label{sec:intro:overview}

It is intended that this first section provide the reader with a bird's eye view
of the direction in which this thesis is headed.
For the sake of clarity, citations, technical detail, and an adequate history
are absent.
These gaps will be addressed in subsequent sections and in Chapter \ref{ch:defs}
as we cover the necessary background material in greater detail.
On that note, we begin.

Given a new model of computation, an effective way to quantify the expressive
power of that model is to compare it to other more fundamental models.
These comparisons derive meaning from the fact that fundamental models of
computation often capture important notions such as efficient computation in the
physical world or the difficulty of solving certain computational problems.
For example, it is widely believed that any problem that can be solved
efficiently by a device built based upon Newtonian physics (such as a desktop
computer) can also be solved by a randomized polynomial-time Turing machine and
vice versa.

An \emph{interactive proof system} is a more exotic model of computation in
which a randomized polynomial-time Turing machine (a \emph{verifier}) is endowed
with the ability to interact with an entity who is computationally unbounded but
not necessarily trustworthy (a \emph{prover}).
Given an input string $x$, the prover uses his unlimited computational power to
attempt to convince the verifier to accept $x$, while the verifier tries to
determine the validity of the prover's argument.
At the end of the interaction, the verifier accepts $x$ if he believes the
prover and rejects $x$ if he does not.

A given set $L$ of strings (a \emph{language}) is said to have an
interactive proof system if there exists a verifier $V$ satisfying the following
standard completeness and soundness conditions:
\begin{description}

\item[Completeness.]

There exists a prover $P$ that can convince $V$ to accept any string $x \in L$
with high probability.

\item[Soundness.]

No prover can convince $V$ to accept any string $x \not \in L$ except with small
probability.

\end{description}

A \emph{refereed game} is another model of computation that generalizes
interactive proof systems in that the verifier interacts with not just one, but
two provers.
In this model, the provers use their unlimited computational power to compete
with each other: one prover (the \emph{yes-prover}) attempts to convince the
verifier to accept $x$, while the other prover (the \emph{no-prover}) attempts
to convince the verifier to reject $x$.
At the end of the interaction, the verifier decides whether to accept or reject
the input $x$, effectively deciding which of the provers wins the game.

Similar to interactive proof systems, a language $L$ is said to have a refereed
game if there exists a verifier $V$ satisfying the following slightly modified
completeness and soundness conditions:
\begin{description}

\item[Completeness.]

There exists a yes-prover $Y$ that can convince $V$ to accept any string
$x \in L$ with high probability, regardless of the no-prover.

\item[Soundness.]

There exists a no-prover $N$ that can convince $V$ to reject any string
$x \not \in L$ with high probability, regardless of the yes-prover.

\end{description}

At a glance, these two models of computation may seem obscure and uninteresting.
However, it can be shown that they characterize two very fundamental models of
computation in the following sense:
\begin{itemize}
\item
A language has an interactive proof system if and only if it can be decided by a
polynomial-space Turing machine.
\item
A language has a refereed game if and only if it can be decided by a
deterministic exponential-time Turing machine.
\end{itemize}
These surprising characterizations bring to light a very deep connection between
exotic interaction-based models and space- and time-bounded Turing machines.

Given the recent proliferation of the theory of quantum computation, it is
natural to consider the possible effects of quantum computers on models such as
interactive proof systems and refereed games.
How does the power of these models change if the verifier and provers are
permitted to perform quantum computations and exchange quantum messages?

Perhaps the most striking contrast between the quantum and nonquantum
(\emph{classical}) cases is that any quantum interactive proof system can be
simulated by another quantum interactive proof system in which the verifier and
prover exchange only three messages.
With classical interactive proof systems, all evidence suggests that such a
simulation does not exist, be it with three or any other fixed number of
messages.
What is it about quantum information that permits this strange shortening of
interactions?

In this thesis, we turn our attention to quantum refereed games, which have not
been previously studied.
We focus on a restricted variant of the quantum refereed game model that we call
\emph{short quantum games} and we prove both a lower bound and an upper bound on
the expressive power of these games.

For the lower bound, we prove that any language having a quantum interactive
proof system also has a quantum refereed game with the following protocol: the
yes-prover sends a quantum state to the verifier, who then processes this state
and forwards it to the no-prover.
The no prover performs a quantum measurement on this state and sends a single
classical bit of information (either a 0 or a 1) back to the verifier, who
finally accepts or rejects based upon this bit.
In order to prove the correctness of our short quantum game, we establish
the existence of a quantum measurement that reliably distinguishes between
quantum states chosen from two disjoint convex sets.

For the upper bound, we consider a slightly looser restriction of the quantum
refereed game model in which the verifier exchanges several quantum messages
with only the yes-prover and then exchanges several more quantum messages with
only the no-prover before deciding whether to accept or reject.
We show that any language having a quantum refereed game obeying this protocol
can be decided by a deterministic exponential-time Turing machine.
Our proof uses the ellipsoid method, which is a polynomial-time algorithm that
determines the emptiness of a convex set given implicitly by a separation
oracle.

\section{Complexity Theory}
\label{sec:intro:complexity}

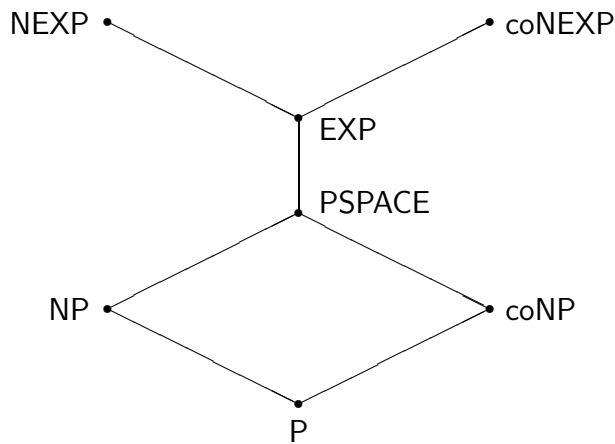
\begin{figure}
\begin{center}
\input{introcls.latex}
\end{center}
\caption{Relationships among fundamental complexity classes}
\label{fig:introcls}
\end{figure}

In this section we review in greater detail certain concepts from complexity
theory that lead to the notion of competitive interaction as a model of
computation and we survey known results pertaining to that model.
We assume at the onset that the reader is familiar with fundamental notions
such as languages, Turing machines, polynomial-time computability, completeness,
and the following fundamental complexity classes:
\begin{itemize}

\item
The polynomial-time classes $\cls{P}$\label{P}, $\cls{NP}$, and
$\cls{coNP}$\label{coNP};

\item
The exponential-time classes $\cls{EXP}$\label{EXP}, $\cls{NEXP}$\label{NEXP},
and $\cls{coNEXP}$\label{coNEXP}; and

\item
The polynomial-space class $\cls{PSPACE}$\label{PSPACE}.

\end{itemize}
Figure \ref{fig:introcls} depicts the known relationships among these complexity
classes.
In that figure, class $\cls{A}$ contains class $\cls{B}$ if $\cls{A}$
can be reached from $\cls{B}$ by following a path of only upwardly sloped edges.

We now establish some relevant notation that will be used throughout this
thesis.
We let $\mathit{poly}$ denote the set of polynomial-time computable functions
$f : \mathbb{N} \to \mathbb{N} \setminus \set{0}$ for which there exists a
polynomial $p$ such that $f(n) \leq p(n)$ for all $n \in \mathbb{N}$.
The sets $2^{-\mathit{poly}}$ and $\mathit{poly}^{-1}$ are derived from
$\mathit{poly}$ as follows.
A polynomial-time computable function $\varepsilon : \mathbb{N} \to [0,1]$
is in $2^{-\mathit{poly}}$ if there exists $f \in \mathit{poly}$ such that
$\varepsilon(n) = 2^{-f(n)}$ for all $n \in \mathbb{N}$.
Similarly, $\varepsilon$ is in $\mathit{poly}^{-1}$ if there exists
$f \in \mathit{poly}$ such that
$\varepsilon(n) = \frac{1}{f(n)}$ for all $n \in \mathbb{N}$.

All strings are over the binary alphabet $\set{0,1}$.
We let $\set{0,1}^n$ denote the set of strings of length $n$ and we let
$\set{0,1}^*$ denote the set of all finite strings.
For any $x \in \set{0,1}^*$, $|x|$ denotes the length of $x$.

\subsection{Nondeterminism}

The notion of nondeterminism and the discovery in the 1970's of
$\cls{NP}$-complete problems \cite{Cook71, Karp72, Levin73} drew unprecedented
attention to the field of computational complexity theory.
Since then, several characterizations of $\cls{NP}$ have been found
\cite{Fagin74, AroraS98, AroraL+98}
and several generalizations of nondeterminism have been explored
\cite{Stockmeyer76, ChandraK+81, Babai85, GoldwasserM+89}.
One of the simpler characterizations views nondeterministic computation as
deterministic verification of a short proof.
Specifically, a language $L \subseteq \set{0,1}^*$ is in $\cls{NP}$\label{NP} if
and only if there exists $p \in \mathit{poly}$ and a deterministic
polynomial-time Turing machine $M$ such that, for all input strings
$x \in \set{0,1}^*$:
\begin{itemize}
\item
If $x \in L$ then there exists $y \in \set{0,1}^{p(|x|)}$ such that $M$ accepts
the pair $(x,y)$.
\item
If $x \not \in L$ then $M$ rejects the pair $(x,y)$ for all
$y \in \set{0,1}^{p(|x|)}$.
\end{itemize}
The Turing machine $M$ in this characterization can be viewed as a
\emph{verifier}.
The string $y$ submitted to $M$ can be viewed as a proof of the claim,
``$x$ is in $L$.''
Informally, the conditions of this characterization tell us that every $x \in L$
has a proof $y$ that can be used by the verifier to verify this claim in
deterministic polynomial-time.
Moreover, if $x \not \in L$ then no proof could possibly convince the verifier
otherwise.

\subsection{Interaction}
\label{subsec:intro:complexity:interaction}

The notion of interaction was introduced in 1985 by Babai \cite{Babai85} and by
Goldwasser, Micali, and Rackoff \cite{GoldwasserM+89} as a generalization of
nondeterminism that extends the verifier-proof analogy by allowing a two-way
dialogue between the verifier and the mysterious supplier-of-proofs.
Specifically, a \emph{prover} is an entity with unlimited computational power
whose goal is always to convince the verifier to accept the input string $x$.
The verifier may ask questions of the prover, perform randomized polynomial-time
computations, and ask additional questions of the prover based upon these
computations and upon answers to previous questions.
At some point the verifier must end the interaction and decide whether or not to
accept $x$.

Such an interaction is called an \emph{interactive proof system}.
It follows from the fact that the verifier is restricted to polynomial-time
computation that the amount and size of the messages exchanged between the
verifier and prover must be polynomial in $|x|$.
Because the verifier may also invoke randomization, it is plausible that his
decision to accept or reject $x$ could differ between independent executions of
the same interaction.
Hence, an allowance is made for unlucky coin tosses that cause the verifier to
erroneously accept or reject $x$ with some small probability.

In this thesis we pay considerable attention to the error probability associated
with different forms of interaction.
In particular, we are interested in any possible differences between the
probability of a false negative (\emph{completeness}) and of a false positive
(\emph{soundness}).
We also consider a more general case in which these probabilities might even
vary as a function of the input length $|x|$.
However, in order to prevent the polynomial-time verifier from accessing
hard-to-compute values encoded in the error probability, we restrict our
attention to error probabilities that are polynomial-time computable.

All these ideas are formalized in the following definition.
For any polynomial-time computable functions $c,s : \mathbb{N} \to [0,1]$,
a language $L \subseteq \set{0,1}^*$ is said to have an interactive proof system
if there exists a randomized polynomial-time verifier $V$ that satisfies the
following completeness and soundness conditions for all input strings
$x \in \set{0,1}^*$:
\begin{description}
\item[Completeness.]
If $x \in L$ then there exists a prover $P$ that convinces $V$ to accept $x$
with probability at least $1 - c(|x|)$.
\item[Soundness.]
If $x \not \in L$ then no prover can convince $V$ to accept $x$ with probability
greater than $s(|x|)$.
\end{description}
The functions $c$ and $s$ are called the \emph{completeness error} and
\emph{soundness error} respectively.
Finally, we let $\cls{IP}(c,s)$\label{IP} denote the complexity class of
languages that have interactive proof systems with completeness error $c$ and
soundness error $s$.

\subsection{Competitive Interaction}

Many different models resembling competitive interaction have been
studied in the context of game theory since the 1950's, but competitive
interaction was not considered as a generalization of interactive proof systems
until 1990 \cite{FeigeS+90}.
In this generalization, the verifier interacts with not one, but two provers
with unlimited computational power.
As mentioned in Section \ref{sec:intro:overview}, these two provers use their
power to compete with each other: one prover, called the \emph{yes-prover},
attempts to convince the verifier to accept the input string $x$, while the
other prover, called the \emph{no-prover}, attempts to convince the verifier to
reject $x$.

As before, the verifier may perform randomized polynomial-time computations.
He may also ask questions of each of the provers and base future questions upon
randomized computations and answers to previous questions.
At the end of the interaction, the verifier decides whether or not to accept
$x$.
Such an interaction is called a \emph{refereed game} because it can be viewed as
a game between the two provers in which the verifier acts as a referee by
ensuring that the provers obey the rules of the game and by announcing a winner
at the end.

More formally, for any polynomial-time computable functions
$c,s : \mathbb{N} \to [0,1]$, a language
$L \subseteq \set{0,1}^*$ is said to have a refereed game if there
exists a randomized polynomial-time verifier $V$ that satisfies the following
completeness and soundness conditions for all input strings $x \in \set{0,1}^*$:
\begin{description}
\item[Completeness.]
If $x \in L$ then there exists a yes-prover $Y$ that convinces $V$ to accept $x$
with probability at least $1 - c(|x|)$, regardless of the no-prover's strategy.
\item[Soundness.]
If $x \not \in L$ then there exists a no-prover $N$ that convinces $V$ to reject
$x$ with probability at least $1 - s(|x|)$, regardless of the yes-prover's
strategy.
\end{description}
As with interactive proof systems, the functions $c$ and $s$ are the
completeness error and soundness error respectively.
Finally, we let $\cls{RG}(c,s)$\label{RG} denote the complexity class of
languages that have refereed games with completeness error $c$ and soundness
error $s$.

It is also of interest to consider refereed games in which the verifier
exchanges just one round of messages with the provers.
In particular, these \emph{one-round} refereed games obey the following
protocol: a message from the verifier to each of the provers, followed by their
responses to the referee, followed by the referee's decision.
We let $\cls{RG}_1(c,s)$\label{RG1} denote the complexity class of languages
that have one-round refereed games with completeness error $c$ and soundness
error $s$.

\subsection{Reasonable Error}
\label{subsec:intro:complexity:error}


In Section \ref{subsec:intro:complexity:interaction} we prevented the verifier
from accessing hard-to-compute error probabilities by requiring that they be
polynomial-time computable.
But even interactions with polynomial-time computable error probabilities can
still have undesirable properties.

Suppose, for example, that the probability of correctly accepting or rejecting
an input $x \in \set{0,1}^*$ is exponentially close to $\frac{1}{2}$ in $|x|$.
It is clear that such an interaction can be simulated with exponential accuracy
by a verifier who ignores all provers and accepts or rejects based solely upon
the result of a coin flip.
Of course, an interaction with this property is not very interesting.

With this example in mind, we say that polynomial-time computable functions
$c,s : \mathbb{N} \to [0,1]$ are \emph{reasonable} if there exists
$\gamma \in \mathit{poly}^{-1}$ such that
$$1 - c(n) - s(n) \geq \gamma(n)$$
for all $n \in \mathbb{N}$.
As $c$ and $s$ are polynomial-time computable, the verifier can compute them and
bias his final decision so that the completeness and soundness error of his
biased decision are both bounded below $\frac{1}{2}$ by at least an inverse
polynomial as desired.

Many (but not all) results concerning interaction are known to hold only when
the functions $c$ and $s$ are reasonable.
We mention the reasonability condition explicitly whenever it is required.

\subsection{Known Results}
\label{subsec:intro:complexity:results}

Several inclusions follow immediately from the definitions of $\cls{IP}(c,s)$
and $\cls{RG}(c,s)$.
For example, because an interactive proof system is just a refereed game without
a no-prover, it is clear that $\cls{IP}(c,s) \subseteq \cls{RG}(c,s)$.

Also, these classes are easily seen to be \emph{robust} with respect to error in
the sense that any interaction with reasonable error can be simulated by another
interaction with exponentially small error.
More formally, for every $\varepsilon \in 2^{-\poly}$ and every reasonable
$c,s : \mathbb{N} \to [0,1]$ we have
$$
\cls{IP}(c,s) \subseteq \cls{IP}(\varepsilon,\varepsilon)
\qquad \textrm{and} \qquad
\cls{RG}(c,s) \subseteq \cls{RG}(\varepsilon,\varepsilon).
$$
To prove these containments it suffices to note that any interaction can
be repeated independently many times in succession.
If the verifier for such a repeated interaction bases his decision upon a
majority vote of the outcomes of each of the repetitions then it follows from
Chernoff bounds that the error of the repeated interaction decreases
exponentially in the number of repetitions.

Of course, sequential repetition necessarily increases the number of rounds in
an interaction and so this simple error reduction technique does not apply to
bounded-round interactions such as one-round refereed games.
Fortunately, as we shall soon see, $\cls{RG}_1(c,s)$ can still be shown to be
robust with respect to error.

An even stronger robustness result is known to hold for $\cls{IP}(c,s)$.
In particular, an interactive proof system with reasonable error can be
simulated by another interactive proof system with zero completeness error
\cite{Schoning89, BalcazarD+90}.
It follows that
$$\cls{IP}(c,s) \subseteq \cls{IP}(0,\varepsilon)$$
for every $\varepsilon \in 2^{-\poly}$ and every reasonable
$c,s : \mathbb{N} \to [0,1]$.
By contrast, it is not known whether zero error
(completeness or soundness) can be achieved for refereed games.
In light of these robustness results, we define the following shorthand
notations:
\begin{itemize}

\item
$\cls{IP}$ is the complexity class of all languages
$L \subseteq \set{0,1}^*$ such that
$L \in \cls{IP}(0,\varepsilon)$ for every
$\varepsilon \in 2^{-\poly}$.

\item
$\cls{RG}$ is the complexity class of all languages
$L \subseteq \set{0,1}^*$ such that
$L \in \cls{RG}(\varepsilon,\varepsilon)$ for every
$\varepsilon \in 2^{-\poly}$.

\item
$\cls{RG}_1$ is the complexity class of all languages
$L \subseteq \set{0,1}^*$ such that
$L \in \cls{RG}_1(\varepsilon,\varepsilon)$ for every
$\varepsilon \in 2^{-\poly}$.

\end{itemize}

The interactive proof system model has a rich history.
However, a comprehensive survey of that history is tangential to the scope of
this thesis and so we summarize only the results most relevant to our purpose.

Because interaction is a generalization of nondeterminism, it is clear that
$\cls{NP} \subseteq \cls{IP}$.
The full extent of the power of interaction was not fully known until 1990
when Lund, Fortnow, Karloff, and Nisan developed the arithmetization technique
\cite{LundF+92} that was used in Reference \cite{Shamir92} to show
$$\cls{IP} = \cls{PSPACE}$$
(see also Reference \cite{Shen92}).
This surprising characterization of polynomial-space computation has
prompted further study of models of computation based upon interaction, one
example of which is refereed games.

The refereed game model and several variations thereof were studied in
References \cite{Reif84,FeigeS+90,FeigeS92,KollerM92,FeigenbaumK+95} among
others.
Many results concerning refereed games can be gleaned from results in game
theory.
For example, in game theoretic terms, refereed games correspond to two-person
games of \emph{incomplete information} because messages exchanged between one
prover and the verifier are kept secret from the other prover.
In 1992, Koller and Megiddo \cite{KollerM92} gave a deterministic algorithm that
solves two-player games of incomplete information in time polynomial in the size
of an induced structure known as a \emph{game tree}.
The game tree induced by a refereed game on input
$x \in \set{0,1}^*$ is easily shown to have size at most exponential in $|x|$,
from which it follows that $\cls{RG} \subseteq \cls{EXP}$.

Feige and Kilian \cite{FeigeK97} used a variant of arithmetization to prove the
reverse inclusion, implying
$$\cls{RG} = \cls{EXP}.$$
They also proved the inclusions $\cls{RG}_1(c,s) \subseteq \cls{PSPACE}$ and
$\cls{PSPACE} \subseteq \cls{RG}_1(\varepsilon,\varepsilon)$ for every
$\varepsilon \in 2^{-\poly}$ and every reasonable $c,s : \mathbb{N} \to [0,1]$.
These inclusions imply the aforementioned robustness of one-round refereed games
as well as the characterization
$$\cls{RG}_1 = \cls{PSPACE}.$$

\section{Quantum Information and Computation}
\label{sec:intro:quantum}

In this section we describe the framework of quantum information.
Our description is not intended to be comprehensive, but merely to refresh the
reader with the aspects of quantum information that are relevant to this thesis
and to establish notation.

We assume familiarity with fundamental concepts from linear algebra such as
complex numbers, vectors, vector spaces, matrices, and basic
matrix-related concepts such as matrix multiplication and the trace of a matrix.
Of course, familiarity with quantum information and quantum circuits is an
asset.

\subsection{Linear Algebra}
\label{subsec:intro:quantum:linalg}

For any positive integer $M$, elements of the vector space $\mathbb{C}^M$ are
identified with $M$-dimensional column vectors in the usual way and are denoted
by lowercase Roman letters such as $u,v,w$, etc.
For any vector $v \in \mathbb{C}^M$, the \emph{conjugate-transpose} of $v$ is
denoted $v^*$, which is an $M$-dimensional row vector.
For any vectors $v,w \in \mathbb{C}^M$, the \emph{standard inner product}
between $v$ and $w$, denoted $\inner{v,w}$, is given by
$$\Inner{v,w} = v^*w.$$
The norm induced on $\mathbb{C}^M$ by the standard inner product is the
\emph{Euclidean norm}.
For $v \in \mathbb{C}^M$, this norm is denoted $\norm{v}$ and given by
$$\Norm{v} = \sqrt{\Inner{v,v}}.$$

The vector space $\mathbb{C}^M$ endowed with the standard inner product is a
\emph{Hilbert space}.
Moreover, every Hilbert space in this thesis is assumed to
take this form for some positive integer $M$.
Hilbert spaces are denoted by uppercase script letters such as $\hilb{F},
\hilb{G}, \hilb{H}$, etc.

We note at this point that it is a common practice in quantum information to use
Dirac notation to describe column and row vectors.
For example, the vector $v \in \hilb{H}$ would be denoted $\ket{v}$, the
corresponding row vector $v^*$ would be denoted $\bra{v}$, and the inner product
$\inner{v,w}$ would be denoted $\inner{v|w}$ using this notation.
However, Dirac notation is found to be cumbersome for our purposes and so we
break from convention by restricting its use in the following manner:
Dirac notation is used in this thesis only to describe the vector
$\ket{0_\hilb{H}} \in \hilb{H}$, which always denotes the first element in the
standard orthonormal basis for $\hilb{H}$.
In other words, the vector $\ket{0_\hilb{H}}$ always denotes the vector with a 1
in the first entry an all other entries equal to zero.
We sometimes write $\ket{0}$ when the Hilbert space $\hilb{H}$ is clear from the
context.

For any Hilbert spaces $\hilb{F}$ and $\hilb{G}$ of dimensions $M$ and $N$
respectively, we let $\mapset{L}{\hilb{F},\hilb{G}}$ denote the set of linear
mappings from $\hilb{F}$ to $\hilb{G}$.
Elements of $\mapset{L}{\hilb{F},\hilb{G}}$ are identified with $N \times M$
matrices in the usual way (with respect to the standard bases for $\hilb{F}$ and
$\hilb{G}$) and are denoted by uppercase Roman letters such as $A,B,C$, etc.
For any matrix $A \in \mapset{L}{\hilb{F}, \hilb{G}}$ we let
$A[i,j] \in \mathbb{C}$ denote the $[i,j]$ entry of $A$.
The \emph{spectral norm}
of $A$, denoted $\norm{A}$, is given by
$$\Norm{A} = \sup_{v \in \hilb{F} \setminus \Set{0}} \frac{\Norm{Av}}
{\Norm{v}}.$$
As with vectors, the \emph{conjugate-transpose} of $A$ is denoted $A^*$, which
is an element of $\mapset{L}{\hilb{G},\hilb{F}}$.
As a natural extension of the standard inner product for vectors, the
\emph{Hilbert-Schmidt inner product} between any pair of matrices
$A,B \in \mapset{L}{\hilb{F},\hilb{G}}$, denoted $\inner{A,B}$, is given by
$$\Inner{A,B} = \ptr{}{A^* B}.$$

Two Hilbert spaces $\hilb{F}_1, \hilb{F}_2$ of dimensions $M_1, M_2$ can be
combined via the \emph{Kronecker product} to form a larger Hilbert space
$\hilb{F}_1 \otimes \hilb{F}_2$ of dimension $M_1 M_2$.
The Kronecker product is also defined on vectors and matrices so that for $v_1
\in \hilb{F}_1, v_2 \in \hilb{F}_2$ we have
$$v_1 \otimes v_2 \in \hilb{F}_1 \otimes \hilb{F}_2$$
and for $A_1 \in \mapset{L}{\hilb{F}_1, \hilb{G}_1}, A_2 \in \mapset{L}{\hilb{F}
_2, \hilb{G}_2}$ we have
$$A_1 \otimes A_2 \in \mapset{L}{\hilb{F}_1 \otimes \hilb{F}_1, \hilb{G}_1
\otimes \hilb{G}_2}.$$
The Kronecker product satisfies many convenient and intuitive properties that we
do not list here.

We write $\mapset{L}{\hilb{F}}$ as a shorthand notation for
$\mapset{L}{\hilb{F},\hilb{F}}$ and we say that a matrix
$A \in \mapset{L}{\hilb{F}}$ \emph{acts on} $\hilb{F}$.
We let $I_\hilb{F} \in \mapset{L}{\hilb{F}}$ denote the identity matrix acting
on $\hilb{F}$, which we often abbreviate to $I$ when the Hilbert space $\hilb{F}
$ is clear from the context.
Often in this thesis we multiply matrices acting on a certain Hilbert space with
matrices or vectors from a larger Hilbert space.
In these cases we implicitly assume that the smaller matrix is extended to the
larger Hilbert space by taking the Kronecker product with the identity.
For example, if $A \in \mapset{L}{\hilb{F}}$,
$B \in \mapset{L}{\hilb{F} \otimes \hilb{G}}$, and
$v \in \hilb{F} \otimes \hilb{G}$ then $AB$ and $Av$ always mean
$(A \otimes I_\hilb{G})B$ and $(A \otimes I_\hilb{G})v$ respectively.

The \emph{partial trace} is a linear mapping
$$\tr_\hilb{G} : \mapset{L}{ \hilb{F} \otimes \hilb{G} } \to
\mapset{L}{ \hilb{F} }$$
defined for all $A \in \mapset{L}{\hilb{F}}, B \in \mapset{L}{\hilb{G}}$ as
$$\ptr{\hilb{G}}{A \otimes B} = \ptr{}{B} A$$
and extending to all of $\mapset{L}{\hilb{F} \otimes \hilb{G}}$ by
linearity.
The partial trace is in some sense complimentary to the Kronecker product in
that the Kronecker product combines two matrices acting on separate Hilbert
spaces to form one matrix acting on one larger Hilbert space.
In contrast, the partial trace takes as input a matrix acting on a larger
Hilbert space and produces a matrix that acts on a smaller Hilbert space.

A matrix $A \in \mapset{L}{\hilb{F}}$ is \emph{unitary} if $A^*A=I$,
\emph{Hermitian} if $A = A^*$, and \emph{positive semidefinite} if $v^*Av$ is a
nonnegative real number for every $v \in \hilb{F}$.
We define the following subsets of $\mapset{L}{\hilb{F}}$:
\begin{itemize}

\item
The set $\mapset{U}{\hilb{F}} \subset \mapset{L}{\hilb{F}}$ contains all unitary
matrices in $\mapset{L}{\hilb{F}}$.

\item
The set $\mapset{H}{\hilb{F}} \subset \mapset{L}{\hilb{F}}$ contains all
Hermitian matrices in $\mapset{L}{\hilb{F}}$.

\item
The set $\mapset{Pos}{\hilb{F}} \subset \mapset{H}{\hilb{F}}$ contains all
positive semidefinite matrices in $\mapset{L}{\hilb{F}}$.

\item
The set $\mapset{D}{\hilb{F}} \subset \mapset{Pos}{\hilb{F}}$ contains all
positive semidefinite matrices $A$ with $\ptr{}{A} = 1$.

\end{itemize}
Elements of $\mapset{D}{\hilb{F}}$ are called \emph{density matrices} and are
typically denoted by lowercase Greek letters such as $\rho, \xi$, etc.

\subsection{Qubits}
\label{subsec:intro:quantum:qubits}

A \emph{qubit} is a fundamental unit of quantum information described as
follows.
Any collection of $m$ qubits has a corresponding Hilbert space $\hilb{F}$ of
dimension $2^m$.
Any density matrix $\rho \in \mapset{D}{\hilb{F}}$ completely describes some
\emph{state} of those $m$ qubits.
Furthermore, any physically realizable state of those $m$ qubits is uniquely
described by some density matrix, so it makes sense to refer to any such
$\rho$ as a ``state'' of those $m$ qubits.

If $\rho \in \mapset{D}{\hilb{F}}$ is given by $\rho = vv^*$ for some vector $v
\in \hilb{F}$ then $\rho$ is called a \emph{pure state}.
It must be the case that $\norm{v}=1$ (that is, $v$ is a \emph{unit vector}) and
$\rho$ is completely described by $v$, so it makes sense to refer to any unit
vector $v$ as a ``pure state'' of those $m$ qubits.
Note that the unit vector $v \in \hilb{F}$ describing a pure state $vv^* \in
\mapset{D}{\hilb{F}}$ is not unique because the vector $u = \omega v$ satisfies
$vv^* = uu^*$ for any $\omega \in \mathbb{C}$ with $\abs{\omega} = 1$.

Let $\hilb{G}$ be a Hilbert space of dimension $2^n$ corresponding to a
collection of $n$ qubits.
Then the Hilbert space corresponding to the combined collection of $m+n$ qubits
is $\hilb{F} \otimes \hilb{G}$ and has dimension $2^{m+n}$.
If $\rho \in \mapset{D}{\hilb{F} \otimes \hilb{G}}$ is any state of
those $m+n$ qubits then $\ptr{\hilb{G}}{\rho} \in \mapset{D}{\hilb{F}}$ always
describes the state of the first $m$ qubits and
$\ptr{\hilb{F}}{\rho} \in \mapset{D}{\hilb{G}}$ always describes the state of
the remaining $n$ qubits.
Although it is the case that
$\rho = \ptr{\hilb{G}}{\rho} \otimes \ptr{\hilb{F}}{\rho}$ whenever
$\ptr{\hilb{G}}{\rho}$ or $\ptr{\hilb{F}}{\rho}$ is a pure state, this equality
does not hold for every $\rho \in \mapset{D}{\hilb{F} \otimes \hilb{G}}$.

\subsection{Quantum Circuits}
\label{subsec:intro:quantum:circuits}

The model of quantum computation that provides a foundation for quantum
interaction is the \emph{quantum circuit model}.
All quantum circuits in this thesis are assumed to be composed of a finite
number of \emph{quantum gates}, each of which is chosen from some finite
\emph{universal set} of quantum gates.
We do not discuss the details of quantum gates or universality, as those details
are not required to understand the material presented in this thesis.
In lieu of such a discussion, we refer the interested reader to References
\cite{AharonovK+98, Berthiaume97, Kitaev97, Yao93}.

For any quantum circuit $Q$ acting on $m$ qubits with corresponding Hilbert
space $\hilb{F}$, there is a unitary matrix $U \in \mapset{U}{\hilb{F}}$
associated with $Q$.
This matrix models the action of $Q$ upon its $m$ input qubits in the state
$\rho \in \mapset{D}{\hilb{F}}$ so that the state of those $m$ qubits after $Q$
is applied becomes $U \rho U^* \in \mapset{D}{\hilb{F}}$.
This formalism extends without complication to any positive semidefinite matrix
$\rho \in \mapset{Pos}{\hilb{F}}$.
If $\rho = vv^*$ for some pure state $v \in \hilb{F}$ then the resulting state
is the pure state $Uv \in \hilb{F}$ and this formalism extends without
complication to any nonzero vector $v \in \hilb{F}$.

Two additional facts concerning quantum circuits warrant attention.
First, it is important to note that we lose no generality by allowing only
unitary quantum circuits because any physically realizable quantum process can
be simulated by unitary circuits as described in Reference \cite{AharonovK+98}.
We elaborate on this simulation in Section
\ref{subsec:QIPinSQG:CI:MixedCircuits}.

Second, the universality condition placed on our set of admissible quantum gates
implies the following fact: for any unitary matrix $V \in \mapset{U}{\hilb{F}}$
and any real $\varepsilon > 0$, there is a quantum circuit with associated
unitary matrix $U \in \mapset{U}{\hilb{F}}$ that satisfies
$$\norm{U - V} < \varepsilon.$$
Although the spectral norm of the difference between $U$ and $V$ is not always a
convenient measure of the ``distance'' between two matrices, in this case it
allows us a simple way to infer that any desired unitary matrix $V$ can be
approximated as closely as desired by quantum circuits considered in this
thesis.

\subsection{Measurement}
\label{subsec:intro:quantum:measurement}

So far, we have discussed qubits and quantum circuits that act on qubits.
It is now time to discuss \emph{measurements}, which allow us to extract
classical information from qubits and enable us to solve real-world problems
using quantum information.

Quantum measurements have several formalizations, each differing in their
simplicity and generality.
Although knowledge of only the most basic notion of measurement is required
throughout most of this thesis, the results of Chapter \ref{ch:QIPinSQG} make
use of the extended generality offered by more complex formalizations.
Hence, we introduce in this subsection the most general form of quantum
measurement, since the added complication of this form is insignificant anyway.

Let $\hilb{F}$ be a Hilbert space corresponding to some collection of qubits and
let $\Gamma$ be a finite set of \emph{outcomes}.
A \emph{quantum measurement} of those qubits with outcomes in $\Gamma$ is
defined by a set of matrices
$$\Set{A_\tau : \tau \in \Gamma} \subset \mapset{L}{\hilb{F}}$$
satisfying
$$\sum_{\tau \in \Gamma} A_\tau^* A_\tau = I_\hilb{F}.$$
When such a measurement is performed on qubits in some state $\rho \in
\mapset{D}{\hilb{F}}$, the outcome of the measurement is $\tau$ with probability
$\ptr{}{A_\tau \rho A_\tau^*}$ for each $\tau \in \Gamma$.
Conditioned on the outcome $\tau$, the state of the qubits becomes
$$\frac{ A_\tau \rho A_\tau^* }{ \ptr{}{A_\tau \rho A_\tau^*} } \in \mapset{D}
{\hilb{F}}$$
once the measurement is complete.
If $\rho = vv^*$ for some pure state $v \in \hilb{F}$ then the outcome of the
measurement is $\tau$ with probability $\norm{A_\tau v}^2$ and the resulting
state is the pure state
$$\frac{A_\tau v}{\Norm{A_\tau v}} \in \hilb{F}.$$

Often, we do not care about the state of the qubits once the measurement is
complete.
Because the probability of outcome $\tau$ is
$$\Norm{A_\tau v}^2 = v^* A_\tau^* A_\tau v$$
for pure states and
$$\ptr{}{A_\tau \rho A_\tau^*} = \ptr{}{A_\tau^* A_\tau \rho}
= \inner{A_\tau^* A_\tau, \rho }$$
for general states, it follows that the quantum measurement is completely
specified in this case by the set
$$\Set{ E_\tau : \tau \in \Gamma } \subset \mapset{Pos}{\hilb{F}}$$
of positive semidefinite matrices defined by $E_\tau = A_\tau^* A_\tau$ for each
$\tau \in \Gamma$ and hence satisfying
$$\sum_{\tau \in \Gamma} E_\tau = I_\hilb{F}.$$
Any measurement expressed in this way is called a \emph{positive operator-valued
measurement} (POVM) for historical reasons.

It is often convenient to specify only a POVM with the understanding that we do
not care about the state of the qubits once the measurement is complete---this
is the formalism of quantum measurements used in Chapter \ref{ch:QIPinSQG}.
In all other chapters, the quantum measurements we discuss have outcomes in
$\Gamma = \set{ \mathrm{accept}, \mathrm{reject} }$ and take the following
form.
For any Hilbert space $\hilb{F}$ corresponding to $m$ qubits, one of those
qubits is designated as the \emph{output qubit}.
Let $\hilb{F} = \hilb{O} \otimes \hilb{F}'$ where $\hilb{O}$ is a
two-dimensional Hilbert space corresponding to the output qubit and $\hilb{F}'$
corresponds to the remaining $m-1$ qubits.
We fix the binary POVM
$$\Set{ \Pi_{\mathrm{accept}}, \Pi_{\mathrm{reject}} } \subset \mapset{Pos}
{\hilb{F}}$$
so that
\begin{eqnarray*}
\Pi_\mathrm{reject}
&=& \ket{0_\hilb{O}} \bra{0_\hilb{O}} \otimes I_{\hilb{F}'}, \\
\Pi_{\mathrm{accept}}
&=& I_\hilb{F} - \Pi_\mathrm{reject}.
\end{eqnarray*}
This measurement is called the \emph{standard measurement} of the output qubit
of $\hilb{F}$.

\subsection{Quantum Algorithms}

We now describe how qubits, quantum circuits, and quantum measurements combine
to form the quantum circuit model of computation.
Let $\rho$ be a quantum state and let $Q$ be a quantum circuit with associated
unitary matrix $U$.
If desired, the input state $\rho$ may be chosen so that it uniquely encodes an
input string $x$ to some computational problem.
The circuit $Q$ is applied to $\rho$ and the output qubit of the resulting state
$U \rho U^*$ is measured according to the standard measurement, which indicates
acceptance or rejection of $\rho$ and hence of the input string $x$.

By definition, quantum circuits act on a fixed number of qubits.
In order to use quantum circuits to decide a language $L \subseteq \set{0,1}^*$
of arbitrarily large strings, it is typical to specify a family of quantum
circuits.
In this thesis, a \emph{family} is a set
$$\set{ Q_x : x \in \set{0,1}^* }$$
of quantum circuits indexed by input strings.
Because the quantum gates in $Q_x$ can depend upon $x$, there is no need to
encode $x$ in the input state $\rho$.
Instead, we may fix once and for all a convenient pure state $\ket{0}$ upon
which all quantum circuits in the family can be assumed to act.

Families of quantum circuits do not yet form a realistic model of computation
because we have not restricted the amount of computation that is used to
construct the circuits in a family.
In order to make the model realistic, we must introduce a \emph{uniformity
constraint}.
In particular, a family of quantum circuits is said to be \emph{polynomial-time
uniformly generated} if there exists a deterministic polynomial-time Turing
machine that, given input $x \in \set{0,1}^*$, outputs a description of the
quantum circuit $Q_x$.

For any language $L \subseteq \set{0,1}^*$, it is widely agreed that the
informal statement ``$L$ has an efficient solution on a quantum computer'' is
adequately formalized by the condition that there exist a polynomial-time
uniformly generated family $\set{ Q_x : x \in \set{0,1}^* }$ of quantum circuits
with associated unitary matrices
$$\Set{ U_x : x \in \set{0,1}^* }$$
such that, for every $x \in \set{0,1}^*$, $Q_x$ correctly accepts or rejects
$\ket{0}$ according to whether or not $x \in L$ with high probability.
More specifically,
\begin{eqnarray*}
\Norm{ \Pi_{\mathrm{accept}} U_x \ket{0} }^2 \geq 2/3 &\iff& x \in L, \\
\Norm{ \Pi_{\mathrm{reject}} U_x \ket{0} }^2 \geq 2/3 &\iff& x \not \in L.
\end{eqnarray*}
The class of languages with this property is denoted $\cls{BQP}$\label{BQP} and
is considered to be the quantum analogue of the complexity class $\cls{P}$.

%% file: introcls.latex
\setlength{\unitlength}{3947sp}%
\begingroup\makeatletter\ifx\SetFigFont\undefined%
\gdef\SetFigFont#1#2#3#4#5{%
  \reset@font\fontsize{#1}{#2pt}%
  \fontfamily{#3}\fontseries{#4}\fontshape{#5}%
  \selectfont}%
\fi\endgroup%
\begin{picture}(4067,2837)(151,-1015)
{\thinlines
\put(901,1739){\circle*{50}}
}%
{\put(2101,1139){\circle*{50}}
}%
{\put(3301,1739){\circle*{50}}
}%
{\put(2101,539){\circle*{50}}
}%
{\put(3301,-61){\circle*{50}}
}%
{\put(901,-61){\circle*{50}}
}%
{\put(2101,-661){\circle*{50}}
}%
{\put(901,1739){\line( 2,-1){1200}}
}%
{\put(2101,1139){\line( 2, 1){1200}}
}%
{\put(2101,1139){\line( 0,-1){600}}
}%
{\put(901,-61){\line( 2, 1){1200}}
}%
{\put(2101,539){\line( 2,-1){1200}}
}%
{\put(2101,-661){\line( 2, 1){1200}}
}%
{\put(901,-61){\line( 2,-1){1200}}
}%
\put(791,1664){\makebox(0,0)[r]{\smash{{\SetFigFont{12}{14.4}{\familydefault}{\mddefault}{\updefault}{$\cls{NEXP}$}%
}}}}
\put(791,-136){\makebox(0,0)[r]{\smash{{\SetFigFont{12}{14.4}{\familydefault}{\mddefault}{\updefault}{$\cls{NP}$}%
}}}}
\put(2101,-901){\makebox(0,0){\smash{{\SetFigFont{12}{14.4}{\familydefault}{\mddefault}{\updefault}{$\cls{P}$}%
}}}}
\put(3400,1664){\makebox(0,0)[l]{\smash{{\SetFigFont{12}{14.4}{\familydefault}{\mddefault}{\updefault}{$\cls{coNEXP}$}%
}}}}
\put(2226,1000){\makebox(0,0)[l]{\smash{{\SetFigFont{12}{14.4}{\familydefault}{\mddefault}{\updefault}{$\cls{EXP}$}%
}}}}
\put(2226,539){\makebox(0,0)[l]{\smash{{\SetFigFont{12}{14.4}{\familydefault}{\mddefault}{\updefault}{$\cls{PSPACE}$}%
}}}}
\put(3400,-136){\makebox(0,0)[l]{\smash{{\SetFigFont{12}{14.4}{\familydefault}{\mddefault}{\updefault}{$\cls{coNP}$}%
}}}}
\end{picture}%

%% file: defs.tex
\chapter{Preliminaries} \label{ch:defs}

In this chapter we provide formal definitions of the quantum interactive proof
system and quantum refereed game models of computation as well as several
collections of complexity classes based upon these models.
We then summarize what is currently known of these models and state the
contributions of this thesis.

\section{Formalizations of Quantum Interaction}
\label{sec:defs:formalizations}

Quantum interactive proof systems were introduced by Watrous in 1999
\cite{Watrous03} and it is that formalization of the model that we reproduce
here.
Although quantum refereed games had not yet been considered prior to the work of
the present thesis, the formalization of that model is a straightforward
extension of the quantum interactive proof system model.

Quantum interactions consist of a \emph{verifier} and one or more \emph{provers}.
For any function $r : \mathbb{N} \to \mathbb{N} \setminus \set{0}$, an
\emph{$r$-round prover} $P$ is a mapping on input strings $x \in \set{0,1}^*$
where
$$P(x) = (P_1,\dots,P_{r(|x|)})$$
is an $r(|x|)$-tuple of quantum circuits, each of which acts upon the same
number of qubits.
No restrictions are placed on the complexity of the prover's circuits, which
captures the notion that the prover has unlimited computational power---each of
the prover's circuits can be viewed as an arbitrary unitary operation on its
input qubits.

Similarly, an \emph{$r$-round verifier} $V$ is a mapping on input strings
$x \in \set{0,1}^*$ where
$$V(x) = (V_0,\dots,V_{r(|x|)})$$
is an $(r(|x|)+1)$-tuple of quantum circuits, each of which acts upon the same
number of qubits.
Unlike a prover, however, we require that the verifier's circuits be generated
by a polynomial-time Turing machine on input $x$.
This uniformity constraint captures the notion that the verifier's computational
power is limited and implicitly restricts the quantity $r(|x|)$ so that
$r \in \poly$ as one might expect.
We often abbreviate $r(|x|)$ to $r$ for easier readability.

\subsection{Quantum Interactive Proof Systems}
\label{subsec:defs:formalizations:QIPs}

A \emph{quantum interactive proof system} has a verifier $V$ and a prover $P$.
The qubits upon which each of the circuits in the prover's $r$-tuple acts are
partitioned into two sets: one set of qubits is private to the prover and the
other is shared with the verifier.
These shared qubits act as a quantum channel between the verifier and the
prover.
The Hilbert spaces corresponding to the private and shared qubits of the prover
are denoted $\hilb{P}$ and $\hilb{M}$ respectively.

Similarly, the qubits upon which each of the circuits in the verifier's $(r+1)$-
tuple acts are partitioned into two sets: one set of qubits is private to the
verifier and the other is shared with the prover.
The Hilbert spaces corresponding to the private and shared qubits of the
verifier are denoted $\hilb{V}$ and $\hilb{M}$ respectively.

\begin{figure}
\begin{center}
\input{QIP.latex}
\end{center}
\caption{Quantum circuit for a two-round quantum interactive proof}
\label{fig:qip}
\end{figure}
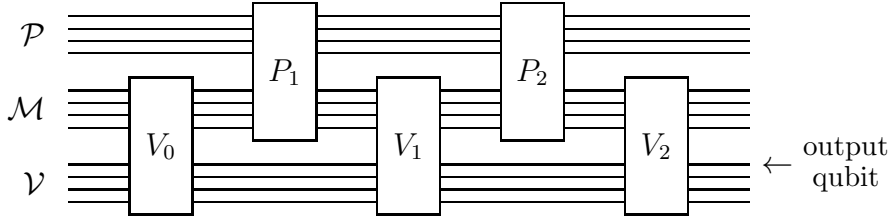

For any input string $x \in \set{0,1}^*$ we create a composite circuit
$(V,P)(x)$ by concatenating the circuits
$$V_0, P_1, V_1, \dots, V_{r-1}, P_r, V_r$$
in sequence, each circuit acting only upon the sets of qubits stipulated
previously.
Such a circuit is illustrated in Figure \ref{fig:qip} for the case $r=2$.
The Hilbert space upon which $(V,P)(x)$ acts is denoted
$$\hilb{S} = \hilb{P} \otimes \hilb{M} \otimes \hilb{V}.$$
The quantum interactive proof system is implemented by applying the circuit
$(V,P)(x)$ to the initial pure state $\ket{0_\hilb{S}} \in \hilb{S}$.
Hence, the pure state of the system after $(V,P)(x)$ is applied is precisely
$$V_r P_r V_{r-1} \cdots V_1 P_1 V_0 \ket{0_\hilb{S}} \in \hilb{S}.$$
Acceptance is dictated by a standard measurement of the output qubit of
$\hilb{S}$, which is assumed to belong to the verifier.
In particular, $(V,P)(x)$ accepts $x$ with probability
$$\Norm{ \Pi_\mathrm{accept} V_r P_r V_{r-1} \cdots V_1 P_1 V_0
\ket{0_\hilb{S}} }^2$$
and rejects $x$ with probability
$$\Norm{ \Pi_\mathrm{reject} V_r P_r V_{r-1} \cdots V_1 P_1 V_0
\ket{0_\hilb{S}} }^2$$
where $\Pi_\mathrm{accept}, \Pi_\mathrm{reject} \in \mapset{Pos}{\hilb{S}}$ are
as defined in Section \ref{subsec:intro:quantum:measurement}.

We now define a collection of complexity classes based upon quantum interactive
proof systems.
For any polynomial-time computable functions $c,s : \mathbb{N} \to [0,1]$, the
class $\cls{QIP}(c,s)$\label{QIP} consists of all languages
$L \subseteq \set{0,1}^*$ for which there exists an $r$-round verifier $V$ that
satisfies the following completeness and soundness conditions:
\begin{description}

\item[Completeness.]
There exists an $r$-round prover $P$ such that, for all $x \in L$, $(V,P)(x)$
rejects $x$ with probability at most $c(|x|)$.
In other words, there exist unitary matrices
$P_1,\dots,P_r \in \mapset{U}{\hilb{P} \otimes \hilb{M}}$ such that
$$\Norm{ \Pi_\mathrm{reject} V_r P_r V_{r-1} \cdots V_1 P_1 V_0
\ket{0_\hilb{S}} }^2 \leq c(|x|).$$

\item[Soundness.]
\begin{sloppypar}
For all $r$-round provers $P$ and all $x \not \in L$, $(V,P)(x)$ accepts $x$
with probability at most $s(|x|)$.
In other words, for all unitary matrices
$P_1,\dots,P_r \in \mapset{U}{\hilb{P} \otimes \hilb{M}}$, we have
$$\Norm{ \Pi_\mathrm{accept} V_r P_r V_{r-1} \cdots V_1 P_1 V_0
\ket{0_\hilb{S}} }^2 \leq s(|x|).$$
\end{sloppypar}

\end{description}
As with classical interactive proof systems, the functions $c$ and $s$ are
called the \emph{completeness error} and \emph{soundness error} respectively.
We often abbreviate $c(|x|)$ to $c$ and $s(|x|)$ to $s$ for easier readability.

\subsection{Quantum Refereed Games}

A \emph{quantum refereed game} has a verifier $V$ and two provers $Y$ and $N$.
As with quantum interactive proof systems, the qubits upon which each of the
circuits in the provers' $r$-tuples acts are partitioned into two sets: one set
of qubits is private to that prover and the other is shared with the verifier.
These shared qubits act as a quantum channel between the verifier and that
prover.

For clarity, $Y$ is called a \emph{yes-prover} and $N$ is called a
\emph{no-prover}.
This distinction is purely a notational convenience: the Hilbert spaces
corresponding to the private and shared qubits of a yes-prover are denoted
$\hilb{Y}$ and $\hilb{M}_Y$ respectively, whereas the Hilbert spaces
corresponding to the private and shared qubits of a no-prover are denoted
$\hilb{N}$ and $\hilb{M}_N$ respectively.

In a quantum refereed game, the qubits upon which each of the circuits in the
verifier's $(r+1)$-tuple acts are partitioned into three sets: one set of
qubits, with corresponding Hilbert space $\hilb{V}$, is private to the verifier
and the two remaining sets have corresponding Hilbert spaces $\hilb{M}_Y$ and
$\hilb{M}_N$ and are shared with the yes- and no-provers respectively.

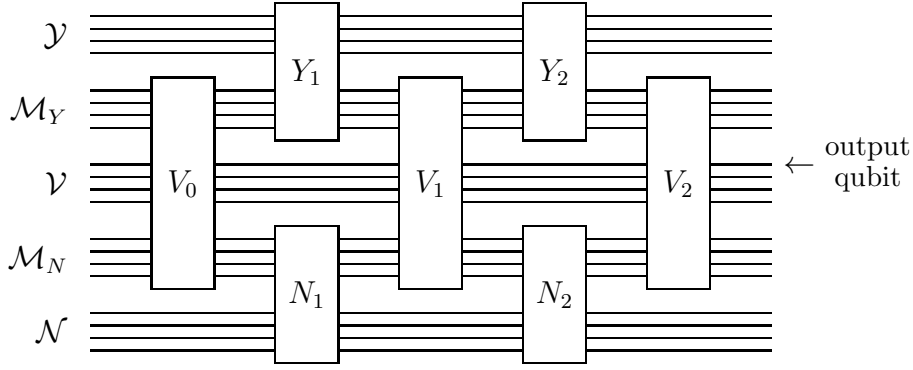
\begin{figure}
\begin{center}
\input{QRG.latex}
\end{center}
\caption{Quantum circuit for a two-round quantum refereed game}
\label{fig:qrg}
\end{figure}

For any input string $x \in \set{0,1}^*$ we create a composite circuit
$(V,Y,N)(x)$ by concatenating the circuits
$$V_0, N_1, Y_1, V_1, \dots, V_{r-1}, N_r, Y_r, V_r$$
in sequence, each circuit acting only upon the sets of qubits stipulated
previously.
Such a circuit is illustrated in Figure \ref{fig:qrg} for the case $r=2$.
The Hilbert space upon which $(V,Y,N)(x)$ acts is denoted
$$\hilb{S} = \hilb{Y} \otimes \hilb{M}_Y \otimes \hilb{V} \otimes \hilb{M}_N
\otimes \hilb{N}.$$
Although $\hilb{S}$ also denotes the Hilbert space for quantum interactive proof
systems, any ambiguity is always resolved by context.
The quantum refereed game is implemented by applying the circuit
$(V,Y,N)(x)$ to the initial pure state $\ket{0_\hilb{S}} \in \hilb{S}$.
Hence, the pure state of the system after $(V,Y,N)(x)$ is applied is precisely
$$V_r Y_r N_r V_{r-1} \cdots V_1 Y_1 N_1 V_0 \ket{0_\hilb{S}} \in \hilb{S}.$$
As with quantum interactive proof systems, acceptance is dictated by a standard
measurement of the output qubit of $\hilb{S}$, which is assumed to belong to the
verifier.
In particular, $(V,Y,N)(x)$ accepts $x$ with probability
$$\Norm{ \Pi_\mathrm{accept} V_r Y_r N_r V_{r-1}
\cdots V_1 Y_1 N_1 V_0 \ket{0_\hilb{S}} }^2$$
and rejects $x$ with probability
$$\Norm{ \Pi_\mathrm{reject} V_r Y_r N_r V_{r-1}
\cdots V_1 Y_1 N_1 V_0 \ket{0_\hilb{S}} }^2.$$

We now define a collection of complexity classes based upon quantum refereed
games.
For any polynomial-time computable functions $c,s : \mathbb{N} \to [0,1]$, the
class $\cls{QRG}(c,s)$\label{QRG} consists of all languages
$L \subseteq \set{0,1}^*$ for which there exists an $r$-round verifier $V$ that
satisfies the following completeness and soundness conditions:
\begin{description}

\item[Completeness.]
There exists an $r$-round yes-prover $Y$ such that, for all
$r$-round no-provers $N$ and all $x \in L$, $(V,Y,N)(x)$ rejects $x$ with
probability at most $c(|x|)$.
In other words, there exist unitary matrices
$Y_1,\dots,Y_r \in \mapset{U}{\hilb{Y} \otimes \hilb{M}_Y}$ such that
$$\Norm{ \Pi_\mathrm{reject} V_r Y_r N_r V_{r-1}
\cdots V_1 Y_1 N_1 V_0 \ket{0_\hilb{S}} }^2 \leq c(|x|)$$
for all unitary matrices
$N_1,\dots,N_r \in \mapset{U}{\hilb{M}_N \otimes \hilb{N}}$.

\item[Soundness.]
There exists an $r$-round no-prover $N$ such that, for all
$r$-round yes-provers $Y$ and all $x \not \in L$, $(V,Y,N)(x)$ accepts $x$
with probability at most $s(|x|)$.
In other words, there exist unitary matrices
$N_1,\dots,N_r \in \mapset{U}{\hilb{M}_N \otimes \hilb{N}}$ such that
$$\Norm{ \Pi_\mathrm{accept} V_r Y_r N_r V_{r-1}
\cdots V_1 Y_1 N_1 V_0 \ket{0_\hilb{S}} }^2 \leq s(|x|)$$
for all unitary matrices
$Y_1,\dots,Y_r \in \mapset{U}{\hilb{Y} \otimes \hilb{M}_Y}$.

\end{description}
As with quantum interactive proof systems, the functions $c$ and $s$ are the
completeness error and soundness error respectively.

\subsection{Short Quantum Games}

\begin{figure}
\begin{center}
\input{SQG.latex}
\end{center}
\caption{Quantum circuit for a short quantum game} \label{fig:sqg}
\end{figure}
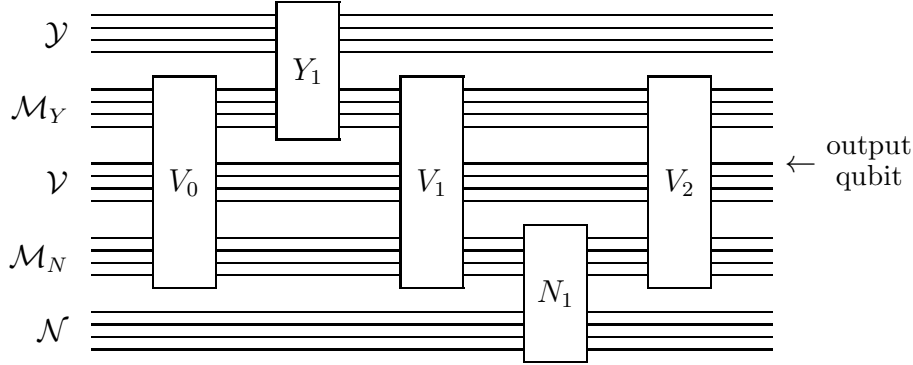

For most of this thesis we restrict our attention to a specific class of quantum
refereed games that we call \emph{short quantum games}.
A short quantum game has a two-round verifier $V$, a one-round yes-prover $Y$,
and a one-round no-prover $N$.
In these games, the composite circuit $(V,Y,N)'(x)$ is created by concatenating
the circuits
$$V_0, Y_1, V_1, N_1, V_2$$
in sequence.
In other words, short quantum games are one-round quantum games in which the
verifier is permitted to process the yes-prover's response before sending a
message to the no-prover.
Figure \ref{fig:sqg} illustrates the quantum circuit for a short quantum game.

For polynomial-time computable $c,s : \mathbb{N} \to [0,1]$, we define the
collection $\cls{SQG}(c,s)$\label{SQG} of complexity classes by restricting the
definition of $\cls{QRG}(c,s)$ to short quantum games.
We also define the collection $\cls{SQG}_*(c,s)$\label{SQGstar} of complexity
classes by further restricting the definition of $\cls{SQG}(c,s)$ to short
quantum games in which the verifier cannot send a message to the yes-prover
(in other words, the verifier's first circuit is empty, so that $V_0=I$).

\section{Remarks and Contributions}
\label{sec:defs:remarks}

This section contains a summary of what is currently known of quantum
interactive proof systems and quantum refereed games.
The contributions of this thesis are stated in Section
\ref{subsec:defs:contributions}.

\subsection{Known Results}
\label{subsec:defs:remarks:results}

Several inclusions follow immediately from the definitions in Section
\ref{sec:defs:formalizations}.
For example, because a quantum interactive proof system is just a quantum
refereed game without a no-prover, it is clear that
$$\cls{QIP}(c,s) \subseteq \cls{QRG}(c,s).$$
Furthermore, we have
$$\cls{SQG}_*(c,s) \subseteq \cls{SQG}(c,s) \subseteq \cls{QRG}(c,s),$$
as any shorter quantum refereed game can be simulated by a longer quantum
refereed game.

Also, it is clear that any classical interaction can be simulated by a
quantum interaction in which the verifier simply measures every message he
receives from any prover (thus collapsing each message to a completely classical
state) and otherwise behaves in a classical manner.
In other words, we have $\cls{IP}(c,s) \subseteq \cls{QIP}(c,s)$,
$\cls{RG}(c,s) \subseteq \cls{QRG}(c,s)$, and
$\cls{RG}_1(c,s) \subseteq \cls{SQG}(c,s)$.
It is instructive to note that the relation
$\cls{RG}_1(c,s) \subseteq \cls{SQG}_*(c,s)$ is not immediately seen to hold, as
$\cls{SQG}_*(c,s)$ does not permit games in which the verifier sends a message
to the yes-prover.

The method of sequential repetition used in Section
\ref{subsec:intro:complexity:results} to demonstrate the robustness of classical
interaction can be applied without complication in the quantum setting.
That is, for every $\varepsilon \in 2^{-\poly}$ and every reasonable
$c,s : \mathbb{N} \to [0,1]$ we have
$$
\cls{QIP}(c,s) \subseteq \cls{QIP}(\varepsilon,\varepsilon)
\qquad \textrm{and} \qquad
\cls{QRG}(c,s) \subseteq \cls{QRG}(\varepsilon,\varepsilon).
$$
Similar to the classical case, quantum interactive proof systems with reasonable
error can be assumed to have zero completeness error \cite{KitaevW00}, yielding
the inclusion
$$\cls{QIP}(c,s) \subseteq \cls{QIP}(0,\varepsilon).$$
In light of this robustness, we define the following shorthand notations:
\begin{itemize}

\item
$\cls{QIP}$ is the complexity class of all languages
$L \subseteq \set{0,1}^*$ such that
$L \in \cls{QIP}(0,\varepsilon)$ for every
$\varepsilon \in 2^{-\poly}$.

\item
$\cls{QRG}$ is the complexity class of all languages
$L \subseteq \set{0,1}^*$ such that
$L \in \cls{QRG}(\varepsilon,\varepsilon)$ for every
$\varepsilon \in 2^{-\poly}$.

\item
$\cls{SQG}$ is the complexity class of all languages
$L \subseteq \set{0,1}^*$ such that
$L \in \cls{SQG}(\varepsilon,\varepsilon)$ for every
$\varepsilon \in 2^{-\poly}$.

\item
$\cls{SQG}_*$ is the complexity class of all languages
$L \subseteq \set{0,1}^*$ such that
$L \in \cls{SQG}_*(\varepsilon,\varepsilon)$ for every
$\varepsilon \in 2^{-\poly}$.

\end{itemize}

The well-known characterization $\cls{IP} = \cls{PSPACE}$ mentioned in Section
\ref{subsec:intro:complexity:results} implies
$$\cls{PSPACE} \subseteq \cls{QIP}.$$
Watrous gave a three-message quantum interactive proof system
demonstrating this containment \cite{Watrous03}.
The existence of a three-message quantum interactive proof system for
$\cls{PSPACE}$ contrasts with the classical case, wherein it is widely believed
that the verifier and prover must exchange a nonconstant number of messages in
order to decide $\cls{PSPACE}$
(see Section \ref{subsec:end:open:parallelization}).

This result was later extended to show that any quantum interactive proof system
can be simulated by a three-message quantum interactive proof system and that
these simulations are robust with respect to error \cite{KitaevW00}.
This extension gives rise to a natural complete promise problem for $\cls{QIP}$
known as \textsc{close-images}, which can in turn be reduced to an
exponential-size instance of the semidefinite programming problem
\cite{KitaevW00}.
As semidefinite programs can be solved in deterministic polynomial time, it
follows from the $\cls{QIP}$-completeness of \textsc{close-images} that
$$\cls{QIP} \subseteq \cls{EXP}.$$
In a later talk on quantum coin-flipping, Kitaev gave a semidefinite program
that directly simulates many-message quantum interactive proof systems
\cite{Kitaev02}, thus yielding a more direct proof of the containment of
$\cls{QIP}$ in $\cls{EXP}$.

Although quantum refereed games were not considered prior to this work, it is
nonetheless appropriate to mention several facts in this subsection.
First, it is clear that the characterizations $\cls{RG} = \cls{EXP}$ and
$\cls{RG}_1 = \cls{PSPACE}$ imply
$$
\cls{EXP} \subseteq \cls{QRG}
\qquad \textrm{and} \qquad
\cls{PSPACE} \subseteq \cls{SQG}.
$$

Also, Kitaev's variant \cite{Kitaev02} of the semidefinite program found in
Reference \cite{KitaevW00} is easily extended to yield
$\cls{QRG} \subseteq \cls{NEXP}$ \cite{Gutoski05}.
Due to the symmetric nature of quantum refereed games, it is clear that
$\cls{QRG}$ is closed under complement, from which it follows that $\cls{QRG}$
is also contained in $\cls{coNEXP}$.
In other words, we have
$$\cls{QRG} \subseteq \cls{NEXP} \cap \cls{coNEXP}.$$
In Section \ref{subsec:end:open:QRGeqEXP} we we discuss possible implications of
the curious fact that $\cls{QRG}$ contains $\cls{EXP}$ and is in turn contained
in $\cls{NEXP} \cap \cls{coNEXP}$.
Chapter \ref{ch:end} also offers a diagram of relationships among the complexity
classes considered in this thesis.

\subsection{Contributions of this Thesis}
\label{subsec:defs:contributions}

We prove in this thesis the following two relationships among the complexity
classes defined in Section \ref{sec:defs:formalizations}:
\begin{itemize}

\item
$\cls{QIP} \subseteq \cls{SQG}_*$ (Chapter \ref{ch:QIPinSQG})

\item
$\cls{SQG} \subseteq \cls{EXP}$ (Chapter \ref{ch:SQGinEXP})

\end{itemize}
The first result is proven in Reference \cite{GutoskiW05}, while the second is
proven in Reference \cite{Gutoski05}.
It is those proofs that we reproduce in Chapters \ref{ch:QIPinSQG} and
\ref{ch:SQGinEXP}.

Two of the intermediate results that are proven in order to obtain the
containment $\cls{QIP} \subseteq \cls{SQG}_*$ might be of independent interest
and so we also list them here:
\begin{itemize}

\item
For any two disjoint closed convex sets
$\mathcal{A}_0, \mathcal{A}_1 \subseteq \mapset{D}{\hilb{H}}$ of quantum states,
there exists a binary POVM such that, for any state
$\rho \in \mathcal{A}_0 \cup \mathcal{A}_1$,
the POVM will correctly determine whether $\rho \in \mathcal{A}_0$ or
$\rho \in \mathcal{A}_1$ with probability proportional to the minimal trace
distance between $\mathcal{A}_0$ and $\mathcal{A}_1$.

\item
$\cls{SQG}(c,s) \subseteq \cls{SQG}(kc,s^k) \cap \cls{SQG}(c^k,ks)$ for
any $c,s : \mathbb{N} \to [0,1]$ and any $k \in \mathit{poly}$.
A similar containment holds with $\cls{SQG}_*$ in place of $\cls{SQG}$.

\end{itemize}
The first of these intermediate results is a generalization of the well-known
fact that two quantum states can be distinguished with probability proportional
to their trace distance.
It can also be viewed as a quantitative version of the well-known separation
theorems in convex analysis.

The second result indicates a partial robustness of short quantum games with
respect to error.
In particular, it states that the completeness (soundness) error can be made
exponentially small at the possible cost of soundness (completeness).
Error reduction results seem to be more elusive in quantum interaction than in
classical interaction and this result represents a first step toward that end
for short quantum games.

The containment $\cls{SQG} \subseteq \cls{EXP}$ builds upon the
semidefinite program for $\cls{QIP}$ \cite{KitaevW00, Kitaev02}.
Hence, we offer a rigorous analysis of that semidefinite program in this thesis
as a precursor to our result.
This precursor leads to several extensions of the containment
$\cls{QIP} \subseteq \cls{EXP}$.
For example, we show that semidefinite programming can be used to simulate a
quantum interactive proof system in deterministic exponential time even if the
verifier's circuits are generated uniformly in \emph{exponential} time, so long
as they still act on only a polynomial number of qubits.
In particular, the verifier can exchange an exponential number of messages with
the prover and can use an exponential number of gates in his quantum circuits
without raising the power of the model beyond $\cls{EXP}$.

The containment $\cls{QRG} \subseteq \cls{NEXP}$ is obtained by
nondeterministically guessing a yes-prover and solving the induced quantum
interactive proof system using the aforementioned semidefinite program
\cite{Gutoski05}.
Most of the extensions of that semidefinite program also apply to this
containment concerning quantum refereed games.
The only exception is that we still require that the verifier exchange only a
polynomial number of rounds of messages with the provers.
This strange restriction is brought on by the fact that a polynomial bound on
the number of qubits required by the provers in a quantum refereed game is known
to hold only if a polynomial number of messages are exchanged
(see Section \ref{subsec:4:bounds:proverqubits}).

We prove $\cls{SQG} \subseteq \cls{EXP}$ by a repeated use of semidefinite
programming in concert with the ellipsoid method for convex feasibility.
Once again, many of the aforementioned extensions also apply to this
containment.
For example, our method can simulate a short quantum game in deterministic
exponential time even if the verifier is permitted to exchange an exponential
number of messages with the yes-prover, followed by an exponential number of
messages with the no-prover.

%% file: QIP.latex
\setlength{\unitlength}{2047sp}%
\begingroup\makeatletter\ifx\SetFigFont\undefined%
\gdef\SetFigFont#1#2#3#4#5{%
  \reset@font\fontsize{#1}{#2pt}%
  \fontfamily{#3}\fontseries{#4}\fontshape{#5}%
  \selectfont}%
\fi\endgroup%
\begin{picture}(9567,2574)(226,-2023)
\thinlines
{\put(601,-511){\line( 1, 0){750}}
}%
{\put(601,-661){\line( 1, 0){750}}
}%
{\put(601,-811){\line( 1, 0){750}}
}%
{\put(601,-961){\line( 1, 0){750}}
}%
{\put(601,-1411){\line( 1, 0){750}}
}%
{\put(601,-1561){\line( 1, 0){750}}
}%
{\put(601,-1711){\line( 1, 0){750}}
}%
{\put(601,-1861){\line( 1, 0){750}}
}%
{\put(2101,-511){\line( 1, 0){750}}
}%
{\put(2101,-661){\line( 1, 0){750}}
}%
{\put(2101,-811){\line( 1, 0){750}}
}%
{\put(2101,-961){\line( 1, 0){750}}
}%
{\put(2851,389){\line(-1, 0){2250}}
}%
{\put(2851,239){\line(-1, 0){2250}}
}%
{\put(2851, 89){\line(-1, 0){2250}}
}%
{\put(2851,-61){\line(-1, 0){2250}}
}%
{\put(3601,-511){\line( 1, 0){750}}
}%
{\put(3601,-661){\line( 1, 0){750}}
}%
{\put(3601,-811){\line( 1, 0){750}}
}%
{\put(3601,-961){\line( 1, 0){750}}
}%
{\put(4351,-1411){\line(-1, 0){2250}}
}%
{\put(4351,-1561){\line(-1, 0){2250}}
}%
{\put(4351,-1711){\line(-1, 0){2250}}
}%
{\put(4351,-1861){\line(-1, 0){2250}}
}%
{\put(5101,-511){\line( 1, 0){750}}
}%
{\put(5101,-661){\line( 1, 0){750}}
}%
{\put(5101,-811){\line( 1, 0){750}}
}%
{\put(5101,-961){\line( 1, 0){750}}
}%
{\put(5851,389){\line(-1, 0){2250}}
}%
{\put(5851,239){\line(-1, 0){2250}}
}%
{\put(5851, 89){\line(-1, 0){2250}}
}%
{\put(5851,-61){\line(-1, 0){2250}}
}%
{\put(6601,-511){\line( 1, 0){750}}
}%
{\put(6601,-661){\line( 1, 0){750}}
}%
{\put(6601,-811){\line( 1, 0){750}}
}%
{\put(6601,-961){\line( 1, 0){750}}
}%
{\put(7351,-1411){\line(-1, 0){2250}}
}%
{\put(7351,-1561){\line(-1, 0){2250}}
}%
{\put(7351,-1711){\line(-1, 0){2250}}
}%
{\put(7351,-1861){\line(-1, 0){2250}}
}%
{\put(8851,389){\line(-1, 0){2250}}
}%
{\put(8851,239){\line(-1, 0){2250}}
}%
{\put(8851, 89){\line(-1, 0){2250}}
}%
{\put(8851,-61){\line(-1, 0){2250}}
}%
{\put(8101,-511){\line( 1, 0){750}}
}%
{\put(8101,-661){\line( 1, 0){750}}
}%
{\put(8101,-811){\line( 1, 0){750}}
}%
{\put(8101,-961){\line( 1, 0){750}}
}%
{\put(8101,-1411){\line( 1, 0){750}}
}%
{\put(8101,-1561){\line( 1, 0){750}}
}%
{\put(8101,-1711){\line( 1, 0){750}}
}%
{\put(8101,-1861){\line( 1, 0){750}}
}%
\put(9001,-1505){$\leftarrow$\hspace*{-3mm}\parbox{2cm}{
		\begin{center}\small
			output\\[-0.7mm]
			qubit
		\end{center}}}
{\put(2851,-1111){\framebox(750,1650){$P_1$}}
}%
{\put(5851,-1111){\framebox(750,1650){$P_2$}}
}%
{\put(1351,-2011){\framebox(750,1650){$V_0$}}
}%
{\put(4351,-2011){\framebox(750,1650){$V_1$}}
}%
{\put(7351,-2011){\framebox(750,1650){$V_2$}}
}%
\put(326,-861){\makebox(0,0)[r]{\smash{{\SetFigFont{12}{14.4}{\familydefault}{\mddefault}{\updefault}{$\hilb{M}$}%
}}}}
\put(326,-1761){\makebox(0,0)[r]{\smash{{\SetFigFont{12}{14.4}{\familydefault}{\mddefault}{\updefault}{$\hilb{V}$}%
}}}}
\put(326, 39){\makebox(0,0)[r]{\smash{{\SetFigFont{12}{14.4}{\familydefault}{\mddefault}{\updefault}{$\hilb{P}$}%
}}}}
\end{picture}%

%% file: QRG.latex
\setlength{\unitlength}{2047sp}%
\begingroup\makeatletter\ifx\SetFigFont\undefined%
\gdef\SetFigFont#1#2#3#4#5{%
  \reset@font\fontsize{#1}{#2pt}%
  \fontfamily{#3}\fontseries{#4}\fontshape{#5}%
  \selectfont}%
\fi\endgroup%
\begin{picture}(9567,4374)(226,-3823)
\thinlines
{\put(601,-511){\line( 1, 0){750}}
}%
{\put(601,-661){\line( 1, 0){750}}
}%
{\put(601,-811){\line( 1, 0){750}}
}%
{\put(601,-961){\line( 1, 0){750}}
}%
{\put(601,-1411){\line( 1, 0){750}}
}%
{\put(601,-1561){\line( 1, 0){750}}
}%
{\put(601,-1711){\line( 1, 0){750}}
}%
{\put(601,-1861){\line( 1, 0){750}}
}%
{\put(601,-2311){\line( 1, 0){750}}
}%
{\put(601,-2461){\line( 1, 0){750}}
}%
{\put(601,-2611){\line( 1, 0){750}}
}%
{\put(601,-2761){\line( 1, 0){750}}
}%
{\put(2101,-511){\line( 1, 0){750}}
}%
{\put(2101,-661){\line( 1, 0){750}}
}%
{\put(2101,-811){\line( 1, 0){750}}
}%
{\put(2101,-961){\line( 1, 0){750}}
}%
{\put(2851,389){\line(-1, 0){2250}}
}%
{\put(2851,239){\line(-1, 0){2250}}
}%
{\put(2851, 89){\line(-1, 0){2250}}
}%
{\put(2851,-61){\line(-1, 0){2250}}
}%
{\put(3601,-511){\line( 1, 0){750}}
}%
{\put(3601,-661){\line( 1, 0){750}}
}%
{\put(3601,-811){\line( 1, 0){750}}
}%
{\put(3601,-961){\line( 1, 0){750}}
}%
{\put(4351,-1411){\line(-1, 0){2250}}
}%
{\put(4351,-1561){\line(-1, 0){2250}}
}%
{\put(4351,-1711){\line(-1, 0){2250}}
}%
{\put(4351,-1861){\line(-1, 0){2250}}
}%
{\put(2101,-2311){\line( 1, 0){750}}
}%
{\put(2101,-2461){\line( 1, 0){750}}
}%
{\put(2101,-2611){\line( 1, 0){750}}
}%
{\put(2101,-2761){\line( 1, 0){750}}
}%
{\put(3601,-2311){\line( 1, 0){750}}
}%
{\put(3601,-2461){\line( 1, 0){750}}
}%
{\put(3601,-2611){\line( 1, 0){750}}
}%
{\put(3601,-2761){\line( 1, 0){750}}
}%
{\put(2851,-3211){\line(-1, 0){2250}}
}%
{\put(2851,-3361){\line(-1, 0){2250}}
}%
{\put(2851,-3511){\line(-1, 0){2250}}
}%
{\put(2851,-3661){\line(-1, 0){2250}}
}%
{\put(5101,-511){\line( 1, 0){750}}
}%
{\put(5101,-661){\line( 1, 0){750}}
}%
{\put(5101,-811){\line( 1, 0){750}}
}%
{\put(5101,-961){\line( 1, 0){750}}
}%
{\put(5101,-2311){\line( 1, 0){750}}
}%
{\put(5101,-2461){\line( 1, 0){750}}
}%
{\put(5101,-2611){\line( 1, 0){750}}
}%
{\put(5101,-2761){\line( 1, 0){750}}
}%
{\put(5851,-3211){\line(-1, 0){2250}}
}%
{\put(5851,-3361){\line(-1, 0){2250}}
}%
{\put(5851,-3511){\line(-1, 0){2250}}
}%
{\put(5851,-3661){\line(-1, 0){2250}}
}%
{\put(5851,389){\line(-1, 0){2250}}
}%
{\put(5851,239){\line(-1, 0){2250}}
}%
{\put(5851, 89){\line(-1, 0){2250}}
}%
{\put(5851,-61){\line(-1, 0){2250}}
}%
{\put(6601,-511){\line( 1, 0){750}}
}%
{\put(6601,-661){\line( 1, 0){750}}
}%
{\put(6601,-811){\line( 1, 0){750}}
}%
{\put(6601,-961){\line( 1, 0){750}}
}%
{\put(6601,-2311){\line( 1, 0){750}}
}%
{\put(6601,-2461){\line( 1, 0){750}}
}%
{\put(6601,-2611){\line( 1, 0){750}}
}%
{\put(6601,-2761){\line( 1, 0){750}}
}%
{\put(7351,-1411){\line(-1, 0){2250}}
}%
{\put(7351,-1561){\line(-1, 0){2250}}
}%
{\put(7351,-1711){\line(-1, 0){2250}}
}%
{\put(7351,-1861){\line(-1, 0){2250}}
}%
{\put(8851,389){\line(-1, 0){2250}}
}%
{\put(8851,239){\line(-1, 0){2250}}
}%
{\put(8851, 89){\line(-1, 0){2250}}
}%
{\put(8851,-61){\line(-1, 0){2250}}
}%
{\put(8101,-511){\line( 1, 0){750}}
}%
{\put(8101,-661){\line( 1, 0){750}}
}%
{\put(8101,-811){\line( 1, 0){750}}
}%
{\put(8101,-961){\line( 1, 0){750}}
}%
{\put(8101,-2311){\line( 1, 0){750}}
}%
{\put(8101,-2461){\line( 1, 0){750}}
}%
{\put(8101,-2611){\line( 1, 0){750}}
}%
{\put(8101,-2761){\line( 1, 0){750}}
}%
{\put(8851,-3211){\line(-1, 0){2250}}
}%
{\put(8851,-3361){\line(-1, 0){2250}}
}%
{\put(8851,-3511){\line(-1, 0){2250}}
}%
{\put(8851,-3661){\line(-1, 0){2250}}
}%
{\put(8101,-1411){\line( 1, 0){750}}
}%
{\put(8101,-1561){\line( 1, 0){750}}
}%
{\put(8101,-1711){\line( 1, 0){750}}
}%
{\put(8101,-1861){\line( 1, 0){750}}
}%
\put(9001,-1505){$\leftarrow$\hspace*{-3mm}\parbox{2cm}{
		\begin{center}\small
			output\\[-0.7mm]
			qubit
		\end{center}}}
{\put(1351,-2911){\framebox(750,2550){$V_0$}}
}%
{\put(2851,-1111){\framebox(750,1650){$Y_1$}}
}%
{\put(4351,-2911){\framebox(750,2550){$V_1$}}
}%
{\put(2851,-3811){\framebox(750,1650){$N_1$}}
}%
{\put(5851,-1111){\framebox(750,1650){$Y_2$}}
}%
{\put(5851,-3811){\framebox(750,1650){$N_2$}}
}%
{\put(7351,-2911){\framebox(750,2550){$V_2$}}
}%
\put(326,-861){\makebox(0,0)[r]{\smash{{\SetFigFont{12}{14.4}{\familydefault}{\mddefault}{\updefault}{$\hilb{M}_Y$}%
}}}}
\put(326,-1761){\makebox(0,0)[r]{\smash{{\SetFigFont{12}{14.4}{\familydefault}{\mddefault}{\updefault}{$\hilb{V}$}%
}}}}
\put(326,-2661){\makebox(0,0)[r]{\smash{{\SetFigFont{12}{14.4}{\familydefault}{\mddefault}{\updefault}{$\hilb{M}_N$}%
}}}}
\put(326,-3561){\makebox(0,0)[r]{\smash{{\SetFigFont{12}{14.4}{\familydefault}{\mddefault}{\updefault}{$\hilb{N}$}%
}}}}
\put(326, 39){\makebox(0,0)[r]{\smash{{\SetFigFont{12}{14.4}{\familydefault}{\mddefault}{\updefault}{$\hilb{Y}$}%
}}}}
\end{picture}%

%% file: SQG.latex
\setlength{\unitlength}{2047sp}%
\begingroup\makeatletter\ifx\SetFigFont\undefined%
\gdef\SetFigFont#1#2#3#4#5{%
  \reset@font\fontsize{#1}{#2pt}%
  \fontfamily{#3}\fontseries{#4}\fontshape{#5}%
  \selectfont}%
\fi\endgroup%
\begin{picture}(9492,4374)(226,-3823)
\thinlines
{\put(601,-511){\line( 1, 0){750}}
}%
{\put(601,-661){\line( 1, 0){750}}
}%
{\put(601,-811){\line( 1, 0){750}}
}%
{\put(601,-961){\line( 1, 0){750}}
}%
{\put(601,-1411){\line( 1, 0){750}}
}%
{\put(601,-1561){\line( 1, 0){750}}
}%
{\put(601,-1711){\line( 1, 0){750}}
}%
{\put(601,-1861){\line( 1, 0){750}}
}%
{\put(601,-2311){\line( 1, 0){750}}
}%
{\put(601,-2461){\line( 1, 0){750}}
}%
{\put(601,-2611){\line( 1, 0){750}}
}%
{\put(601,-2761){\line( 1, 0){750}}
}%
{\put(2101,-511){\line( 1, 0){750}}
}%
{\put(2101,-661){\line( 1, 0){750}}
}%
{\put(2101,-811){\line( 1, 0){750}}
}%
{\put(2101,-961){\line( 1, 0){750}}
}%
{\put(2851,389){\line(-1, 0){2250}}
}%
{\put(2851,239){\line(-1, 0){2250}}
}%
{\put(2851, 89){\line(-1, 0){2250}}
}%
{\put(2851,-61){\line(-1, 0){2250}}
}%
{\put(3601,-511){\line( 1, 0){750}}
}%
{\put(3601,-661){\line( 1, 0){750}}
}%
{\put(3601,-811){\line( 1, 0){750}}
}%
{\put(3601,-961){\line( 1, 0){750}}
}%
{\put(4351,-1411){\line(-1, 0){2250}}
}%
{\put(4351,-1561){\line(-1, 0){2250}}
}%
{\put(4351,-1711){\line(-1, 0){2250}}
}%
{\put(4351,-1861){\line(-1, 0){2250}}
}%
{\put(2851,-3211){\line(-1, 0){2250}}
}%
{\put(2851,-3361){\line(-1, 0){2250}}
}%
{\put(2851,-3511){\line(-1, 0){2250}}
}%
{\put(2851,-3661){\line(-1, 0){2250}}
}%
{\put(5101,-2311){\line( 1, 0){750}}
}%
{\put(5101,-2461){\line( 1, 0){750}}
}%
{\put(5101,-2611){\line( 1, 0){750}}
}%
{\put(5101,-2761){\line( 1, 0){750}}
}%
{\put(5851,-3211){\line(-1, 0){2250}}
}%
{\put(5851,-3361){\line(-1, 0){2250}}
}%
{\put(5851,-3511){\line(-1, 0){2250}}
}%
{\put(5851,-3661){\line(-1, 0){2250}}
}%
{\put(5851,389){\line(-1, 0){2250}}
}%
{\put(5851,239){\line(-1, 0){2250}}
}%
{\put(5851, 89){\line(-1, 0){2250}}
}%
{\put(5851,-61){\line(-1, 0){2250}}
}%
{\put(6601,-2311){\line( 1, 0){750}}
}%
{\put(6601,-2461){\line( 1, 0){750}}
}%
{\put(6601,-2611){\line( 1, 0){750}}
}%
{\put(6601,-2761){\line( 1, 0){750}}
}%
{\put(7351,-1411){\line(-1, 0){2250}}
}%
{\put(7351,-1561){\line(-1, 0){2250}}
}%
{\put(7351,-1711){\line(-1, 0){2250}}
}%
{\put(7351,-1861){\line(-1, 0){2250}}
}%
{\put(8851,389){\line(-1, 0){2250}}
}%
{\put(8851,239){\line(-1, 0){2250}}
}%
{\put(8851, 89){\line(-1, 0){2250}}
}%
{\put(8851,-61){\line(-1, 0){2250}}
}%
{\put(8101,-511){\line( 1, 0){750}}
}%
{\put(8101,-661){\line( 1, 0){750}}
}%
{\put(8101,-811){\line( 1, 0){750}}
}%
{\put(8101,-961){\line( 1, 0){750}}
}%
{\put(8101,-2311){\line( 1, 0){750}}
}%
{\put(8101,-2461){\line( 1, 0){750}}
}%
{\put(8101,-2611){\line( 1, 0){750}}
}%
{\put(8101,-2761){\line( 1, 0){750}}
}%
{\put(8851,-3211){\line(-1, 0){2250}}
}%
{\put(8851,-3361){\line(-1, 0){2250}}
}%
{\put(8851,-3511){\line(-1, 0){2250}}
}%
{\put(8851,-3661){\line(-1, 0){2250}}
}%
{\put(4351,-2311){\line(-1, 0){2250}}
}%
{\put(4351,-2461){\line(-1, 0){2250}}
}%
{\put(4351,-2611){\line(-1, 0){2250}}
}%
{\put(4351,-2761){\line(-1, 0){2250}}
}%
{\put(7351,-511){\line(-1, 0){2250}}
}%
{\put(7351,-661){\line(-1, 0){2250}}
}%
{\put(7351,-811){\line(-1, 0){2250}}
}%
{\put(7351,-961){\line(-1, 0){2250}}
}%
{\put(5851,389){\line( 1, 0){750}}
}%
{\put(5851,239){\line( 1, 0){750}}
}%
{\put(5851, 89){\line( 1, 0){750}}
}%
{\put(5851,-61){\line( 1, 0){750}}
}%
{\put(2851,-3211){\line( 1, 0){750}}
}%
{\put(2851,-3361){\line( 1, 0){750}}
}%
{\put(2851,-3511){\line( 1, 0){750}}
}%
{\put(2851,-3661){\line( 1, 0){750}}
}%
{\put(8101,-1411){\line( 1, 0){750}}
}%
{\put(8101,-1561){\line( 1, 0){750}}
}%
{\put(8101,-1711){\line( 1, 0){750}}
}%
{\put(8101,-1861){\line( 1, 0){750}}
}%
\put(9001,-1505){$\leftarrow$\hspace*{-3mm}\parbox{2cm}{
		\begin{center}\small
			output\\[-0.7mm]
			qubit
		\end{center}}}
{\put(1351,-2911){\framebox(750,2550){$V_0$}}
}%
{\put(2851,-1111){\framebox(750,1650){$Y_1$}}
}%
{\put(4351,-2911){\framebox(750,2550){$V_1$}}
}%
{\put(5851,-3811){\framebox(750,1650){$N_1$}}
}%
{\put(7351,-2911){\framebox(750,2550){$V_2$}}
}%
\put(326,-861){\makebox(0,0)[r]{\smash{{\SetFigFont{12}{14.4}{\familydefault}{\mddefault}{\updefault}{$\hilb{M}_Y$}%
}}}}
\put(326,-1761){\makebox(0,0)[r]{\smash{{\SetFigFont{12}{14.4}{\familydefault}{\mddefault}{\updefault}{$\hilb{V}$}%
}}}}
\put(326,-2661){\makebox(0,0)[r]{\smash{{\SetFigFont{12}{14.4}{\familydefault}{\mddefault}{\updefault}{$\hilb{M}_N$}%
}}}}
\put(326,-3561){\makebox(0,0)[r]{\smash{{\SetFigFont{12}{14.4}{\familydefault}{\mddefault}{\updefault}{$\hilb{N}$}%
}}}}
\put(326, 39){\makebox(0,0)[r]{\smash{{\SetFigFont{12}{14.4}{\familydefault}{\mddefault}{\updefault}{$\hilb{Y}$}%
}}}}
\end{picture}%

%% file: QIPinSQG.tex
\chapter{A Lower Bound for Short Quantum Games}
\label{ch:QIPinSQG}

In this chapter we prove that $\cls{QIP} \subseteq \cls{SQG}_*$, which is the
main result of Reference \cite{GutoskiW05}.
In order to prove this containment we exhibit a short quantum game that solves
the $\cls{QIP}$-complete problem \textsc{close-images} with completeness error
$\frac{1}{2}$ and exponentially small soundness error.
To show the correctness of our game, we prove an information-theoretic assertion
that there exists a quantum measurement that reliably distinguishes between
quantum states chosen from two disjoint convex sets.
We then prove a general error reduction technique for short quantum games that
allows us to produce a game for \textsc{close-images} in which both the
completeness error and soundness error are exponentially small.

We start by defining the \textsc{close-images} problem in Section
\ref{sec:QIPinSQG:CI}.
We prove the quantum measurement result in Section \ref{sec:QIPinSQG:separation}
and use that result in Section \ref{sec:QIPinSQG:game} to yield a short quantum
game for \textsc{close-images}.
We finish the chapter with our error reduction result in Section
\ref{sec:QIPinSQG:error}.

\section{The Close-Images Problem}
\label{sec:QIPinSQG:CI}

Before we can define \textsc{close-images}, we must expand our repertoire of
quantum formalism.
In particular, we discuss the simulation of any physically realizable quantum
process by means of a unitary quantum circuit in Section
\ref{subsec:QIPinSQG:CI:MixedCircuits}.
In Section \ref{subsec:QIPinSQG:CI:distance} we consider several different
distance measures for quantum states and we settle on the trace norm as the
distance measure of choice for this thesis.
A formal statement of the \textsc{close-images} problem appears in Section
\ref{subsec:QIPinSQG:CI:CI} along with some comments concerning that problem.

\subsection{Mixed-State Quantum Circuits}
\label{subsec:QIPinSQG:CI:MixedCircuits}

In Section \ref{subsec:intro:quantum:circuits} we described a standard model of
quantum circuits in which every $m$-qubit input state is unitarily mapped to an
$m$-qubit output state.
We also mentioned in that section that any physically realizable quantum process
can be simulated by a unitary quantum circuit.

In particular, suppose $\hilb{F}$ and $\hilb{G}$ are Hilbert spaces
corresponding to the $m$ input qubits and $n$ output qubits of some physical
process
$$\Phi : \mapset{D}{\hilb{F}} \to \mapset{D}{\hilb{G}}.$$
Then there exists a unitary matrix
$U \in \mapset{U}{\hilb{F} \otimes \hilb{G} \otimes \hilb{G}'}$ satisfying
$$\Phi(\rho) = \Ptr{\hilb{F} \otimes \hilb{G}'}
{U \Paren{ \rho \otimes \ket{0_{\hilb{G} \otimes \hilb{G}'}}
\bra{0_{\hilb{G} \otimes \hilb{G}'}} } U^*}$$
for every $\rho \in \mapset{D}{\hilb{F}}$ where $\hilb{G}'$ is a new Hilbert
space with $\dim(\hilb{G}') = \dim(\hilb{G})$.
In consideration with the discussion in Section
\ref{subsec:intro:quantum:circuits}, it follows that the matrix $U$ can be
approximated as closely as desired by a quantum circuit $Q$ acting on $m+2n$
qubits.
In this construction, the vector
$\ket{0_{\hilb{G} \otimes \hilb{G}'}}$
describes the initial pure state of the remaining $2n$ input qubits of $Q$.

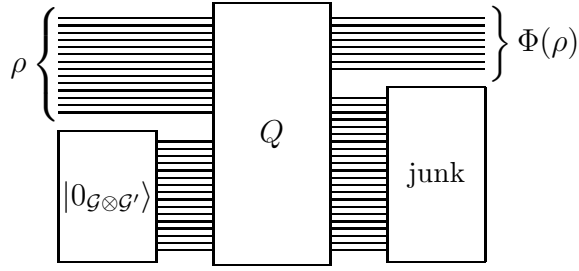
\begin{figure}
\begin{center}
\input{mixed.latex}
\end{center}
\caption{Simulation of the physical process $\Phi$ via a quantum circuit $Q$.}
\label{fig:mixed}
\end{figure}

Figure \ref{fig:mixed} illustrates such a circuit.
This circuit is sometimes called the Stinespring Dilation of $\Phi$ --- its
existence is typically attributed to Choi \cite{Choi75} and
a proof may be found in Kitaev, Shen, and Vyali \cite{KitaevS+02}.
The quantum circuits described in this subsection are called
\emph{mixed-state quantum circuits} in order to differentiate them from the
unitary quantum circuits of Section \ref{subsec:intro:quantum:circuits}.

Because quantum circuits are composed of a finite number of gates chosen from a
finite set of universal gates, it is clear that mixed-state quantum circuits,
like unitary circuits, can be encoded into a finite binary string
$x \in \set{0,1}^*$.
It therefore makes sense, for example, to define languages over $\set{0,1}^*$ in
terms of mixed-state quantum circuits.

\subsection{Distance Measures for Quantum States}
\label{subsec:QIPinSQG:CI:distance}

Given two quantum states $\rho, \xi \in \mapset{D}{\hilb{F}}$, how ``close'' are
they to each other?  In particular, with what reliability can these two states
be distinguished by a measurement?
A natural way to address the first of these questions is in the context of some
norm defined on $\mapset{L}{\hilb{F}}$.
It turns out, as we shall soon see, that the second question can be answered if
we consider the right norm.

It is standard to define the distance between two matrices $\rho$ and $\xi$ as
the norm of the difference $\rho - \xi$.
But different norms induce different distance measures, some of which are more
physically meaningful than others.
For example, we used the spectral norm in Section
\ref{subsec:intro:quantum:circuits} to formalize the notion that any
unitary matrix can be approximated arbitrarily closely by a quantum circuit.
Although the spectral norm was sufficient for that purpose, the quantity
$\norm{\rho - \xi}$ is not known to have much physical meaning.
Hence, all that can be said for certain is that if $\norm{\rho - \xi}$ is small
then $\rho$ and $\xi$ must be ``close'' and that the two are equal if this
quantity is zero.

A norm that is much more useful for quantifying the distinguishability of
quantum states is the \emph{trace norm}, defined for all
$A \in \mapset{L}{\hilb{F}}$ as
$$\TNorm{A} = \Ptr{}{\sqrt{A^* A}}.$$
To make sense of this definition, we point out that
$A^* A$ is positive semidefinite for every $A \in \mapset{L}{\hilb{F}}$ and
that, for every $P \in \mapset{Pos}{\hilb{F}}$, there is a unique
$Q \in \mapset{Pos}{\hilb{F}}$ satisfying $Q^2 = P$.
This matrix $Q$ is called the \emph{square root} of $P$ and is denoted
$Q = \sqrt{P}$.
Thus, the trace norm of an arbitrary matrix $A \in \mapset{L}{\hilb{F}}$ is the
trace of the unique positive semidefinite matrix $\sqrt{A^* A}$.
If $A \in \mapset{H}{\hilb{F}}$ is Hermitian then $\tnorm{A}$ is just the sum of
the absolute values of the eigenvalues of $A$.
In comparison, the spectral norm $\norm{A}$ of a Hermitian matrix $A$ is the
maximum of the absolute values of the eigenvalues of $A$.
The trace norm and the spectral norm are \emph{dual} to each other with respect
to the Hilbert-Schmidt inner product, meaning that the following fact holds (see
Bhatia \cite{Bhatia97}):

\begin{fact}[Duality of the Spectral and Trace Norms]
\label{fact:QIPinSQG:dual}

For every $A \in \mapset{L}{\hilb{F}}$ we have
\begin{align*}
\Norm{A} &= \max \Set{ \Abs{ \Inner{B,A} } : B \in \mapset{L}{\hilb{F}},
\TNorm{B} \leq 1 }, \\
\TNorm{A} &= \max \Set{ \Abs{ \Inner{B,A} } : B \in \mapset{L}{\hilb{F}},
\Norm{B} \leq 1 }.
\end{align*}

\end{fact}

For every $\rho_0, \rho_1 \in \mapset{D}{\hilb{F}}$, the quantity
$\tnorm{ \rho_0 - \rho_1 }$ lies in the interval $[0,2]$.
The trace norm characterizes the distinguishability of $\rho_0$ and
$\rho_1$ in the following sense:
there exists a binary-valued POVM such that if $\rho \in \set{\rho_0, \rho_1}$
is chosen uniformly at random then the POVM correctly determines which of
$\rho_0$ or $\rho_1$ was chosen with probability
$$\frac{1}{2} + \frac{1}{4} \TNorm{\rho_0 - \rho_1}.$$
Furthermore, such a POVM is optimal in the sense that no other quantum
measurement could possibly distinguish between $\rho_0$ and $\rho_1$ with a
higher rate of success.
It is because of this property that the trace norm is often a very convenient
and satisfactory distance measure for quantum states.
The quantity $\tnorm{ \rho_0 - \rho_1 }$ is sometimes called the
\emph{trace distance} between $\rho_0$ and $\rho_1$.

Incidentally, even if $\rho$ were chosen from $\set{\rho_0, \rho_1}$ according
to some arbitrary and unknown distribution, it can still be shown that the same
POVM will correctly distinguish between $\rho_0$ and $\rho_1$ with probability
at least $\frac{1}{2} \TNorm{\rho_0 - \rho_1}$.

Besides the trace norm, several other distance measures exist for quantum
states.
One example of such a measure is the \emph{fidelity}, which is a function
$$F : \mapset{Pos}{\hilb{F}} \times \mapset{Pos}{\hilb{F}} \to \mathbb{R}$$
defined by the expression
$$F(X, Y) = \TNorm{ \sqrt{X} \sqrt{Y} }$$
for every $X,Y \in \mapset{Pos}{\hilb{F}}$.
If $\rho$ and $\xi$ are density matrices then $F(\rho,\xi)$ always lies in the
interval $[0,1]$.
Furthermore, $F(\rho,\xi) = 1$ if and only if $\rho = \xi$ and $F(\rho,\xi) = 0$
if and only if $\rho$ and $\xi$ describe perfectly distinguishable quantum
states.
The fidelity and the trace norm are related by the following inequalities, which
hold for every $\rho, \xi \in \mapset{D}{\hilb{F}}$ (see Fuchs and van de Graaf
\cite{FuchsvdG99}):
\begin{equation}
\label{eq:QIPinSQG:FtoTNorm}
1 - \frac{1}{2} \TNorm{ \rho - \xi }
\leq F( \rho, \xi )
\leq \sqrt{ 1 - \frac{1}{4} \TNorm{ \rho - \xi }^2 }.
\end{equation}
Although the fidelity satisfies many useful properties, we need not consider it
any further because most expressions involving the fidelity can be converted
into expressions involving the trace norm via (\ref{eq:QIPinSQG:FtoTNorm}) and
because the trace norm adequately meets our needs in this thesis.

\subsection{Statement of the Problem}
\label{subsec:QIPinSQG:CI:CI}

For any mixed-state quantum circuit
$Q : \mapset{D}{\hilb{F}} \to \mapset{D}{\hilb{G}}$,
the \emph{image} of $Q$ is the set
$$\Set{ Q(\rho) : \rho \in \mapset{D}{\hilb{F}} } \subseteq
\mapset{D}{\hilb{G}}.$$
Given mixed-state quantum circuits $Q_0$ and $Q_1$, the definition of
\textsc{close-images} promises that the images of $Q_0$ and $Q_1$ either
intersect or are disjoint.

More formally, the \textsc{close-images} problem---parameterized by any desired
function $\varepsilon \in 2^{-\mathit{poly}}$---is defined as in Figure
\ref{fig:QIPinSQG:CI}.
\begin{figure}
\hrulefill
\begin{description}

\item[Problem.] $\textsc{close-images}(\varepsilon)$.

\item[Input.]

Two mixed-state quantum circuits
$Q_0, Q_1 : \mapset{D}{\hilb{F}} \to \mapset{D}{\hilb{G}}$
acting on $m$-qubit states.

\item[Promise.]

Exactly one of the following conditions holds:
\begin{enumerate}

\item
There exist $m$-qubit states $\rho_0, \rho_1 \in \mapset{D}{\hilb{F}}$ such that
$$Q_0(\rho_0) = Q_1(\rho_1).$$

\item
For all $m$-qubit states $\rho_0, \rho_1 \in \mapset{D}{\hilb{F}}$,
$Q_0(\rho_0)$ and $Q_1(\rho_1)$ satisfy
$$\TNorm{ Q_0(\rho_0) - Q_1(\rho_1) } > 2 - \varepsilon(m).$$

\end{enumerate}

\item[Output.]

``Accept'' if condition 1 holds, ``reject'' if condition 2 holds.

\end{description}
\hrulefill
\caption{Definition of \textsc{close-images}}
\label{fig:QIPinSQG:CI}
\end{figure}
This problem was implicitly shown to be $\cls{QIP}$-complete in Reference
\cite{KitaevW00}.
The statement presented in Figure \ref{fig:QIPinSQG:CI} is based upon the
formulation found in Reference \cite{RosgenW05}.
In that paper, condition 2 in the promise is stated using the fidelity instead
of the trace norm.
However, it is more convenient for our purposes to rephrase the problem in terms
of the trace norm.
That this rephrased version is equivalent to the original follows from
(\ref{eq:QIPinSQG:FtoTNorm}).

\section{Distinguishing Convex Sets of States}
\label{sec:QIPinSQG:separation}

In Section \ref{subsec:QIPinSQG:CI:distance} we pointed out that the trace norm
is a distance measure for quantum states that characterizes the
distinguishability of two states $\rho_0, \rho_1 \in \mapset{D}{\hilb{F}}$.
In this section we generalize that notion from single states
$\rho_0, \rho_1 \in \mapset{D}{\hilb{F}}$ to sets of states
$\mathcal{A}_0, \mathcal{A}_1 \subseteq \mapset{D}{\hilb{F}}$.
We motivate discussion of this generalization in Section
\ref{subsec:QIPinSQG:separation:preamble} before we state and prove our result
in Section \ref{subsec:QIPinSQG:separation:proof}.

\subsection{Motivation and Preamble}
\label{subsec:QIPinSQG:separation:preamble}

Let $V$ be any vector space over $\mathbb{R}$ or $\mathbb{C}$ (for example, a
Hilbert space $\hilb{F}$ and the set $\mapset{L}{\hilb{F}}$ are vector spaces
over $\mathbb{C}$).
A set $C \subseteq V$ is \emph{convex} if for every $x,y \in C$ and
$\lambda \in [0,1]$ we have $\lambda x + (1 - \lambda)y \in C$.
It follows from the fact that density matrices have unit trace that
$\mapset{D}{\hilb{F}}$ is a convex subset of $\mapset{L}{\hilb{F}}$.
Because mixed-state quantum circuits act linearly on their input qubits, it
follows that the image of any mixed-state quantum circuit is convex.
Hence, many results pertaining to convexity can be applied to the images of
mixed-state quantum circuits.

In particular, the separation theorems of convex analysis tell us
that between any two disjoint convex sets there exists a hyperplane that
separates them.
Typically, the separation results are stated in terms of the vector space
$\mathbb{R}^n$.
At first, the restriction to real numbers might seem like a problem.
But fortunately, the set $\mapset{H}{\hilb{F}}$ of complex Hermitian
matrices acting on a Hilbert space $\hilb{F}$ of dimension $n$ is readily shown
to be isomorphic to $\mathbb{R}^{n^2}$ and can therefore be regarded as a vector
space over $\mathbb{R}$.
Because density matrices are always positive semidefinite and hence Hermitian,
it follows that the separation results apply without complication to convex sets
of density matrices such as the image of a quantum circuit.
The separation result of most use to us, recast in terms of
$\mapset{H}{\hilb{F}}$, is stated as follows (see Rockafellar
\cite{Rockafellar70}):

\begin{fact}[Separation Theorem]
\label{fact:QIPinSQG:hyperplane}

Let $\mathcal{A}, \mathcal{B} \subset \mapset{H}{\hilb{F}}$ be disjoint convex
sets with $\mathcal{A}$ compact and $\mathcal{B}$ open.
There exists $H \in \mapset{H}{\hilb{F}}$ and $a \in \mathbb{R}$ such that
$\inner{H,X} \geq a > \inner{H,Y}$ for every $X \in \mathcal{A}$ and
$Y \in \mathcal{B}$.

\end{fact}

By choosing suitable convex sets upon which to apply the Separation Theorem
(Fact \ref{fact:QIPinSQG:hyperplane}),
we can use the the corresponding hyperplane to define a quantum measurement that
distinguishes between disjoint images of two mixed-state quantum circuits.
As one might expect, such a measurement is useful for solving problems such as
\textsc{close-images}, in which disjoint images must be distinguished from
overlapping images.

\subsection{A Generalized Distinguishability Result}
\label{subsec:QIPinSQG:separation:proof}

Our goal in this subsection is to solve the following generalization of the
distinguishability problem.
We are given $\rho \in \mapset{D}{\hilb{F}}$ chosen according to some arbitrary
and unknown distribution from the set $\set{\rho_0, \rho_1}$.
All we know about $\rho_0$ and $\rho_1$ is that $\rho_0 \in \mathcal{A}_0$ and
$\rho_1 \in \mathcal{A}_1$ where
$\mathcal{A}_0, \mathcal{A}_1 \subseteq \mapset{D}{\hilb{F}}$
are convex sets of density matrices.
Our task is to determine which of $\rho_0$ or $\rho_1$ was chosen.
With what probability can we correctly make this distinction?

The answer to this question depends on the minimal trace distance between
$\mathcal{A}_0$ and $\mathcal{A}_1$ in much the same way as the
distinguishability of $\rho_0, \rho_1 \in \mapset{D}{\hilb{F}}$ depends upon the
trace distance between $\rho_0$ and $\rho_1$.
Our formalization of this answer begins with the following theorem.

\begin{thm}[Distinguishability of Sets of States]
\label{thm:QIPinSQG:separate}

Let $\mathcal{A}_0, \mathcal{A}_1 \subseteq \mapset{D}{\hilb{F}}$ be closed
convex sets of density matrices and let $d$ denote the minimum of
$\tnorm{ \rho_0 - \rho_1 }$ over all $\rho_0 \in \mathcal{A}_0$ and all
$\rho_1 \in \mathcal{A}_1$.
There exists a binary-valued POVM such that, for every pair
$\rho_0 \in \mathcal{A}_0$ and $\rho_1 \in \mathcal{A}_1$, if
$\rho$ is chosen uniformly at random from $\set{ \rho_0, \rho_1 }$
then the POVM will correctly determine which of
$\rho_0$ or $\rho_1$ was chosen with probability at least
$\frac{1}{2} + \frac{d}{4}$.

\end{thm}

\begin{proof}

We preface this proof with some comments regarding the quantum measurement that
we are to construct.
The POVM that we apply to $\rho$ will have outcomes in $\Gamma = \set{0,1}$
where outcome $0 \in \Gamma$ indicates that $\rho_0$ was chosen and outcome
$1 \in \Gamma$ indicates that $\rho_1$ was chosen.
In accordance with Section \ref{subsec:intro:quantum:measurement}, the POVM with
outcomes in $\Gamma$ will be given by the set
$\set{E_0, E_1} \subset \mapset{Pos}{\hilb{F}}$ of positive semidefinite
matrices satisfying $E_0 + E_1 = I_\hilb{F}$.

Let $C$ denote the event that our POVM yields the correct outcome.
That is, the event $C$ is said to occur if
$\rho = \rho_0$ and we obtain outcome $0 \in \Gamma$ or if
$\rho = \rho_1$ and we obtain outcome $1 \in \Gamma$.
According to Section \ref{subsec:intro:quantum:measurement}, we have
\begin{align*}
\Pr[C \, | \, \rho = \rho_0] &= \inner{E_0, \rho_0}, \\
\Pr[C \, | \, \rho = \rho_1] &= \inner{E_1, \rho_1}.
\end{align*}
As $\rho$ is chosen uniformly at random, we can combine the previous two
conditional probabilities to obtain
$$\Pr[C]
= \frac{1}{2} \Inner{E_0, \rho_0} + \frac{1}{2} \Inner{E_1, \rho_1}.$$
By similar reasoning it follows that
$$\Pr[\lnot C]
= \frac{1}{2} \Inner{E_1, \rho_0} + \frac{1}{2} \Inner{E_0, \rho_1}$$
and hence
$$\Pr[C] -  \Pr[\lnot C] = \frac{1}{2} \Inner{E_0 - E_1, \rho_0 - \rho_1 }.$$
This expression will be of use later in this proof.

We are now ready to begin the proof in earnest.
If the minimum $d = 0$ then it suffices that our POVM be as good as a random
coin flip.
By choosing $E_0 = E_1 = \frac{1}{2}I$ we obtain
$$\Pr[C] = \frac{1}{2} \ptr{}{\rho_0} + \frac{1}{2} \ptr{}{\rho_1}
= \frac{1}{2},$$
which holds for every $\rho_0, \rho_1 \in \mapset{D}{\hilb{F}}$ as desired.
Hence, for the remainder of this proof we assume that $d > 0$.

We define
$$\mathcal{A} = \mathcal{A}_0 - \mathcal{A}_1 = \Set{ \rho_0 - \rho_1 :
\rho_0 \in \mathcal{A}_0, \rho_1 \in \mathcal{A}_1 }.$$
The set $\mathcal{A} \subset \mapset{H}{\hilb{F}}$ is a closed convex set of
Hermitian matrices such that $\tnorm{X} \geq d$ for every $X \in \mathcal{A}$.
Let
$$\mathcal{B} = \Set{ Y \in \mapset{H}{\hilb{F}} : \TNorm{Y} < d }$$
denote the open ball of radius $d$ with respect to the trace norm.
The sets $\mathcal{A}$ and $\mathcal{B}$ satisfy the conditions of the
Separation Theorem (Fact \ref{fact:QIPinSQG:hyperplane}), and therefore there
exists a hyperplane that separates them.
That is, there exists a Hermitian matrix $H \in \mapset{H}{\hilb{F}}$ and a real
number $a \in \mathbb{R}$ such that
$$\Inner{H,X} \geq a > \Inner{H,Y}$$
for every $X \in \mathcal{A}$ and $Y \in \mathcal{B}$.
Note that because the ball $\mathcal{B}$ is centred at the origin, we have
$Y \in \mathcal{B}$ if and only if $-Y \in \mathcal{B}$ and hence
$a > \Inner{H,Y}$ and $a > \Inner{H,-Y},$
from which it follows that $a > \Inner{H,Y} > -a$ and in particular $a > 0$.

We now use the Hermitian matrix $H$ and the positive reals $a$ and $d$ to
construct our POVM $\set{E_0, E_1}$.
Let $K = \frac{d}{a} H$.
Then $\inner{K,X} \geq d$ for every $X \in \mathcal{A}$ and
$\inner{K,\frac{1}{d} Y} < 1$ for every $Y \in \mathcal{B}$.
As $\frac{1}{d} Y$ ranges over all Hermitian matrices with trace norm smaller
than 1, it follows from the Duality of the Spectral and Trace Norms
(Fact \ref{fact:QIPinSQG:dual}) that $\norm{K} \leq 1$.

Now let $K_+, K_- \in \mapset{Pos}{\hilb{F}}$ be the Jordan decomposition of
$K$, meaning that $K = K_+ - K_-$ and $K_+$ and $K_-$ act on orthogonal
subspaces of $\hilb{F}$.
It follows that
$$\norm{K_+ + K_-} = \norm{K_+ - K_-} = \norm{K} \leq 1$$
and hence $I - K_+ - K_-$ is positive semidefinite.

The matrices $E_0$ and $E_1$ composing our binary-valued POVM are given by
\begin{align*}
E_0 &= K_+ + \frac{1}{2} \Paren{ I - K_+ - K_- }, \\
E_1 &= K_- + \frac{1}{2} \Paren{ I - K_+ - K_- }.
\end{align*}
Of course, $E_0$ and $E_1$ are positive semidefinite and satisfy
\begin{align*}
E_0 + E_1 &= I, \\
E_0 - E_1 &= K.
\end{align*}
We now compute the probability with which the POVM yields the correct outcome.
$$\Pr[C] - \Pr[\lnot C]
= \frac{1}{2} \Inner{K, \rho_0 - \rho_1 }
\geq \frac{d}{2}$$
with the inequality following from the fact that
$K = \frac{d}{a} H$ and $\rho_0 - \rho_1 \in \mathcal{A}$ for every
$\rho_0 \in \mathcal{A}_0$ and $\rho_1 \in \mathcal{A}_1$.
 From here, it is straightforward to solve the system
\begin{align*}
\Pr[C] - \Pr[\lnot C] &\geq \frac{d}{2} \\
\Pr[C] + \Pr[\lnot C] &= 1
\end{align*}
and obtain $\Pr[C] \geq \frac{1}{2} + \frac{d}{4}$ as desired.
\end{proof}

We now use the Distinguishability of Sets of States
(Theorem \ref{thm:QIPinSQG:separate}) to obtain a result that holds
even when $\rho$ is chosen nonuniformly from $\set{ \rho_0, \rho_1 }$.

\begin{cor}
\label{thm:QIPinSQG:separate:cor}

Let $\mathcal{A}_0$, $\mathcal{A}_1$, and $d$ be defined as in the statement of
Theorem \ref{thm:QIPinSQG:separate}.
The binary-valued POVM $\set{ E_0, E_1 }$ from the proof of Theorem
\ref{thm:QIPinSQG:separate} satisfies the following property.
For every pair $\rho_0 \in \mathcal{A}_0$ and $\rho_1 \in \mathcal{A}_1$, if
$\rho$ is chosen from $\set{ \rho_0, \rho_1 }$ according to some arbitrary and
unknown distribution then $\set{ E_0, E_1 }$ will correctly determine which of
$\rho_0$ or $\rho_1$ was chosen with probability at least $\frac{d}{2}$.

\end{cor}

\begin{proof}

As in the proof of Theorem \ref{thm:QIPinSQG:separate},
let $C$ denote the event that $\set{ E_0, E_1 }$ yields the correct outcome.
We start by pointing out that any binary probability distribution over
$\set{ \rho_0, \rho_1 }$ can be expressed as a composition of the uniform
distribution and a zero-entropy distribution.
It will then suffice to show that $\Pr[C] \geq \frac{d}{2}$ under these two
distributions.

Assume for now that $\rho_0$ is the more likely choice---the case in which
$\rho_1$ is the more likely choice will follow by symmetry.
In other words, we assume $\rho$ is chosen from $\set{ \rho_0, \rho_1 }$ so that
$\rho = \rho_0$ with probability $\lambda$ for some
$\lambda \in [\frac{1}{2}, 1]$.

Consider the following composite distribution.
With probability $2 - 2 \lambda$ we choose $\rho$ uniformly at random from
$\set{ \rho_0, \rho_1 }$ and with probability $2 \lambda - 1$ we choose
$\rho = \rho_0$ with certainty.
It follows that, under this composite distribution, $\rho_0$ is chosen with
probability $\lambda$ as desired.

Under the uniform distribution ($\lambda = \frac{1}{2}$),
Theorem \ref{thm:QIPinSQG:separate} tells us that
$$\Pr[C] \geq \frac{1}{2} + \frac{d}{4} \geq \frac{d}{2}.$$
Hence, it suffices to show that $\Pr[C] \geq \frac{d}{2}$ under the
zero-entropy distribution ($\lambda = 1$).
We achieve this bound using the following facts from the proof of Theorem
\ref{thm:QIPinSQG:separate}:
\begin{eqnarray*}
\Pr[C \, | \, \rho = \rho_0] &=& \Inner{ E_0, \rho_0 }, \\
\Pr[\lnot C \, | \, \rho = \rho_0] &=& \Inner{ E_1, \rho_0 }, \\
\Inner{ E_0 - E_1, \rho_0 - \rho_1 } &\geq& d, \\
\Norm{ E_0 - E_1 } &\leq& 1.
\end{eqnarray*}
Since $\rho = \rho_0$ with certainty, it follows from the first two expressions
that
$$\Pr[C] - \Pr[\lnot C] = \Inner{ E_0 - E_1, \rho_0 }.$$
The third expression implies that
$$\Inner{ E_0 - E_1, \rho_0 } \geq d + \Inner{ E_0 - E_1, \rho_1 }.$$
It follows from the fourth expression and from the Duality of the Spectral and
Trace Norms (Fact \ref{fact:QIPinSQG:dual}) that
$$\Abs{ \Inner{ E_0 - E_1, \rho_1 } } \leq \TNorm{ \rho_1 } = 1$$
and hence $\Inner{ E_0 - E_1, \rho_1 } \geq -1$.
Combining all these inequalities, we obtain
$$\Pr[C] - \Pr[\lnot C] \geq d - 1$$
from which it follows that $\Pr[C] \geq \frac{d}{2}$ as desired.
\end{proof}

The most important lesson to take away from the Distinguishability of Sets of
States (Theorem \ref{thm:QIPinSQG:separate}) and Corollary
\ref{thm:QIPinSQG:separate:cor} is that the POVM $\set{ E_0, E_1 }$ depends only
upon $\mathcal{A}_0$ and $\mathcal{A}_1$ and not on any particular pair of
density matrices in those sets.
In other words, the very same quantum measurement can be used to distinguish
between every pair of density matrices chosen from those sets.
This independence is critical to the correctness of our solution to
\textsc{close-images}.

\section{A Short Quantum Game for Close-Images}
\label{sec:QIPinSQG:game}

In this section, we prove that any language with a quantum interactive proof
system also has a short quantum game by solving the $\cls{QIP}$-complete problem
\textsc{close-images} from Section \ref{subsec:QIPinSQG:CI:CI}.
In order to prove membership in $\cls{SQG}_*(c,s)$, we must exhibit a verifier
for a short quantum game who satisfies the completeness and soundness conditions
stated in Section \ref{sec:defs:formalizations}.
Such a verifier receives one message from the yes-prover and then exchanges a
round of messages with the no-prover before deciding whether to accept the
input.

Informally, the verifier we seek obeys the following protocol: given
descriptions of two mixed-state quantum circuits $Q_0$ and $Q_1$, the verifier
receives states $\rho_0$ and $\rho_1$ from the yes-prover, randomly chooses $i
\in \set{0,1}$, applies $Q_i$ to $\rho_i$, and forwards the result to the
no-prover.
The no-prover is then challenged to identify which of $Q_0(\rho_0)$ and
$Q_1(\rho_1)$ was sent to him.
If he succeeds then the verifier assumes that the no-prover can reliably
distinguish between $Q_0(\rho_0)$ and $Q_1(\rho_1)$ and hence the images of
$Q_0$ and $Q_1$ are far apart.
If he fails then the verifier assumes that the no-prover cannot reliably
distinguish between $Q_0(\rho_0)$ and $Q_1(\rho_1)$ because they are equal and
hence the images of $Q_0$ and $Q_1$ intersect.
This argument is formalized in the following theorem.

\begin{thm}
\label{thm:QIPinSQG:game}

$\cls{QIP} \subseteq \cls{SQG}_*(\frac{1}{2},\varepsilon)$ for every
$\varepsilon \in 2^{-\poly}$.

\end{thm}

\begin{proof}

Given any $\varepsilon \in 2^{-\poly}$, it suffices to show that
$\textsc{close-images}(\varepsilon)$ is in
$\cls{SQG}_*(\frac{1}{2},\varepsilon)$.
Let
$$Q_0,Q_1 : \mapset{D}{\hilb{F}} \to \mapset{D}{\hilb{G}}$$
be any given mixed-state quantum circuits acting on $m$ qubits and let
$\mathcal{A}_i$ denote the image of $Q_i$ for $i \in \set{0,1}$.
The sets $\mathcal{A}_0, \mathcal{A}_1 \subseteq \mapset{D}{\hilb{G}}$ are
closed convex sets of density operators.

Consider the verifier for a short quantum game described in Figure
\ref{fig:QIPinSQG:verifier}.
\begin{figure}
\hrulefill
\begin{enumerate}

\item

Receive $m$-qubit registers $\mathsf{X}_0$ and $\mathsf{X}_1$ from the
yes-prover.

\item

Choose $i \in \set{0,1}$ uniformly at random and apply $Q_i$ to register
$\mathsf{X}_i$.
Let the output be contained in a register $\mathsf{Y}$, which is then sent to
the no-prover.

\item

Receive a classical bit $b$ from the no-prover.  Accept if $b \neq i$ and reject
if $b = i$.

\end{enumerate}
\hrulefill
\caption{Verifier's protocol for Theorem \ref{thm:QIPinSQG:game}}
\label{fig:QIPinSQG:verifier}
\end{figure}
If $(Q_0,Q_1)$ is a ``yes'' instance of $\textsc{close-images}(\varepsilon)$
then there exist $\rho_0, \rho_1 \in \mapset{D}{\hilb{F}}$ such that
$Q_0(\rho_0) = Q_1(\rho_1)$.
The strategy for the yes-prover is to prepare the registers $\mathsf{X}_0$ and
$\mathsf{X}_1$ in the states $\rho_0$ and $\rho_1$ respectively and to send them
to the verifier in step 1 of the verifier's protocol.
Because $Q_0(\rho_0) = Q_1(\rho_1)$, the state contained in the register
$\mathsf{Y}$ is independent of $i$, so the no-prover can do no better than a
random guess in step 3.
The verifier will therefore accept with probability at most $\frac{1}{2}$ in
this case.

Let $d$ be the minimum of $\TNorm{ \rho_0 - \rho_1 }$ over all choices of
$\rho_0 \in \mathcal{A}_0$ and $\rho_1 \in \mathcal{A}_1$.
If $(Q_0,Q_1)$ is a ``no'' instance of $\textsc{close-images}(\varepsilon)$
then we are promised that $d > 2 - \varepsilon(m)$.
Regardless of the state of the registers $\mathsf{X}_0$ and $\mathsf{X}_1$ sent
to the verifier by the yes-prover, we must have that the state
$\rho \in \mapset{D}{\hilb{G}}$ of the register $\mathsf{Y}$ sent to the
no-prover is in either $\mathcal{A}_0$ or $\mathcal{A}_1$.
Furthermore, we have
$$\Pr[\rho \in \mathcal{A}_0] = \Pr[\rho \in \mathcal{A}_1] = \frac{1}{2}.$$
Hence, by the Distinguishability of Sets of States
(Theorem \ref{thm:QIPinSQG:separate}) there exists a binary-valued POVM
that correctly determines whether $\rho \in \mathcal{A}_0$ or
$\rho \in \mathcal{A}_1$ with probability at least
$$\frac{1}{2} + \frac{d}{4} > 1 - \frac{\varepsilon(m)}{4}.$$
The strategy for the no-prover is to perform the quantum measurement from
Theorem \ref{thm:QIPinSQG:separate} and send the result to the verifier in step
3, which causes the verifier to reject with probability greater than
$1 - \frac{\varepsilon(m)}{4}$.
\end{proof}

\section{Error Reduction for Short Quantum Games}
\label{sec:QIPinSQG:error}

In this section we prove that short quantum games are at least partially robust
with respect to error in the sense that the completeness error can be made
exponentially small at the possible cost of an increase in the soundness error
and vice versa.
Fortunately, because the short quantum game for \textsc{close-images} in Section
\ref{sec:QIPinSQG:game} has exponentially small soundness error, any increase in
that quantity can be absorbed into the arbitrarily small factor, yielding a
short quantum game for \textsc{close-images} with exponentially small
completeness \emph{and} soundness error.

The error reduction technique we present in this section relies heavily upon
previous results in error reduction for quantum interaction.
We summarize the necessary material in Section
\ref{subsec:QIPinSQG:error:results} before proving our new result in Section
\ref{subsec:QIPinSQG:error:robust}.

\subsection{Parallel Repetition and Transformations}
\label{subsec:QIPinSQG:error:results}

In Section \ref{subsec:defs:remarks:results} we mentioned that any quantum
refereed game with reasonable error can be simulated by another quantum refereed
game with exponentially small error that repeats the initial game many times in
succession and then accepts based upon the outcomes of each of
the repetitions.
However, we pointed out in Section \ref{subsec:intro:complexity:results} that
sequential repetition of this form necessarily increases the number of rounds in
an interaction and so this technique does not apply to bounded-round
interactions such as short quantum games.
In the classical case, this problem was circumvented by identifying one-round
refereed games with $\cls{PSPACE}$.
Unfortunately, no analogous result is known to hold for short quantum games.

A natural approach to the task of error reduction for bounded-round interactions
is to run many copies of the interaction \emph{in parallel} and act as though
the repetitions were sequential, basing the decision to accept accordingly.
This technique, known as \emph{parallel repetition}, is purely classical and has
been successfully applied to classical single- and multi-prover interactive
proof systems (see Raz \cite{Raz98} and the references therein).
A potential problem with this technique is that the provers need not treat each
repetition independently---they might try to correlate the parallel repetitions
(or entangle them in the quantum case) in some devious way such that the
completeness or soundness error does not decrease as desired.

In the quantum setting, the general case of this problem has not been completely
solved.
But for three-message single-prover quantum interactive proof systems with zero
completeness error, Reference \cite{KitaevW00} proves that parallel repetition
followed by a unanimous vote does indeed achieve the exponential reduction in
soundness error that one might expect, regardless of any possible entanglement
by the prover among the parallel copies.

Because we will incorporate parts of the proof of this result into our
reduction, it is necessary to summarize some of the additional formalism upon
which it draws.
Toward that end, recall that $\mapset{L}{\hilb{F}}$ and $\mapset{L}{\hilb{G}}$
denote the sets of linear mappings acting on $\hilb{F}$ and $\hilb{G}$
respectively.
As $\mapset{L}{\hilb{F}}$ and $\mapset{L}{\hilb{G}}$ are themselves vector
spaces, it makes sense to consider the set $\mapset{T}{\hilb{F},\hilb{G}}$ of
linear mappings from $\mapset{L}{\hilb{F}}$ to $\mapset{L}{\hilb{G}}$, also
known as \emph{transformations}.
As one might expect, the Kronecker product extends naturally to transformations.

Recall also that the spectral norm on $\mapset{L}{\hilb{F}}$ is induced by the
Euclidean norm on $\hilb{F}$ by the relation
$$\Norm{A} = \sup_{v \in \hilb{F} \setminus \Set{0}} \frac{\Norm{Av}}
{\Norm{v}}$$
for every $A \in \mapset{L}{\hilb{F}}$.
We can extend the trace norm to $\mapset{T}{\hilb{F},\hilb{G}}$ in a similar
way: for any $T \in \mapset{T}{\hilb{F},\hilb{G}}$ we have
$$\TNorm{T} = \sup_{A \in \mapset{L}{\hilb{F}} \setminus \Set{0}}
\frac{\TNorm{T(A)}}{\TNorm{A}}.$$
However, this extension of the trace norm does not induce an overly desirable
metric on $\mapset{T}{\hilb{F},\hilb{G}}$ in part because its value can change
upon taking the Kronecker product of a transformation with the
identity transformation.
That is, there exist Hilbert spaces $\hilb{H}$ and transformations
$T \in \mapset{T}{\hilb{F},\hilb{G}}$ with
$$\TNorm{T} \neq \TNorm{ T \otimes I_{\mapset{L}{\hilb{H}}} }$$
where $I_{\mapset{L}{\hilb{H}}} \in \mapset{T}{\hilb{H},\hilb{H}}$ is the
identity transformation on $\mapset{L}{\hilb{H}}$.

With this fact in mind, the \emph{diamond norm} of a transformation
$T \in \mapset{T}{\hilb{F},\hilb{G}}$ is defined as
$$\DNorm{T} = \TNorm{ T \otimes I_{\mapset{L}{\hilb{K}}} }$$
where $\dim(\hilb{K}) = \dim(\hilb{F})$.
The diamond norm satisfies several convenient properties.
For one, it is robust with respect to taking the Kronecker product with the
identity.
Another nice property of the diamond norm is that it is multiplicative with
respect to the Kronecker product.
In other words,
$$\DNorm{T_1 \otimes T_2} = \DNorm{T_1} \DNorm{T_2}$$
for any choice of transformations $T_1$ and $T_2$.
Proofs of these and other properties of the diamond norm can be found in
Kitaev, Shen, and Vyali \cite{KitaevS+02}.

Now that we have introduced the diamond norm for transformations, we are ready
to discuss its relevance to quantum interaction.
In what follows, the projection $\Pi_\mathrm{init} \in \mapset{Pos}{\hilb{S}}$
is defined as
$$\Pi_\mathrm{init} = \ket{0_\hilb{S}} \bra{0_\hilb{S}}$$
where $\ket{0_\hilb{S}}$ is the initial pure state of any quantum interaction.
The fact upon which we base our error reduction result is stated as follows
(see Reference \cite[Lemma 7]{KitaevW00}):

\begin{fact}
\label{fact:QIPinSQG:dnorm}

Let $V(x)=(V_0,V_1)$ be a verifier for a one-round quantum interactive proof
system on input $x \in \set{0,1}^*$
(such an interaction consists of a message from the verifier to the prover
followed by the prover's response).
Let $T \in \mapset{T}{\hilb{M} \otimes \hilb{V}, \hilb{M}}$ be a transformation
defined as
$$T(X) = \Ptr{\hilb{V}}{(V_0 \Pi_\mathrm{init}) X (\Pi_\mathrm{accept} V_1) }$$
for every $X \in \mapset{L}{\hilb{M} \otimes \hilb{V}}$.
The maximum probability with which any prover could convince $V$ to accept $x$
is precisely $\dnorm{T}^2$.

\end{fact}

As we shall soon see, if we consider the Kronecker product of $T$ with itself
many times then the multiplicative property of the diamond norm in concert with
Fact \ref{fact:QIPinSQG:dnorm} yields an exponentially small upper bound on the
soundness error of repeated one-round quantum interactions.

\subsection{A Partial Robustness Result}
\label{subsec:QIPinSQG:error:robust}

In this subsection we prove that parallel repetition followed by a unanimous
vote can be used to improve the error bounds for short quantum games by reducing
the problem to error reduction for single-prover quantum interactive proof
systems with three or fewer messages.
The reduction is achieved by fixing a yes- or no-prover $P(x)$ that is
guaranteed to win with a certain probability.
By viewing the verifier-prover pair $(V,P)(x)$ as a new composite verifier, we
are left with what is now effectively a two-message quantum interactive proof
system in which the opposing prover is the lone prover.
We define a verifier-prover pair $(V',P')(x)$ that runs many copies of
$(V,P)(x)$ in parallel and accepts based upon a unanimous vote.
We can then employ Fact \ref{fact:QIPinSQG:dnorm} to prove that the error of the
new game decreases exponentially in the number of repetitions.

We are now prepared to give the main result of this section, whose proof is
based upon the proof of Theorem 6 in Reference \cite{KitaevW00}:

\begin{thm}[Partial Robustness of Short Quantum Games]
\label{thm:QIPinSQG:error}

$$\cls{SQG}(c,s) \subseteq \cls{SQG}(kc,s^k) \cap \cls{SQG}(c^k,ks)$$
for any choice of $c,s : \mathbb{N} \to [0,1]$ and $k \in \poly$.

\end{thm}

\begin{proof}
For brevity we write $k = k(|x|)$, $c = c(|x|)$, and $s = s(|x|)$ where
$x \in \set{0,1}^*$.
For any matrix $A$ and any positive integer $n$, we write
$A^{\otimes n} = A \otimes \dots \otimes A$ as shorthand for the $n$-fold
Kronecker product of $A$ with itself.

We first prove that $\cls{SQG}(c,s) \subseteq \cls{SQG}(kc,s^k)$.
Let $L \in \cls{SQG}(c,s)$ and let $V(x) = (V_0,V_1,V_2)$ be a verifier
witnessing this fact.
Let $V'(x) = (V_0^{\otimes k}, V_1^{\otimes k}, V_2^{\otimes k})$ be a verifier
that runs $k$ copies of the protocol of $V(x)$ in parallel and accepts if and
only if every one of the $k$ copies accepts.
We must show that $V'(x)$ has completeness error at most $kc$ and
soundness error at most $s^k$.

First consider the case $x \in L$.
Let $Y(x) = Y_1$ be a yes-prover that convinces $V(x)$ to accept $x$ with
probability at least $1 - c$.
Let $Y'(x) = Y_1^{\otimes k}$ be a yes-prover that runs $k$ independent copies
of the protocol of $Y(x)$ in parallel.
Then no no-prover can win any one of the $k$ copies with probability greater
than $c$ and so by the union bound we know that the completeness error of the
repeated game is at most $kc$.

Next consider the case $x \not \in L$.
Let $N(x) = N_1$ be a no-prover that convinces $V(x)$ to reject $x$ with
probability at least $1 - s$ and let $\Pi_\mathrm{init}$ be as defined in
Section \ref{subsec:QIPinSQG:error:results}.
As earlier intimated, we may view $(V,N)(x) = (V_0, V_2 N_1 V_1)$ as a new
one-round composite verifier and the yes-prover as the lone prover for some
two-message quantum interactive proof system.
Define the transformation
$$T_s(X) = \Ptr{\hilb{V} \otimes \hilb{M}_N \otimes \hilb{N}}
{ (V_0 \Pi_\mathrm{init}) X (\Pi_\mathrm{accept} V_2 N_1 V_1) }.$$
By Fact \ref{fact:QIPinSQG:dnorm} we know that the maximum probability with
which any prover could convince $(V,N)(x)$ to accept $x$ is $\dnorm{T_s}^2$.
As $(V,N)(x)$ has soundness error at most $s$, we have $\dnorm{T_s}^2 \leq s$.

Now let $N'(x) = N_1^{\otimes k}$ be a no-prover that runs $k$ independent
copies of the protocol of $N(x)$ in parallel.
We now show that no yes-prover can win against $N'(x)$ using verifier $V'(x)$
with probability greater than $s^k$.
Let $\Pi'_\mathrm{init} = \Pi_\mathrm{init}^{\otimes k}$ and
$\Pi'_\mathrm{accept} = \Pi_\mathrm{accept}^{\otimes k}$ be the projections
corresponding to the initial and accepting states of the repeated game.
Define the transformation
$$T'_s(X) =
\Ptr{\Paren{\hilb{V} \otimes \hilb{M}_N \otimes \hilb{N}}^{\otimes k}}
{ (V'_0 \Pi'_\mathrm{init}) X (\Pi'_\mathrm{accept} V'_2 N'_1 V'_1) }.$$
It is clear that $T'_s = T_s^{\otimes k}$.
By Fact \ref{fact:QIPinSQG:dnorm} and the multiplicativity of the diamond norm
it follows that the maximum probability with which any prover could convince
$(V',N')(x)$ to accept $x$ is
$$\DNorm{T'_s}^2 = \DNorm{T_s^{\otimes k}}^2 = \DNorm{T_s}^{2k} \leq s^k,$$
which establishes the desired result.

Due to the symmetric nature of quantum refereed games, we can modify the above
proof to show that $\cls{SQG}(c,s) \subseteq \cls{SQG}(c^k,ks)$.
In particular, define the verifier $V''(x)$ so that he rejects if and only if
all $k$ copies reject.
For the case $x \not \in L$, the proof that $V''(x)$ has soundness error $ks$ is
completely symmetric to the proof that $V'(x)$ has completeness error $kc$.

For the case $x \in L$, we let $Y(x) = Y_1$ be a yes-prover that convinces
$V(x)$ to accept with probability at least $1 - c$.
Let $(V,Y)(x) = (V_1 Y_1 V_0, V_2)$ be a new one-round composite verifier for a
two-message quantum interactive proof system in which the no-prover is the lone
prover.
The two differences here are that the prover's goal is now to convince
$(V,Y)(x)$ to reject $x$ instead of to accept $x$ and that the transformation
$T_s$ is now replaced with the transformation
$$T_c(X) = \Ptr{\hilb{Y} \otimes \hilb{M}_Y \otimes \hilb{V}}
{ (V_1 Y_1 V_0 \Pi_\mathrm{init}) X (\Pi_\mathrm{reject} V_2) }.$$

Fortunately, Fact \ref{fact:QIPinSQG:dnorm} still applies and so the maximum
probability with which any prover could convince $(V,Y)(x)$ to reject $x$ is
precisely $\dnorm{T_c}^2$.
That $V''(x)$ has completeness error $c^k$ follows as before.
\end{proof}

Of course, the Partial Robustness of Short Quantum Games
(Theorem \ref{thm:QIPinSQG:error}) holds in the special case where the
verifier $V(x) = (I,V_1,V_2)$ and so we obtain
$$\cls{SQG}_*(c,s) \subseteq \cls{SQG}_*(kc,s^k) \cap \cls{SQG}_*(c^k,ks)$$
as an easy corollary.
A more important corollary that follows from Theorems \ref{thm:QIPinSQG:game}
and \ref{thm:QIPinSQG:error} is the main result of this chapter:

\begin{cor}
$\cls{QIP} \subseteq \cls{SQG}_*$.
\end{cor}

\begin{proof}
Given a desired error bound $2^{-p}$ where $p \in \poly$, choose
$\varepsilon \in 2^{-\poly}$ so that
$p(n) \varepsilon(n) \leq 2^{-p(n)}$ for all $n \in \mathbb{N}$.
We have $\cls{QIP} \subseteq \cls{SQG}_*(\frac{1}{2},\varepsilon) \subseteq
\cls{SQG}_*(2^{-p},2^{-p})$.
\end{proof}

%% file: mixed.latex
\setlength{\unitlength}{1200sp}%
\begingroup\makeatletter\ifx\SetFigFont\undefined%
\gdef\SetFigFont#1#2#3#4#5{%
  \reset@font\fontsize{#1}{#2pt}%
  \fontfamily{#3}\fontseries{#4}\fontshape{#5}%
  \selectfont}%
\fi\endgroup%
\begin{picture}(6024,5924)(2389,-5773)
\thinlines
\put(4200,-5761){\framebox(2400,5400){$Q$}}
\put(1000,-1711){\line( 1, 0){3200}}
\put(1000,-1861){\line( 1, 0){3200}}
\put(1000,-2011){\line( 1, 0){3200}}
\put(1000,-661){\line( 1, 0){3200}}
\put(1000,-961){\line( 1, 0){3200}}
\put(1000,-1261){\line( 1, 0){3200}}
\put(1000,-1561){\line( 1, 0){3200}}
\put(1000,-811){\line( 1, 0){3200}}
\put(1000,-1411){\line( 1, 0){3200}}
\put(1000,-1111){\line( 1, 0){3200}}
\put(6600,-1711){\line( 1, 0){3200}}
\put(6600,-661){\line( 1, 0){3200}}
\put(6600,-961){\line( 1, 0){3200}}
\put(6600,-1261){\line( 1, 0){3200}}
\put(6600,-1561){\line( 1, 0){3200}}
\put(6600,-811){\line( 1, 0){3200}}
\put(6600,-1411){\line( 1, 0){3200}}
\put(6600,-1111){\line( 1, 0){3200}}
\put(3000,-3361){\line( 1, 0){1200}}
\put(3000,-3211){\line( 1, 0){1200}}
\put(3000,-3511){\line( 1, 0){1200}}
\put(3000,-3811){\line( 1, 0){1200}}
\put(3000,-4111){\line( 1, 0){1200}}
\put(3000,-4411){\line( 1, 0){1200}}
\put(3000,-3661){\line( 1, 0){1200}}
\put(3000,-4261){\line( 1, 0){1200}}
\put(3000,-3961){\line( 1, 0){1200}}
\put(3000,-4561){\line( 1, 0){1200}}
\put(3000,-4861){\line( 1, 0){1200}}
\put(3000,-5161){\line( 1, 0){1200}}
\put(3000,-5461){\line( 1, 0){1200}}
\put(3000,-4711){\line( 1, 0){1200}}
\put(3000,-5311){\line( 1, 0){1200}}
\put(3000,-5011){\line( 1, 0){1200}}
\put(6600,-3361){\line( 1, 0){1200}}
\put(6600,-3211){\line( 1, 0){1200}}
\put(6600,-3061){\line( 1, 0){1200}}
\put(6600,-2911){\line( 1, 0){1200}}
\put(6600,-2761){\line( 1, 0){1200}}
\put(6600,-2611){\line( 1, 0){1200}}
\put(6600,-3511){\line( 1, 0){1200}}
\put(6600,-3811){\line( 1, 0){1200}}
\put(6600,-4111){\line( 1, 0){1200}}
\put(6600,-4411){\line( 1, 0){1200}}
\put(6600,-3661){\line( 1, 0){1200}}
\put(6600,-4261){\line( 1, 0){1200}}
\put(6600,-3961){\line( 1, 0){1200}}
\put(6600,-4561){\line( 1, 0){1200}}
\put(6600,-4861){\line( 1, 0){1200}}
\put(6600,-5161){\line( 1, 0){1200}}
\put(6600,-5461){\line( 1, 0){1200}}
\put(6600,-4711){\line( 1, 0){1200}}
\put(6600,-5311){\line( 1, 0){1200}}
\put(6600,-5011){\line( 1, 0){1200}}
\put(6600,-2461){\line( 1, 0){1200}}
\put(1000,-2161){\line( 1, 0){3200}}
\put(1000,-2311){\line( 1, 0){3200}}
\put(6600,-2311){\line( 1, 0){1200}}
\put(1000,-2461){\line( 1, 0){3200}}
\put(1000,-2611){\line( 1, 0){3200}}
\put(1000,-5700){\framebox(2000,2700){$\ket{0_{\hilb{G} \otimes \hilb{G}'}}$}}
\put(7800,-5700){\framebox(2000,3600){\small junk}}
\put(0,-1775){$\rho\left\{\rule[-6mm]{0mm}{0mm}\right.$}
\put(9800,-1350){$\left.\rule[-4mm]{0mm}{0mm}\right\}\Phi(\rho)$}
\end{picture}

%% file: SQGinEXP.tex
\chapter{An Upper Bound for Short Quantum Games}
\label{ch:SQGinEXP}

Given that $\cls{QIP}$ is contained in both $\cls{EXP}$ \cite{KitaevW00} and
$\cls{SQG}_*$ (Chapter \ref{ch:QIPinSQG}), it is natural to wonder how
complexity classes based upon short quantum games relate to $\cls{EXP}$.
In this chapter we prove that $\cls{SQG} \subseteq \cls{EXP}$, which is the main
contribution of Reference \cite{Gutoski05}.

In order to prove this containment, we build upon previously known techniques
for simulating quantum interaction with classical computation.
In particular, Kitaev sketched an alternate proof \cite{Kitaev02} of the
containment $\cls{QIP} \subseteq \cls{EXP}$ \cite{KitaevW00}.
We provide the first complete formalization of that proof and offer
several extensions, one of which is that $\cls{QRG} \subseteq \cls{NEXP}$.

Finally, we show that $\cls{SQG} \subseteq \cls{EXP}$ by employing the
accumulated results in a separation oracle for use with the ellipsoid method for
convex feasibility.
In fact, the containment $\cls{SQG} \subseteq \cls{EXP}$ is a special case of a
stronger result proven in this chapter.

\section{The Opt Problem}
\label{sec:4:opt}

In this section we define the computational problem \textsc{opt} based upon some
observations regarding quantum interactive proof systems.
We show that \textsc{opt} can be reduced to a semidefinite program and hence
admits a deterministic polynomial-time solution.

\subsection{Optimization, Transcripts, and Consistency}
\label{subsec:4:opt:transcript}

Let $c,s : \mathbb{N} \to [0,1]$, let $L \in \cls{QIP}(c,s)$, let
$V(x) = (V_0,\dots,V_r)$ be an $r$-round verifier witnessing this fact, and
consider the following optimization problem
\begin{align}
\label{eqn:4:dumbopt}
\textrm{maximize} \quad
& \Norm{ \Pi_\mathrm{accept} V_r P_r V_{r-1} \cdots V_1 P_1 V_0 \ket{0} }^2 \\
\textrm{subject to} \quad
& P_1,\dots,P_r \in \mapset{U}{\hilb{P} \otimes \hilb{M}}. \nonumber
\end{align}
By definition, if $x \in L$ then the optimal value of this problem is at least
$1 - c$, whereas if $x \not \in L$ then the optimal value of this problem
is at most $s$.
Hence, if $c$ and $s$ are reasonable then $L$ can be decided by solving this
problem.

However, the problem (\ref{eqn:4:dumbopt}) in its stated form is
incompatible with standard optimization algorithms.
In this subsection, we define the notion of a ``transcript'' of a quantum
interactive proof system and we identify every prover with such a transcript.
In so doing, we reduce the optimization problem (\ref{eqn:4:dumbopt}) to a much
more manageable problem that can be solved using algorithms for semidefinite
programming.

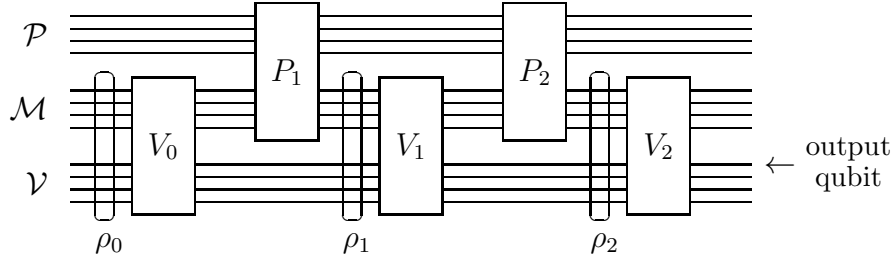
\begin{figure}
\begin{center}
\input{QIPtranscript.latex}
\end{center}
\caption{Transcript of a two-round quantum interactive proof}
\label{fig:QIPtranscript}
\end{figure}

Let $P(x) = (P_1,\dots,P_r)$ be any $r$-round prover and consider the quantum
circuit $(V,P)(x)$ for some input string $x \in \set{0,1}^*$.
For each $i \in \set{0,\dots ,r}$, let
$\rho_i \in \mapset{D}{\hilb{M} \otimes \hilb{V}}$
denote the state of the verifier's qubits immediately before $V_i$ is applied.
The state $\rho_i$ can be viewed as a ``snapshot'' of the verifier's qubits at
the beginning of the $i$th round of interaction.
In this sense, the states $\rho_0,\dots,\rho_r$ indicate a complete
\emph{transcript} of the quantum interactive proof system.
Such a transcript is illustrated in Figure \ref{fig:QIPtranscript} for the case
$r=2$.

What can be said about transcripts?
Two observations follow immediately from the definition of a quantum interactive
proof system.
First, it must be the case that
$$\rho_0
= \ket{0_{\hilb{M} \otimes \hilb{V}}}\bra{0_{\hilb{M} \otimes \hilb{V}}},$$
as the initial pure state of the entire system is always $\ket{0_\hilb{S}}$.
Second, the probability with which $V(x)$ accepts $x$ is given by
$$
\Ptr{}{ \Pi_\mathrm{accept} V_r \rho_r V_r^* \Pi_\mathrm{accept}^* }
=
\Inner{ V_r^* \Pi_\mathrm{accept}^* \Pi_\mathrm{accept} V_r, \rho_r }
$$
in accordance with the rules for quantum measurement discussed in Section
\ref{subsec:intro:quantum:measurement}.

\begin{figure}
\begin{center}
\input{QIPtranscript2.latex}
\end{center}
\caption{Selected states in a two-round quantum interactive proof}
\label{fig:QIPtranscript2}
\end{figure}
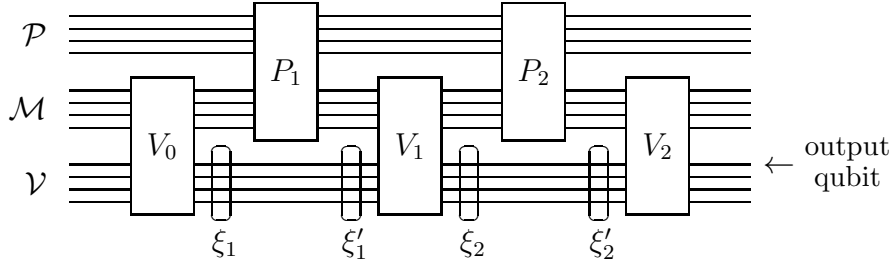

As a third observation, consider for each $i \in \set{1,\dots,r}$ the states
$\xi_i,\xi'_i \in \mapset{D}{\hilb{V}}$ of the verifier's private qubits
immediately before and after the prover's circuit $P_i$ is applied.
These states are illustrated in Figure \ref{fig:QIPtranscript2} for the case
$r=2$.

It is clear from Figures \ref{fig:QIPtranscript} and \ref{fig:QIPtranscript2}
that $\xi'_i$ is obtained from $\rho_i$ by discarding the message qubits.
That is,
$$\xi'_i = \Ptr{\hilb{M}}{\rho_i}.$$
Similarly, $\xi_i$ is obtained from $\rho_{i-1}$ by applying $V_{i-1}$ and
discarding the message qubits.  In other words,
$$\xi_i = \Ptr{\hilb{M}}{V_{i-1} \rho_{i-1} V_{i-1}^*}.$$
Finally, since the prover circuit $P_i$ cannot act on the verifier's private
qubits, it follows that $\xi_i = \xi'_i$
(this fact is also made intuitively evident in Figure \ref{fig:QIPtranscript2}).
Hence, we claim that
$$
\Ptr{\hilb{M}}{\rho_{i}} = \Ptr{\hilb{M}}{V_{i-1} \rho_{i-1} V_{i-1}^*}
\qquad \forall \ i \in \set{1,\dots,r}.
$$

As a generalization of these observations, let $\hilb{F}$ and $\hilb{G}$ be
Hilbert
spaces, let $X_1,\dots,X_r \in \mapset{Pos}{\hilb{F} \otimes \hilb{G}}$ be
positive semidefinite matrices, and let
$A_0,\dots,A_{r-1} \in \mapset{L}{\hilb{F} \otimes \hilb{G}}$ be arbitrary
matrices.
We say that the list $X_1,\dots,X_r$ is \emph{$\hilb{G}$-consistent} with
$A_0,\dots,A_{r-1}$ if
$$
\Ptr{\hilb{G}}{X_{i+1}} = \Ptr{\hilb{G}}{A_i X_i A_i^*}
\qquad \forall \ i \in \set{0,\dots,r-1}
$$
where
$X_0=\ket{0_{\hilb{F} \otimes \hilb{G}}} \bra{0_{\hilb{F} \otimes \hilb{G}}}$.
Our goal then is to prove that a list of density matrices
$\rho_1,\dots,\rho_r$ indicates a valid transcript for a quantum interactive
proof system with verifier $V(x)$ if and only if it is $\hilb{M}$-consistent
with $V_0,\dots,V_{r-1}$.

We require some additional formalism in order to accomplish this goal.
For any positive semidefinite matrix $X \in \mapset{Pos}{\hilb{F}}$
and any vector $v \in \hilb{F} \otimes \hilb{G}$,
$v$ is called a \emph{purification} of $X$ if
$$X = \Ptr{\hilb{G}}{vv^*}.$$
The purifications of $X$ are related to one another as indicated by the
following fact (see Hughston, Jozsa, and Wootters \cite{HughstonJW93}):

\begin{fact}[Unitary Equivalence of Purifications]
\label{fact:4:purification}
A purification $v \in \hilb{F} \otimes \hilb{G}$ of
$X \in \mapset{Pos}{\hilb{F}}$ exists if and only if
$\dim(\hilb{G}) \geq \rank(X)$.
Moreover, purifications of $X$ are \emph{unitarily equivalent} in the
sense that if $u,v \in \hilb{F} \otimes \hilb{G}$ are both purifications of $X$
then there exists a unitary matrix $U \in \mapset{U}{\hilb{G}}$ such that
$(I_\hilb{F} \otimes U)u = v.$
\end{fact}

Intuitively, we make use of the Unitary Equivalence of Purifications
(Fact \ref{fact:4:purification}) in the following manner.
For any $i \in \set{1,\dots,r}$, let $u,u' \in \hilb{S}$ be purifications of the
states $\xi_i, \xi'_i \in \mapset{D}{\hilb{V}}$ in Figure
\ref{fig:QIPtranscript2}.
The vectors $u$ and $u'$ can be thought of as the pure states of the entire
system corresponding to the snapshots $\xi_i$ and $\xi'_i$.
As $\xi_i = \xi'_i$, it follows that $u$ and $u'$ are purifications of the same
state and hence by the Unitary Equivalence of Purifications
(Fact \ref{fact:4:purification}) there exists a unitary matrix
$P_i \in \mapset{U}{\hilb{P} \otimes \hilb{M}}$ such that
$$u' = (P_i \otimes I_\hilb{V}) u.$$
The unitary matrix $P_i$ indicates precisely the actions that the prover must
take during the $i$th round of the interaction in order to take the pure state
of the entire system from $u$ to $u'$.
Given any transcript $\rho_0,\dots,\rho_r$, we can construct in this manner
unitary matrices $P_1,\dots,P_r \in \mapset{U}{\hilb{P} \otimes \hilb{M}}$
corresponding to a prover $P(x)$ who gives rise to that transcript.

We now formalize this intuition.
Because our result will be applied in different contexts later in this chapter,
we state it in its full generality and then follow its proof with a corollary
that relates it to quantum interactive proof systems.

\begin{lm}[Consistency Characterization]
\label{lm:4:norminner}

Let $A_0,\dots,A_{r-1} \in \mapset{L}{\hilb{F} \otimes \hilb{G}}$.
For every Hilbert space $\hilb{H}$ and every
$U_1,\dots,U_r \in \mapset{U}{\hilb{G} \otimes \hilb{H}}$ there exist
$X_1,\dots,X_r \in \mapset{Pos}{\hilb{F} \otimes \hilb{G}}$ that are
$\hilb{G}$-consistent with $A_0,\dots,A_{r-1}$ such that
\begin{equation}
\label{eqn:4:norminner}
\Norm{A U_r A_{r-1} \cdots A_1 U_1 A_0
\ket{0_{\hilb{F} \otimes \hilb{G} \otimes \hilb{H}}}}^2
= \Inner{A^* A, X_r}
\quad \forall \ A \in \mapset{L}{\hilb{F} \otimes \hilb{G}}.
\end{equation}
Conversely, if $\dim(\hilb{H}) \geq \dim(\hilb{F} \otimes \hilb{G})$
then for every
$X_1,\dots,X_r \in \mapset{Pos}{\hilb{F} \otimes \hilb{G}}$
$\hilb{G}$-consistent with $A_0,\dots,A_{r-1}$ there exist
$U_1,\dots,U_r \in \mapset{U}{\hilb{G} \otimes \hilb{H}}$ such that
(\ref{eqn:4:norminner}) holds.

\end{lm}

\begin{proof}

We start by proving the first statement.
Define $u_0,\dots,u_r \in \hilb{F} \otimes \hilb{G} \otimes \hilb{H}$ and
$X_1,\dots,X_r \in \mapset{Pos}{\hilb{F} \otimes \hilb{G}}$ as follows:
for every $i \in \set{0,\dots,r-1}$, let $u_{i+1} = U_{i+1} A_i u_i$ with
$u_0 = \ket{0}$ and let $X_{i+1} = \ptr{\hilb{H}}{ u_{i+1} u_{i+1}^* }$.
Then for any $A \in \mapset{L}{\hilb{F} \otimes \hilb{G}}$ we have
$$\Norm{A U_r A_{r-1} \cdots A_1 U_1 A_0 \ket{0}}^2
= \Norm{A u_r}^2 = \Inner{A^* A, u_r u_r^*}
= \Inner{A^* A, X_r}.$$
It remains only to show the $\hilb{G}$-consistency of
$X_1,\dots,X_r$ with $A_0,\dots,A_{r-1}$.
For every $i \in \set{0,\dots,r-1}$ we have
\begin{align*}
\ptr{\hilb{G}}{X_{i+1}}
&= \ptr{\hilb{G} \otimes \hilb{H}}{ u_{i+1} u_{i+1}^* } \\
&= \ptr{\hilb{G} \otimes \hilb{H}}{ U_{i+1} A_i u_i u_i^* A_i^* U_{i+1}^* } \\
&= \ptr{\hilb{G}}{ A_i \ptr{\hilb{H}}{ u_i u_i^*} A_i } \\
&= \ptr{\hilb{G}}{ A_i X_i A_i }.
\end{align*}

To prove the converse, let $\hilb{H}$ be a Hilbert space with
$\dim(\hilb{H}) \geq \dim(\hilb{F} \otimes \hilb{G})$ and let
$X_1,\dots,X_r \in \mapset{Pos}{\hilb{F} \otimes \hilb{G}}$ be
$\hilb{G}$-consistent with $A_0,\dots,A_{r-1}$.
Let $u_0,\dots,u_r \in \hilb{F} \otimes \hilb{G} \otimes \hilb{H}$ be
purifications of $X_0,\dots,X_r$ with $u_0 = \ket{0}$.
These purifications are guaranteed to exist by
the Unitary Equivalence of Purifications (Fact \ref{fact:4:purification})
because
$$\dim(\hilb{H}) \geq \dim(\hilb{F} \otimes \hilb{G}) \geq \rank(X)$$
for any $X \in \mapset{L}{\hilb{F} \otimes \hilb{G}}$.
For every $i \in \set{0,\dots,r-1}$, it follows that $A_i u_i$ is a purification
of $A_i X_i A_i$.
As $\ptr{\hilb{G}}{X_{i+1}} = \ptr{\hilb{G}}{A_i X_i A_i^*}$,
the Unitary Equivalence of Purifications (Fact \ref{fact:4:purification})
implies that there exists some unitary matrix
$U_{i+1} \in \mapset{U}{\hilb{G} \otimes \hilb{H}}$ such that
$u_{i+1} = U_{i+1} A_i u_i$.
Again, we have
$$\Norm{ A U_r A_{r-1} \cdots A_1 U_1 A_0 \ket{0} }^2
= \Norm{ A u_r }^2 = \Inner{A^* A, u_r u_r^*}
= \Inner{ A^* A, X_r }$$
for any $A \in \mapset{L}{\hilb{F} \otimes \hilb{G}}$.
\end{proof}

The Consistency Characterization (Lemma \ref{lm:4:norminner}) provides us with
the ability to convert from a prover $P(x)$ to a transcript
$\rho_1,\dots,\rho_r$ and vice versa.
We use that ability in the following corollary to reformulate the optimization
problem (\ref{eqn:4:dumbopt}).

\begin{cor}
\label{cor:4:qip2opt}

Let $c,s : \mathbb{N} \to [0,1]$, let $L \in \cls{QIP}(c,s)$, and
let $V(x) = (V_0,\dots,V_r)$ be a verifier witnessing this fact.
Consider the following optimization problem
\begin{align}
\label{eqn:4:qipopt}
\textrm{maximize} \quad
& \Inner{ V_r^* \Pi_\mathrm{accept}^* \Pi_\mathrm{accept} V_r, \rho_r } \\
\textrm{subject to} \quad
& \rho_1,\dots,\rho_r \in \mapset{D}{\hilb{M} \otimes \hilb{V}} \nonumber \\
& \rho_1,\dots,\rho_r \ \hilb{M}\textrm{-consistent with } V_0,\dots,V_{r-1}.
\nonumber
\end{align}
If $x \in L$ then the optimal value of this problem is at least $1 - c$ and
if $x \not \in L$ then the optimal value of this problem is at most $s$.

\end{cor}

\begin{proof}


If $x \in L$ then by definition there exist
$P_1,\dots,P_r \in \mapset{U}{\hilb{P} \otimes \hilb{M}}$ such that
$$
\Norm{ \Pi_\mathrm{accept} V_r P_r V_{r-1} \cdots V_1 P_1 V_0 \ket{0} }^2
\geq 1 - c.
$$
By the Consistency Characterization (Lemma \ref{lm:4:norminner})
there exists a transcript
$\rho_1,\dots,\rho_r$
that is $\hilb{M}$-consistent with $V_0,\dots,V_{r-1}$ such that
$$
\Inner{ V_r^* \Pi_\mathrm{accept}^* \Pi_\mathrm{accept} V_r, \rho_r }
= \Norm{ \Pi_\mathrm{accept} V_r P_r V_{r-1} \cdots V_1 P_1 V_0 \ket{0} }^2,
$$
from which the first claim of the corollary follows.

Now suppose $x \not \in L$ and let
$\rho_1,\dots,\rho_r$
be any transcript that is $\hilb{M}$-consistent with $V_0,\dots,V_{r-1}$.
By the Consistency Characterization (Lemma \ref{lm:4:norminner})
there exists a prover
$P(x) = (P_1,\dots,P_r)$ such that
$$
\Inner{ V_r^* \Pi_\mathrm{accept}^* \Pi_\mathrm{accept} V_r, \rho_r }
= \Norm{ \Pi_\mathrm{accept} V_r P_r V_{r-1} \cdots V_1 P_1 V_0 \ket{0} }^2.
$$
By definition, the quantity on the right is at most $s$.
\end{proof}

As Corollary \ref{cor:4:qip2opt} suggests, any language in $\cls{QIP}$
can be decided by solving the optimization problem (\ref{eqn:4:qipopt}).
Indeed, this entire section is dedicated to solving that problem efficiently.

Like the Consistency Characterization (Lemma \ref{lm:4:norminner}),
our solution to (\ref{eqn:4:qipopt}) will be used in several different contexts
later in this chapter.
Hence, we name the problem \textsc{opt} and restate it in full generality in
Figure \ref{fig:4:opt}.
\begin{figure}
\hrulefill
\begin{description}

\item[Problem.] \textsc{opt}.

\item[Input.]

Matrices $A_0,\dots,A_r \in \mapset{L}{\hilb{F} \otimes \hilb{G}}$
and an accuracy parameter $\varepsilon > 0$.

\item[Output.]

A list
$X_1,\dots,X_r \in \mapset{Pos}{\hilb{F} \otimes \hilb{G}}$
of positive semidefinite matrices that is
$\hilb{G}$-consistent with $A_0,\dots,A_{r-1}$ such that
$$\inner{A_r^* A_r, X_r} > \inner{A_r^* A_r, Z_r} - \varepsilon$$
for every list $Z_1,\dots,Z_r \in \mapset{Pos}{\hilb{F} \otimes \hilb{G}}$
that is $\hilb{G}$-consistent with $A_0,\dots,A_{r-1}$.

\end{description}
\hrulefill
\caption{Definition of \textsc{opt}}
\label{fig:4:opt}
\end{figure}
It is assumed that the real and imaginary parts of all input numbers to
\textsc{opt} are represented in binary notation.

As a final note, we observe that the optimization problem (\ref{eqn:4:qipopt})
appearing in the statement of Corollary \ref{cor:4:qip2opt} can be phrased as
an instance of \textsc{opt} with input matrices
$V_0,\dots,V_{r-1},\Pi_\mathrm{accept} V_r$
and a suitably small accuracy parameter $\varepsilon$ that depends only on the
completeness error $c$ and soundness error $s$.

\subsection{Semidefinite Programming}
\label{subsec:4:opt:sdp}

Our proof that \textsc{opt} admits a deterministic polynomial-time solution will
rely upon existing polynomial-time algorithms for semidefinite programming.
Hence, we offer a brief summary of semidefinite programming in this subsection.

\emph{Semidefinite programming} is derived from
\emph{linear programming}, which is the name given to the problem of maximizing
a linear function subject to a finite number of linear constraints.
Given a Hilbert space $\hilb{F}$,
a Hermitian matrix $H \in \mapset{H}{\hilb{F}}$,
matrices $A_1,\dots,A_m \in \mapset{L}{\hilb{F}}$,
and scalars $\alpha_1,\dots,\alpha_m \in \mathbb{C}$,
a \emph{semidefinite program} over $\mathbb{C}$ has the form
\begin{align*}
\textrm{maximize} \quad  & \Inner{H,X} \\
\textrm{subject to} \quad & \Inner{A_i,X} = \alpha_i
\textrm{ for all } i \in \Set{1,\dots,m} \\
& X \in \mapset{Pos}{\hilb{F}}.
\end{align*}
The \emph{feasible set} $\mathcal{A} \subseteq \mapset{Pos}{\hilb{F}}$ is
defined by
$$\mathcal{A} = \Set{ X \in \mapset{Pos}{\hilb{F}} : \Inner{A_i,X} = \alpha_i
\textrm{ for all } i \in \Set{1,\dots,m} }.$$
The goal is to find a matrix $X \in \mathcal{A}$ such that $\inner{H,X}$
is maximized over all $X \in \mathcal{A}$.
Because both $H$ and $X$ are Hermitian, it follows that $\Inner{H,X}$ is always
real and so it makes sense to consider its maximal value.

The semidefinite programming problem that we use is stated in Figure
\ref{fig:4:sdp}.
\begin{figure}
\hrulefill
\begin{description}

\item[Problem.] \textsc{sdp}.

\item[Input.]

A Hermitian matrix $H \in \mapset{H}{\hilb{F}}$,
matrices $A_1,\dots,A_m \in \mapset{L}{\hilb{F}}$ and
scalars $\alpha_1,\dots,\alpha_m \in \mathbb{C}$
defining the feasible set $\mathcal{A}$,
a feasible solution $X_\mathrm{init} \in \mathcal{A}$,
a positive real number $b$ such that $\Norm{X} \leq b$
for every $X \in \mathcal{A}$,
and an accuracy parameter $\varepsilon > 0$.

\item[Output.]
$X \in \hilb{A}$ such that
$\inner{H,X} > \inner{H,Z} - \varepsilon$
for every $Z \in \mathcal{A}$.

\end{description}
\hrulefill
\caption{Definition of \textsc{sdp}}
\label{fig:4:sdp}
\end{figure}
As with \textsc{opt}, it is assumed in this problem that the real and imaginary
parts of all input numbers are represented in binary notation.
The \textsc{sdp} problem can be solved in time polynomial in the bit length of
the input data using interior point methods
(see, for example, Nesterov and Nemirovskii \cite{NesterovN94}).

\subsection{A Semidefinite Program for Opt}
\label{subsec:4:opt:opt2sdp}

Our goal in this subsection is to prove that \textsc{opt} can be reduced to
\textsc{sdp} and therefore has a deterministic polynomial-time solution.
We accomplish this goal by formalizing the reduction that appears implicitly in
Reference \cite{Kitaev02}.
The main idea is to ``stack'' the positive semidefinite matrices
$X_1,\dots,X_r \in \mapset{Pos}{\hilb{F} \otimes \hilb{G}}$
into one large block-diagonal matrix.
This stacked matrix will serve as the variable over which \textsc{sdp} is
to optimize.

Given matrices $A_0,\dots,A_{r-1} \in \mapset{L}{\hilb{F} \otimes \hilb{G}}$,
we construct linear equality constraints on the stacked matrix variable
that characterize $\hilb{G}$-consistency with $A_0,\dots,A_{r-1}$.
Since the equality conditions describing the $\hilb{G}$-consistency of
$X_1,\dots,X_r$ are already linear in those matrices, this construction is
largely a technical exercise that expresses those conditions in a way that is
compatible with the stacked matrix variable.

Toward that end, we introduce some new notation.
For any Hilbert space $\hilb{H}$ we let
$E_\hilb{H}^{i,j} \in \mapset{L}{\hilb{H}}$ denote the matrix with
all entries equal to zero except for a 1 in the $[i,j]$ entry.
It follows that $B[i,j] = \inner{E_\hilb{H}^{i,j},B}$ for any
$B \in \mapset{L}{\hilb{H}}$.

For any positive integer $n$ we let $\hilb{H}^{\oplus n}$ denote the Hilbert
space with dimension $n \dim(\hilb{H})$.
For any matrices $B_1,\dots,B_n \in \mapset{L}{\hilb{H}}$ we let
$(B_1,\dots,B_n) \in \mapset{L}{\hilb{H}^{\oplus n}}$ denote the block-diagonal
matrix
$$
\left(
\begin{array}{ccc}
B_1 & & 0 \\
& \ddots & \\
0 & & B_n
\end{array}
\right).
$$

Letting $\hilb{R} = (\hilb{F} \otimes \hilb{G})^{\oplus r+1}$, we start by
describing linear equality constraints that ensure every feasible solution
$X \in \mathcal{A} \subset \mapset{Pos}{\hilb{R}}$
is a block-diagonal matrix of the form $(X_0,\dots,X_r)$ for some
$X_0,\dots,X_r \in \mapset{Pos}{\hilb{F} \otimes \hilb{G}}$.
For this task, the ``brute force'' method of simply forcing every
off-block-diagonal entry to zero works just fine.
In other words, we require that
$$\Inner{E_\hilb{R}^{i,j},X} = 0$$
for all suitably chosen $i$ and $j$.
Using this same brute force technique, we set every entry of $X_0$ to indicate
the matrix $\ket{0} \bra{0}$.

We require additional notation before we can proceed to the
$\hilb{G}$-consistency constraints.
Define
$$
\Xi_k : \Mapset{L}{(\hilb{F} \otimes \hilb{G})^{\oplus 2}}
\to \mapset{L}{\hilb{R}}
$$
for all $k \in \set{0,\dots,r-1}$ so that, given
$C = (C_1,C_2) \in \mapset{L}{(\hilb{F} \otimes \hilb{G})^{\oplus 2}}$,
we have
$$
\Xi_k(C) =
\left(
\begin{array}{ccc}
0 & \cdots & 0 \\
\vdots &
\begin{array}{cc}
C_1 & 0 \\
0 & C_2
\end{array}
& \vdots \\
0 & \cdots & 0
\end{array}
\right)
\begin{array}{l}
\\
\leftarrow \textrm{ block $k$} \\
\leftarrow \textrm{ block $k+1$} \\
\\
\end{array}
$$
That is, $C$ is embedded into the all-zero matrix so that if $X=(X_0,\dots,X_r)$
is block-diagonal then
$$\Inner{ \Xi_k(C), X } = \Inner{ C, (X_k,X_{k+1}) }.$$

We also define
$$T^{i,j} : \mapset{L}{\hilb{F} \otimes \hilb{G}} \to
\Mapset{L}{(\hilb{F} \otimes \hilb{G})^{\oplus 2}}$$
for all $i,j \in \set{1,\dots,\dim(\hilb{F})}$ so that, given
$A \in \mapset{L}{\hilb{F} \otimes \hilb{G}}$, we have
$$
T^{i,j}(A) =
\left(
\begin{array}{cc}
A^* \Paren{ E_\hilb{F}^{i,j} \otimes I_\hilb{G} } A & 0 \\
0 & -E_\hilb{F}^{i,j} \otimes I_\hilb{G}
\end{array}
\right).
$$
We now prove the following lemma.

\begin{lm}[$\hilb{G}$-Consistency Constraints]
\label{lm:4:sdpconstraints}

Let $X=(X_0,\dots,X_r) \in \mapset{Pos}{\hilb{R}}$ be a block-diagonal matrix
with $X_0 = \ket{0} \bra{0}$.
Then the list $X_1,\dots,X_r$ is $\hilb{G}$-consistent with $A_0,\dots,A_{r-1}$
if and only if $X$ satisfies
$$\Inner{ \Xi_k(T^{i,j}(A_k)), X } = 0$$
for all $i,j \in \set{1,\dots,\dim(\hilb{F})}$ and all
$k \in \set{0,\dots,r-1}$.

\end{lm}

\begin{proof}

We have
\begin{align*}
\Inner{ \Xi_k(T^{i,j}(A_k)), X }
&= \Inner{ T^{i,j}(A_k), (X_k,X_{k+1}) } \\
&= \Inner{ A_k^* \Paren{ E_\hilb{F}^{i,j} \otimes I_\hilb{G} } A_k, X_k }
- \Inner{ E_\hilb{F}^{i,j} \otimes I_\hilb{G}, X_{k+1} } \\
&= \Inner{ E_\hilb{F}^{i,j} \otimes I_\hilb{G}, A_k X_k A_k^* }
- \Inner{ E_\hilb{F}^{i,j} \otimes I_\hilb{G}, X_{k+1} } \\
&= \Inner{ E_\hilb{F}^{i,j}, \ptr{\hilb{G}}{A_k X_k A_k^*} }
- \Inner{ E_\hilb{F}^{i,j}, \ptr{\hilb{G}}{X_{k+1}} } \\
&= \ptr{\hilb{G}}{A_k X_k A_k^*}[i,j] - \ptr{\hilb{G}}{X_{k+1}}[i,j].
\end{align*}
Of course, $\ptr{\hilb{G}}{A_k X_k A_k^*} = \ptr{\hilb{G}}{X_{k+1}}$ if and only
if their entrywise difference is zero, from which the lemma follows.
\end{proof}

The $\hilb{G}$-Consistency Constraints (Lemma \ref{lm:4:sdpconstraints}) are
based upon similar constraints found in Reference \cite{KitaevW05}.
We have thus established a polynomial number of linear equality constraints that
characterize $\hilb{G}$-consistency.
Our next task is to bound the feasible set
$\mathcal{A} \subset \mapset{Pos}{\hilb{R}}$ of matrices that satisfy these
constraints.

\begin{lm}[$\hilb{G}$-Consistency Constraint Bound]
\label{lm:4:feasiblebound}
Let $X=(X_0,\dots,X_r) \in \mapset{Pos}{\hilb{R}}$ be a block-diagonal matrix
with $X_0 = \ket{0} \bra{0}$ such that the list $X_1,\dots,X_r$ is
$\hilb{G}$-consistent with $A_0,\dots,A_{r-1}$.
Then
$$
\norm{X} \leq \max_{i \in \set{0,\dots,r}}
\Set{\prod_{j=0}^{i-1} \Norm{A_j}^2 }.
$$
\end{lm}

\begin{proof}
It is clear that $\norm{X_0}=1$ and that $\norm{X}$ is just the maximum of
$\norm{X_i}$ over all $i \in \set{0,\dots,r}$.
Hence, it remains only to bound $\norm{X_i}$ for $i \geq 1$.
Let $\hilb{H}$ be a Hilbert space with
$\dim(\hilb{H}) = \dim(\hilb{F} \otimes \hilb{G})$.
By the Consistency Characterization (Lemma \ref{lm:4:norminner}),
there exist
$U_1,\dots,U_i \in \mapset{U}{\hilb{G} \otimes \hilb{H}}$ such that
$$
\Norm{A U_i A_{i-1} \cdots A_1 U_1 A_0
\ket{0_{\hilb{F} \otimes \hilb{G} \otimes \hilb{H}}}}^2
= \Inner{A^* A, X_i}
\quad \forall \ A \in \mapset{L}{\hilb{F} \otimes \hilb{G}}.
$$
In particular,
$$\Norm{I U_i A_{i-1} \cdots A_1 U_1 A_0 \ket{0}}^2
= \Inner{I^* I, X_i} = \ptr{}{X_i} \geq \Norm{X_i}$$
where the final inequality follows from the fact that $X_i$ is positive
semidefinite.
That
$$\Norm{I U_i A_{i-1} \cdots A_1 U_1 A_0 \ket{0}}^2 \leq
\prod_{j=0}^{i-1} \Norm{A_j}^2$$
follows from the fact that $U_1,\dots,U_i$ are unitary.
\end{proof}

We have now developed the tools needed to prove the following theorem.

\begin{thm}
\label{thm:4:opt2sdp}

\textsc{opt} can be solved in time polynomial in the bit length of the input
data.

\end{thm}

\begin{proof}

The proof is by reduction to \textsc{sdp}.
Given inputs $A_0,\dots,A_r$ and $\varepsilon$ to \textsc{opt}, we construct
inputs to \textsc{sdp} as follows:
\begin{itemize}

\item

The error parameter $\varepsilon$ is passed unchanged from \textsc{opt} to
\textsc{sdp}.

\item

The objective matrix $H \in \mapset{H}{\hilb{R}}$ is the block-diagonal matrix
$(0,\dots,0,A_r^* A_r)$.

\item

The linear equality constraints are the $\hilb{G}$-Consistency Constraints
(Lemma \ref{lm:4:sdpconstraints}).

\item

The bound $b$ for all feasible solutions is given by the
$\hilb{G}$-Consistency Constraint Bound (Lemma \ref{lm:4:feasiblebound}).
Note that if $A_0,\dots,A_{r-1}$ are unitary then we have $b=1$.

\item

The initial feasible solution $X_\mathrm{init} \in \mathcal{A}$ can be taken to
be the block-diagonal matrix $(X_0,\dots,X_r)$ where $X_{i+1} = A_i X_i A_i^*$
for every $i \in \set{0,\dots,r-1}$ with $X_0 = \ket{0} \bra{0}$.
This feasible solution corresponds to a prover who always acts trivially upon
his qubits.

\end{itemize}
\end{proof}

\section{Some Upper Bounds}
\label{sec:4:bounds}

In this section we use our polynomial-time solution to \textsc{opt}
(Theorem \ref{thm:4:opt2sdp}) to prove the upper bounds
$\cls{QIP} \subseteq \cls{EXP}$ and $\cls{QRG} \subseteq \cls{NEXP}$.
Indeed, these containments are special cases of stronger results proven in this
section.

Several details must be considered before we can formalize these containments.
For example, it is prudent to discuss numerical error introduced by
finite-precision approximations of continuous quantities.
In the case of quantum refereed games, we also require a tractable bound on the
number of qubits used by the provers.

\subsection{Roundoff Error}
\label{subsec:4:bounds:roundoff}

Let $L \in \cls{QIP}$ and let $V(x) = (V_0,\dots,V_r)$ be a verifier witnessing
this fact.
In Section \ref{subsec:4:opt:transcript} we pointed out that $L$ can be
decided by solving \textsc{opt} with input matrices
$V_0,\dots,V_{r-1},\Pi_\mathrm{accept} V_r$
and a small enough accuracy parameter $\varepsilon$.
However, it is often the case that the unitary matrices
$V_0,\dots,V_r \in \mapset{U}{\hilb{M} \otimes \hilb{V}}$
associated with the verifier's quantum circuits contain entries that are
complicated algebraic expressions involving irrational numbers.
Because \textsc{opt} was defined to accept input matrices whose entries are
expressed in binary notation, it follows that the best we can do is approximate
$V_0,\dots,V_r$ with finite-precision matrices
$\tilde V_0,\dots,\tilde V_r \in \mapset{L}{\hilb{M} \otimes \hilb{V}}$.

Our intuition tells us that, by choosing a suitable level of precision with
which to express $\tilde V_0,\dots,\tilde V_r$, the induced verifier
$\tilde V(x)$ will always have reasonable completeness error and soundness
error.
Moreover, we expect that matrices $\tilde V_0,\dots,\tilde V_r$ with the
required level of precision can be computed efficiently, so that $L$ can still
be decided in exponential time.
Indeed, these intuitions are correct.
This subsection is dedicated to arguing that accurate enough approximations can
be computed efficiently.
For simplicity, we restrict our discussion to quantum interactive proof systems,
but much of the discussion in this subsection transfers to quantum refereed
games without complication.

In Section \ref{subsec:intro:quantum:circuits} we stipulated that all quantum
circuits in this thesis are composed of quantum gates chosen from some finite
universal set.
We take it as given that the unitary matrices associated with the quantum gates
in this universal set can all be computed so that each entry is accurate to $t$
bits of precision in time polynomial in the bit length of $t$.
Of course, the unitary matrix associated with any single quantum gate is readily
extended to a larger Hilbert space by taking the Kronecker product with the
identity matrix as usual.

Since the verifier's quantum circuits are generated uniformly in polynomial
time, it follows that each of the $r(|x|)$ circuits is composed of at most
$g(|x|)$ quantum gates for some $r,g \in \poly$.
For each $k \in \set{0,\dots,r}$ and $l \in \set{1,\dots,g}$ let
$U_{k,l} \in \mapset{U}{\hilb{M} \otimes \hilb{V}}$ be the unitary
matrix associated with the $l$th gate in the verifier's $k$th quantum circuit,
extending to $\hilb{M} \otimes \hilb{V}$ so that
$V_k = U_{k,g} \cdots U_{k,1}$.

Next, suppose that $\tilde U_{k,l}$ is an approximation of $U_{k,l}$ such that
each entry of $\tilde U_{k,l}$ is accurate to $t$ bits of precision and let
$\tilde V_k = \tilde U_{k,g} \cdots \tilde U_{k,1}$.
Finally, let $P(x) = (P_1,\dots,P_r)$ be any prover.
The probability $p$ with which $(V,P)(x)$ accepts $x$ is precisely
$$
p = \norm{ \Pi_\mathrm{accept} V_r P_r V_{r-1} \cdots V_1 P_1 V_0 \ket{0} }^2
$$
and the probability $\tilde p$ with which $(\tilde V,P)(x)$ accepts $x$ is
precisely
$$
\tilde p = \norm{ \Pi_\mathrm{accept} \tilde V_r P_r \tilde V_{r-1} \cdots
\tilde V_1 P_1 \tilde V_0 \ket{0} }^2.
$$
Our goal is to prove an upper bound on the difference $\abs{p - \tilde p}$ in
terms of $t$.

Toward that end, let $\delta > 0$ and let $A, \tilde A \in \mapset{L}{\hilb{F}}$
be any matrices whose entrywise difference is at most $\delta$.
In other words,
$$\Abs{ A[i,j] - \tilde A[i,j] } < \delta$$
for every $i,j \in \set{1,\dots,\dim(\hilb{F})}$.
It is not difficult to show that
$$\norm{ A - \tilde A } < \dim(\hilb{F}) \delta.$$
The following lemma allows us to deduce the accuracy required of our
approximations.

\begin{lm}
\label{lm:4:precision}

Let $\delta > 0$ and let
$A_1,\dots,A_m,\tilde A_1,\dots,\tilde A_m$
be any matrices such that the product $A_m \cdots A_1$ is defined and, for all
$i \in \set{1,\dots,m}$,
$\tilde A_i$ has the same dimensions as $A_i$,
$\norm{A_i - \tilde A_i} < \delta$, and
$\norm{A_i}, \norm{\tilde A_i} \leq 1$.
Then
$$\Abs{ \norm{ A_m \cdots A_1 } -
\norm{ \tilde A_m \cdots \tilde A_1 } } < m\delta$$
and
$$\Abs{ \norm{ A_m \cdots A_1 }^2 -
\norm{ \tilde A_m \cdots \tilde A_1 }^2 } < 2m\delta.$$

\end{lm}

\begin{proof}

We have
$$\Abs{ \norm{ A_m \cdots A_1 } -
\norm{ \tilde A_m \cdots \tilde A_1 } } \leq
\norm{ A_m \cdots A_1 - \tilde A_m \cdots \tilde A_1 }.$$
By repeated application of the triangle inequality, this quantity is at most
\begin{align*}
&\sum_{i=1}^m
\Norm{ A_m \cdots A_i \tilde A_{i-1} \cdots \tilde A_1 -
A_m \cdots A_{i+1} \tilde A_i \cdots \tilde A_1 } \\
&= \sum_{i=1}^m
\Norm{A_m \cdots A_{i+1} (A_i - \tilde A_i) \tilde A_{i-1} \cdots \tilde A_1} \\
&\leq \sum_{i=1}^m
\Norm{ A_m \cdots A_{i+1} } \norm{ A_i - \tilde A_i }
\norm{ \tilde A_{i-1} \cdots \tilde A_1 } \\
&\leq \sum_{i=1}^m \norm{ A_i - \tilde A_i } < m\delta.
\end{align*}
The lemma follows from the fact that
$$\Abs{a^2 - b^2} = \Abs{a - b}(a + b) < 2 \Abs{a - b}$$
whenever $a$ and $b$ are real numbers in the interval $[0,1]$.
\end{proof}

Since the verifier's quantum circuits are generated uniformly in polynomial
time, it follows that they each act on at most $q(|x|)$ qubits for some
$q \in \poly$, so that $\dim(\hilb{M} \otimes \hilb{V}) = 2^q$.
As each entry of $\tilde U_{k,l}$ is accurate to $t$ bits of precision, we have
$$\Abs{ U_{k,l}[i,j] - \tilde U_{k,l}[i,j] } < 2^{-t}$$
for every $i,j \in \set{1,\dots,2^q}$ and hence
$$\norm{U_{k,l} - \tilde U_{k,l}} < 2^{q-t}.$$
It follows from Lemma \ref{lm:4:precision} that
$$\abs{p - \tilde p} < (r+1) g 2^{q-t+1}.$$
Hence, we can compute an exponentially close approximation $\tilde p$ of $p$ by
choosing a suitable $t \in \poly$ and approximating each matrix $U_{k,l}$ to
$t(|x|)$ bits of precision.
In fact, this result holds even if $r$ and $g$ grow exponentially in $|x|$.

\subsection{An Extension of $\cls{QIP} \subseteq \cls{EXP}$}
\label{sec:4:bounds:QIPinEXP}

We are now ready to prove the upper bound $\cls{QIP} \subseteq \cls{EXP}$.
Indeed, we prove that the containment holds under the following relaxations of
the definition of $\cls{QIP}$:
\begin{itemize}

\item
The verifier may exchange an exponential number of messages with the prover.

\item
The verifier's quantum circuits may contain an exponential number of gates, so
long as they still act upon at most a polynomial number of qubits.

\item
The completeness error and soundness error may be exponentially close to
$\frac{1}{2}$ in $|x|$.

\end{itemize}

We define a \emph{strong} verifier to be a verifier whose quantum circuits are
generated by an exponential-time Turing machine on input $x$, but they act on at
most a polynomial number of qubits.

\begin{thm}[An Extension of $\cls{QIP} \subseteq \cls{EXP}$]
\label{thm:4:QIPinEXP}

Let $c,s : \mathbb{N} \to [0,1]$ be any polynomial-time computable functions
satisfying
$1 - c(n) - s(n) > 0$
for all $n \in \mathbb{N}$.
Any language $L \subseteq \set{0,1}^*$ that can be decided by a quantum
interactive proof system with a strong verifier having completeness error $c$
and soundness error $s$ is in $\cls{EXP}$.

\end{thm}

\begin{proof}

We assume without loss of generality that $c(n), s(n) < \frac{1}{2}$ for all
$n \in \mathbb{N}$, as any verifier who can compute $c(|x|)$ and $s(|x|)$ can
also bias his final decision to satisfy this condition.
Let $\varepsilon = \min \Set{\frac{1}{2} - c, \frac{1}{2} - s}$.
It follows from the fact that $c$ and $s$ are polynomial-time computable that
$\varepsilon \in 2^{-\poly}$.

Figure \ref{fig:4:QIPalg} describes a deterministic exponential-time
algorithm that decides $L$.
\begin{figure}
\hrulefill
\begin{enumerate}

\item
Run the exponential-time Turing machine that generates the verifier's
quantum circuits on input string $x \in \set{0,1}^*$.
Let $V(x) = (V_0,\dots,V_r)$ denote the unitary matrices associated with these
circuits.

\item
Compute an approximation $\tilde V(x) = (\tilde V_0,\dots, \tilde V_r)$ of
$V(x)$ satisfying
$$\norm{V_i - \tilde V_i} < \frac{\varepsilon}{4(r+1)}$$
for every $i \in \set{0,\dots,r}$.

\item
Solve \textsc{opt} with input matrices
$\tilde V_0,\dots,\tilde V_{r-1}, \Pi_\mathrm{accept} \tilde V_r$ and accuracy
parameter $\frac{\varepsilon}{2}$.
Let $\varpi$ denote the optimal value indicated by this solution.
Accept $x$ if $\varpi > \frac{1}{2}$, otherwise reject $x$.

\end{enumerate}
\hrulefill
\caption{An exponential-time algorithm for $L \in \cls{QIP}$}
\label{fig:4:QIPalg}
\end{figure}
To see that this algorithm is correct, let $p$ (respectively $\tilde p$)
denote the maximum probability with which $V(x)$ (respectively $\tilde V(x)$)
can be made to accept $x$.
By our choice of accuracy parameter for \textsc{opt} we have
$\abs{\varpi - \tilde p} < \frac{\varepsilon}{2}$
and by Lemma \ref{lm:4:precision} we have
$$\abs{\tilde p - p} < 2(r+1) \frac{\varepsilon}{4(r+1)}
= \frac{\varepsilon}{2},$$
from which it follows that
$\abs{\varpi - p} < \varepsilon$.
By definition, if $x \in L$ then $p \geq \frac{1}{2} + \varepsilon$ and hence
$\varpi > \frac{1}{2}$.
Conversely, if $x \not \in L$ then $p \leq \frac{1}{2} - \varepsilon$ and hence
$\varpi < \frac{1}{2}$.

It remains only to verify that this algorithm runs in exponential time.
According to Section \ref{subsec:4:bounds:roundoff}, the approximation
$\tilde V(x)$ in step 2 can be computed in exponential time by choosing a
suitable $t \in \poly$ and approximating the unitary matrices associated with
each of the verifier's quantum gates to $t(|x|)$ bits of precision.
As the input matrices to \textsc{opt} in step 3 can be computed in exponential
time and have at most exponential dimension, the desired result follows from the
fact that \textsc{opt} admits a polynomial-time solution
(Theorem \ref{thm:4:opt2sdp}).
\end{proof}

\subsection{Bounding the Number of Prover Qubits}
\label{subsec:4:bounds:proverqubits}

In this subsection we look at extending the proof of
$\cls{QIP} \subseteq \cls{EXP}$ (Theorem \ref{thm:4:QIPinEXP})
to provide an upper bound for $\cls{QRG}$.

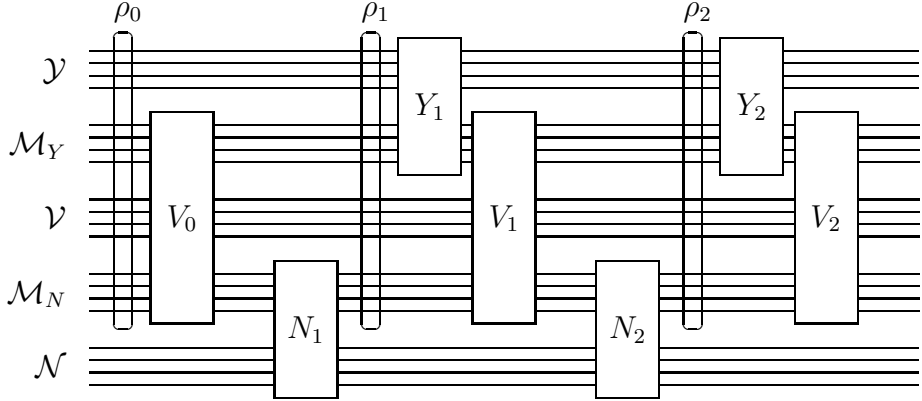
\begin{figure}
\begin{center}
\input{QRGtranscript.latex}
\end{center}
\caption{Transcript of a two-round quantum refereed game}
\label{fig:QRGtranscript}
\end{figure}

We begin by applying the Consistency Characterization
(Lemma \ref{lm:4:norminner}) to quantum refereed games in much the same way as
it was applied to quantum interactive proof systems in Corollary
\ref{cor:4:qip2opt} in Section \ref{subsec:4:opt:transcript}.
In particular, if the yes-prover $Y(x)$ is fixed then the combination of
$V(x)$ and $Y(x)$ can be viewed as a new verifier $(V,Y)(x)$ for an ordinary
quantum interactive proof system in which the no-prover is the only prover.
In this case, the actions of the no-prover can be described by a transcript,
just as with quantum interactive proof systems.
Such a transcript is illustrated in Figure \ref{fig:QRGtranscript}.
In this case, we use the Consistency Characterization
(Lemma \ref{lm:4:norminner}) to convert from a no-prover $N(x)$ to a transcript
$\rho_1,\dots,\rho_r$ and vice versa.

\begin{cor}
\label{cor:4:qrg2opt}

Let $c,s : \mathbb{N} \to [0,1]$, let $L \in \cls{QRG}(c,s)$, let
$V(x) = (V_0,\dots,V_r)$ be a verifier witnessing this fact, and let
$Y(x) = (Y_1,\dots,Y_r)$ be a yes-prover.
Consider the following optimization problem
\begin{align}
\label{eqn:4:qrgopt}
\textrm{maximize} \quad
& \Inner{  Y_r^* V_r^* \Pi_\mathrm{reject}^* \Pi_\mathrm{reject} V_r Y_r,
\rho_r } \\
\textrm{subject to} \quad
& \rho_1,\dots,\rho_r \in
\mapset{D}{\hilb{Y} \otimes \hilb{M}_Y \otimes \hilb{V} \otimes \hilb{M}_N}
\nonumber \\
& \rho_1,\dots,\rho_r \ \hilb{M}_N\textrm{-consistent with }
V_0, V_1 Y_1, \dots,V_{r-1} Y_{r-1}. \nonumber
\end{align}
If $x \in L$ then there exists a yes-prover $Y(x)$ such that the optimal value
of this problem is at most $c$ and if $x \not \in L$ then for every
yes-prover $Y(x)$ the optimal value of this problem is at least $1 - s$.

\end{cor}

\begin{proof}


If $x \not \in L$ then by definition there exist
$N_1,\dots,N_r \in \mapset{U}{\hilb{M}_N \otimes \hilb{N}}$ such that
$$
\Norm{\Pi_\mathrm{reject} V_r Y_r N_r V_{r-1} \cdots V_1 Y_1 N_1 V_0 \ket{0}}^2
\geq 1 - s
$$
for every $Y_1,\dots,Y_r \in \mapset{U}{\hilb{Y} \otimes \hilb{M}_Y}$.
By the Consistency Characterization (Lemma \ref{lm:4:norminner})
there exists a transcript $\rho_1,\dots,\rho_r$
that is $\hilb{M}_N$-consistent with $V_0, V_1 Y_1, \dots,V_{r-1} Y_{r-1}$
such that
$$
\Inner{ Y_r^* V_r^* \Pi_\mathrm{reject}^* \Pi_\mathrm{reject} V_r Y_r, \rho_r }
=
\Norm{\Pi_\mathrm{reject} V_r Y_r N_r V_{r-1} \cdots V_1 Y_1 N_1 V_0 \ket{0}}^2,
$$
from which the second claim of the corollary follows.

Now suppose $x \in L$, let $Y(x) = (Y_1,\dots,Y_r)$ witness this fact, and let
$\rho_1,\dots,\rho_r$ be any transcript that is $\hilb{M}_N$-consistent with
$V_0, V_1 Y_1, \dots,V_{r-1} Y_{r-1}$.
By the Consistency Characterization (Lemma \ref{lm:4:norminner})
there exists a no-prover
$N(x) = (N_1,\dots,N_r)$ such that
$$
\Inner{ Y_r^* V_r^* \Pi_\mathrm{reject}^* \Pi_\mathrm{reject} V_r Y_r, \rho_r }
=
\Norm{\Pi_\mathrm{reject} V_r Y_r N_r V_{r-1} \cdots V_1 Y_1 N_1 V_0 \ket{0}}^2.
$$
By definition, the quantity on the right is at most $c$.
\end{proof}

Corollary \ref{cor:4:qrg2opt} suggests that any language in $\cls{QRG}$ can be
decided by nondeterministically ``guessing'' the yes-prover
$Y(x) = (Y_1,\dots,Y_r)$ and solving the optimization problem
(\ref{eqn:4:qrgopt}).
Given a verifier $V(x)$ and a yes-prover $Y(x)$, this problem can be phrased as
an instance of \textsc{opt} with input matrices
\begin{equation}
\label{eqn:4:qrgoptinstance}
V_0, V_1 Y_1, \dots, V_{r-1} Y_{r-1}, \Pi_\mathrm{reject} V_r Y_r
\end{equation}
and a suitably small accuracy parameter $\varepsilon$ that depends only on the
completeness error $c$ and soundness error $s$.
As \textsc{opt} admits a deterministic polynomial-time solution
(Theorem \ref{thm:4:opt2sdp}), it is tempting to conclude that any language in
$\cls{QRG}$ can also be decided in nondeterministic exponential time.

However, we must take care to ensure that the size of the induced instance of
\textsc{opt} is in fact bounded by an exponential in $|x|$.
The input matrices (\ref{eqn:4:qrgoptinstance}) act upon the Hilbert space
$\hilb{Y} \otimes \hilb{M}_Y \otimes \hilb{V} \otimes \hilb{M}_N$:
as the quantum circuits belonging to any prover are unbounded, it is conceivable
that the yes-prover uses a superpolynomial number of private qubits.
In this case, the dimension of $\hilb{Y}$ and hence of the input matrices
(\ref{eqn:4:qrgoptinstance}) to \textsc{opt} may be superexponential in $|x|$.
In order to achieve the desired upper bound for quantum refereed games,
we require a polynomial bound on the number of qubits used by the yes-prover.

As far as quantum interactive proof systems are concerned, it follows from the
Consistency Characterization (Lemma \ref{lm:4:norminner}) that the
prover's Hilbert space $\hilb{P}$ need only satisfy
$\dim(\hilb{P}) \geq \dim(\hilb{M} \otimes \hilb{V})$.
In other words, any quantum interactive proof system can be simulated by another
quantum interactive proof system in which the prover uses no more qubits than
the verifier.
Unfortunately, this convenient bound is not known to extend to quantum
interactions with multiple provers.

Fortunately, there is a looser polynomial bound that does hold for quantum
interactions with multiple provers.
In particular, the following fact holds (see Kobayashi and Matsumoto
\cite{KobayashiM03}):

\begin{fact}
\label{fact:4:proverqubits}
In any quantum interaction, the number of qubits required by each of the provers
is polynomial in the number of message qubits shared with the verifier and the
number of rounds in the interaction.
\end{fact}

Hence, if both the number of message qubits and the number of rounds in the
interaction are polynomial then any prover in a quantum refereed game can
be assumed to use a polynomial number of qubits.
In particular, the Hilbert space $\hilb{Y}$ corresponding to the yes-prover's
private qubits has dimension at most exponential in $|x|$ as desired.

\subsection{An Extension of $\cls{QRG} \subseteq \cls{NEXP}$}
\label{subsec:4:bounds:QRGinNEXP}

In this subsection we prove the upper bound $\cls{QRG} \subseteq \cls{NEXP}$.
Indeed, we prove that the containment holds under the following relaxations of
the definition of $\cls{QRG}$:
\begin{itemize}

\item
The verifier's quantum circuits may contain an exponential number of gates, so
long as they still act upon at most a polynomial number of qubits.

\item
The completeness error and soundness error may be exponentially close to
$\frac{1}{2}$ in $|x|$.

\end{itemize}
It is interesting to note that the containment $\cls{QRG} \subseteq \cls{NEXP}$
is not known to hold for quantum refereed games with a superpolynomial number of
rounds.
By contrast, we showed in Theorem \ref{thm:4:QIPinEXP} that the
containment $\cls{QIP} \subseteq \cls{EXP}$ holds even when the number of rounds
is exponential.
As explained in Section \ref{subsec:4:bounds:proverqubits}, this strange
discrepancy is brought on by the conditions of Fact
\ref{fact:4:proverqubits}.

\begin{thm}[An extension of $\cls{QRG} \subseteq \cls{NEXP}$]
\label{thm:4:QRGinNEXP}

Let $c,s : \mathbb{N} \to [0,1]$ be any polynomial-time computable functions
satisfying
$1 - c(n) - s(n) > 0$
for all $n \in \mathbb{N}$ and let $r \in \poly$.
Any language $L \subseteq \set{0,1}^*$ that can be decided by a quantum refereed
game with a strong $r$-round verifier having completeness error $c$ and
soundness error $s$ is in $\cls{NEXP}$.

\end{thm}

\begin{proof}

As in the proof of Theorem \ref{thm:4:QIPinEXP}, we assume without loss
of generality that $c,s < \frac{1}{2}$ and we let $\varepsilon \in 2^{-\poly}$
be defined by $\varepsilon = \min \Set{\frac{1}{2} - c, \frac{1}{2} - s}$.

The nondeterministic step of our solution is to guess the unitary matrices
$Y(x) = (Y_1,\dots,Y_r)$ belonging to the yes-prover and compute and
approximation $\tilde Y(x) = (\tilde Y_1,\dots,\tilde Y_r)$ of $Y(x)$ satisfying
$$\norm{Y_i - \tilde Y_i} < \frac{\varepsilon}{4(2r+1)}$$
for every $i \in \set{1,\dots,r}$.
This step can be accomplished in many different ways.
For example, any unitary matrix $U$ is given by $\exp(iH)$ for some Hermitian
matrix $H$.
To guess the entries of $U$ accurate to $t$ bits of precision, we first choose a
suitable $t' > t$, guess the entries of $H$ to $t'$ bits of precision, and run
any stable algorithm that computes the matrix exponential
(see, for instance, Golub and Van Loan \cite{GolubV89}).

The deterministic step of our solution is to run the algorithm of Figure
\ref{fig:4:QIPalg} in Theorem \ref{thm:4:QIPinEXP} with the following changes:
\begin{itemize}

\item
The approximation $\tilde V(x)$ in step 2 must satisfy
$$\norm{V_i - \tilde V_i} < \frac{\varepsilon}{4(2r+1)}$$
for every $i \in \set{0,\dots,r}$.

\item
In step 3 we solve \textsc{opt} with input matrices
$
\tilde V_0, \tilde V_1 \tilde Y_1, \dots, \tilde V_{r-1} \tilde Y_{r-1},
\Pi_\mathrm{reject} \tilde V_r \tilde Y_r
$
and we reject $x$ if $\varpi > \frac{1}{2}$, otherwise we accept $x$.

\end{itemize}

To see that this algorithm is correct, let $p$ denote the maximum probability
with which $V(x)$ can be made to reject $x$ given yes-prover $Y(x)$.
That $\abs{\varpi - p} < \varepsilon$ follows just as in Theorem
\ref{thm:4:QIPinEXP}.
By definition, if $x \in L$ then there exists a yes-prover $Y(x)$ such that
$p \leq \frac{1}{2} - \varepsilon$ and hence $\varpi < \frac{1}{2}$.
Conversely, if $x \not \in L$ then for all yes-provers we have
$p \geq \frac{1}{2} + \varepsilon$ and hence $\varpi > \frac{1}{2}$.

That this nondeterministic algorithm runs in exponential time follows as in the
proof of Theorem \ref{thm:4:QIPinEXP}.
\end{proof}

\section{The Ellipsoid Method and Short Quantum Games}
\label{sec:4:SQG}

In this section we prove $\cls{SQG} \subseteq \cls{EXP}$.
Indeed, this containment is a special case of a stronger result proven in this
section.

Given a verifier for a short quantum game, we construct a convex set of matrices
that is nonempty if and only if there exists a winning yes-prover for that game.
The desired result is achieved by providing an algorithm that determines in
exponential time whether or not this set is empty.

\subsection{The Set of Winning Yes-Provers}
\label{subsec:4:SQG:start}

\begin{figure}
\begin{center}
\input{SQGtranscript.latex}
\end{center}
\caption{Transcript of a short quantum game}
\label{fig:SQGtranscript}
\end{figure}
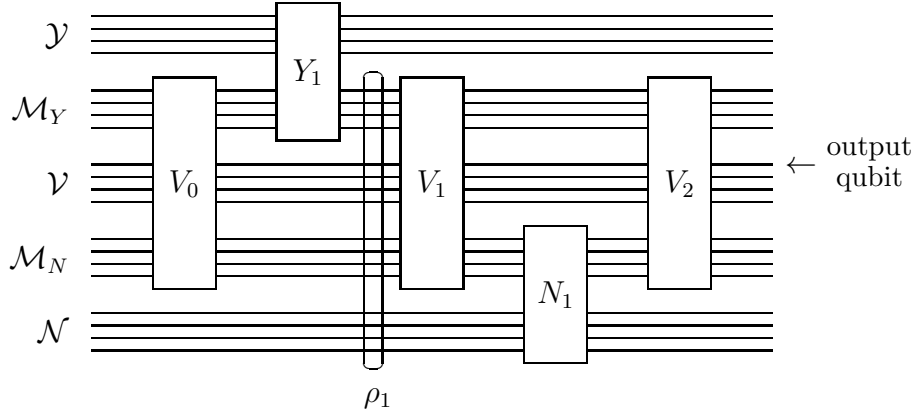

The key fact that we exploit in order to put $\cls{SQG}$ inside $\cls{EXP}$ is
that the no-prover does not become involved in any short quantum game until the
verifier has finished exchanging messages with the yes-prover.
Because of this fact, the actions of the yes-prover prior to the no-prover's
involvement can be completely described by a transcript, just as with quantum
interactive proof systems.
As the verifier exchanges only one message with the yes-prover, the transcript
under consideration consists only of the state $\rho_1 \in
\mapset{D}{\hilb{M}_Y \otimes \hilb{V} \otimes \hilb{M}_N \otimes \hilb{N}}$
illustrated in Figure \ref{fig:SQGtranscript}.

We begin by applying the Consistency Characterization
(Lemma \ref{lm:4:norminner}) to short quantum games in a way that identifies the
set of winning yes-provers $Y(x) = (Y_1)$ with a set of winning transcripts
$\rho_1$.

\begin{cor}
\label{cor:4:sqg2opt}

Let $c,s : \mathbb{N} \to [0,1]$ satisfy
$1 - c(n) - s(n) > 0$
for all $n \in \mathbb{N}$, let $L \in \cls{SQG}(c,s)$, and
let $V(x) = (V_0,V_1,V_2)$ be a verifier witnessing this fact.
Define $\mapset{Win}{V,c}$ to be the set of all
$\rho_1 \in
\mapset{D}{\hilb{M}_Y \otimes \hilb{V} \otimes \hilb{M}_N \otimes \hilb{N}}$
such that $\rho_1$ is $\hilb{M}_Y$-consistent with $V_0$ and
$$
\Inner{
V_1^* N_1^* V_2^* \Pi_\mathrm{reject}^*
\Pi_\mathrm{reject} V_2 N_1 V_1, \rho_1 } \leq c
\qquad \forall \ N_1 \in \mapset{U}{\hilb{M}_N \otimes \hilb{N}}.
$$
Then $\mapset{Win}{V,c}$ is nonempty if and only if $x \in L$.

\end{cor}

\begin{proof}


If $x \in L$ then by definition there exists
$Y_1 \in \mapset{U}{\hilb{Y} \otimes \hilb{M}_Y}$ such that
$$
\Norm{\Pi_\mathrm{reject} V_2 N_1 V_1 Y_1 V_0 \ket{0}}^2
\leq c
$$
for every $N_1 \in \mapset{U}{\hilb{M}_N \otimes \hilb{N}}$.
By the Consistency Characterization (Lemma \ref{lm:4:norminner})
there exists a transcript
$\rho_1$
that is $\hilb{M}_Y$-consistent with $V_0$ such that
$$
\Inner{ V_1^* N_1^* V_2^* \Pi_\mathrm{reject}^*
\Pi_\mathrm{reject} V_2 N_1 V_1, \rho_1 }
=
\Norm{\Pi_\mathrm{reject} V_2 N_1 V_1 Y_1 V_0 \ket{0}}^2,
$$
from which it follows that $\rho_1 \in \mapset{Win}{V,c}$.

If $x \not \in L$ then let $N(x) = (N_1)$ witness this fact and let
$\rho_1$
be any transcript that is $\hilb{M}_Y$-consistent with $V_0$.
By the Consistency Characterization (Lemma \ref{lm:4:norminner})
there exists a yes-prover
$Y(x) = (Y_1)$ such that
$$
\Inner{ V_1^* N_1^* V_2^* \Pi_\mathrm{reject}^*
\Pi_\mathrm{reject} V_2 N_1 V_1, \rho_1 }
=
\Norm{\Pi_\mathrm{reject} V_2 N_1 V_1 Y_1 V_0 \ket{0}}^2.
$$
By definition, the quantity on the right is at least $1 - s > c$, from
which it follows that $\rho_1 \not \in \mapset{Win}{V,c}$
\end{proof}

As Corollary \ref{cor:4:sqg2opt} suggests, any language $L \in \cls{SQG}$
can be decided by an algorithm that decides the emptiness of the set
$\mapset{Win}{V,c}$.
But how can the emptiness of this set be decided efficiently?

In order to answer that question, we point out that once a yes-prover has been
fixed the short quantum game essentially becomes a two-message quantum
interactive proof system in which the no-prover is the only prover.
As demonstrated in Section \ref{sec:4:opt}, such an interaction can be decided
by solving \textsc{opt}.
Therefore, if we are given a candidate transcript $\rho_1$ for some yes-prover
$Y(x)$ then we can use our solution to \textsc{opt} to decide whether there
exists a no-prover that wins against $Y(x)$.
If we find that no such no-prover exists then we can safely conclude that
$\rho_1 \in \mapset{Win}{V,c}$ and hence $x \in L$.

Unfortunately, if we find that there \emph{does} exist a no-prover that wins
against $Y(x)$ then we cannot immediately conclude that $x \not \in L$ because
it might also be the case that $x \in L$ but $Y(x)$ is a bad yes-prover who does
not properly witness this fact.

However, all is not lost: using our solution to \textsc{opt} it is easy to
recover the unitary matrix $N_1 \in \mapset{U}{\hilb{M}_N \otimes \hilb{N}}$
belonging to a no-prover who wins against $Y(x)$.
This unitary matrix satisfies
$$
\Inner{ V_1^* N_1^* V_2^* \Pi_\mathrm{reject}^*
\Pi_\mathrm{reject} V_2 N_1 V_1, \rho_1 }
> c,
$$
but by the definition of $\mapset{Win}{V,c}$ we have
$$
\Inner{ V_1^* N_1^* V_2^* \Pi_\mathrm{reject}^*
\Pi_\mathrm{reject} V_2 N_1 V_1, \rho }
\leq c \quad \forall \ \rho \in \mapset{Win}{V,c}
$$
and hence $N_1$ is in some sense a witness to the fact that
$\rho_1 \not \in \mapset{Win}{V,c}$.
In particular, it follows from these inequalities that the matrix
$V_1^* N_1^* V_2^* \Pi_\mathrm{reject}^* \Pi_\mathrm{reject} V_2 N_1 V_1$
and the scalar $c$ define a hyperplane that separates $\rho_1$ from
$\mapset{Win}{V,c}$.

To summarize the ideas presented thus far, we have that if a given density
matrix $\rho_1$ is an element of $\mapset{Win}{V,c}$ then this fact can be
verified efficiently by solving \textsc{opt}.
Otherwise, if $\rho_1 \not \in \mapset{Win}{V,c}$ then we can use our solution
to \textsc{opt} to construct a hyperplane separating $\rho_1$ from
$\mapset{Win}{V,c}$.
Later we will see that these two abilities can be used to efficiently decide the
emptiness of $\mapset{Win}{V,c}$ via convex feasibility methods such as the
ellipsoid method.

\subsection{Double Quantum Interactive Proof Systems}
\label{subsec:4:SQG:doubleQIP}

Before we formalize the ideas presented in Section \ref{subsec:4:SQG:start}, it
is instructive to note that those ideas apply to a slightly larger subclass of
quantum refereed games.
Suppose, for example, that $\rho$ is the state of the verifier's qubits
after exchanging not just one, but several messages with the yes-prover.
Presumably, the same method can still be used to determine whether $\rho$
indicates a winning yes-prover.
The only complication is that it must be possible for the yes-prover to somehow
get the verifier's qubits into the state $\rho$.
Fortunately, the Consistency Characterization (Lemma \ref{lm:4:norminner}) tells
us precisely when this task is possible.

Of course, once $\rho$ is given and the yes-prover is fixed there is no reason
to restrict the induced quantum interactive proof system to only two
messages---our solution to \textsc{opt} will easily handle the case in which the
verifier exchanges many messages with the no-prover.
The only complication here is how to generate a separating hyperplane using the
output of \textsc{opt}.

With that extension in mind, consider a short quantum game in which the verifier
exchanges not just one message with each prover, but $r_1$ messages with the
yes-prover followed by $r_2$ messages with the no-prover before making his
decision.
One can think of a quantum refereed game of this strange form as two consecutive
quantum interactive proof systems---one with the yes-prover, then one with the
no-prover.
Hence, we give the name \emph{double quantum interactive proof system} to
quantum refereed games that obey this protocol and we let
$\cls{DQIP}(c,s)$\label{DQIP} denote the complexity class of languages that have
double quantum interactive proof systems with completeness error $c$ and
soundness error $s$.

Like short quantum games, it is still the case with double quantum interactive
proof systems that the no-prover does not become involved until the verifier has
finished exchanging messages with the yes-prover.
Hence, it is still the case that actions of the yes-prover can be completely
described by a transcript.
This time, however, the transcript consists of several states
$\rho_1,\dots,\rho_{r_1}$
instead of just a single state.

We apply the Consistency Characterization (Lemma \ref{lm:4:norminner}) to double
quantum interactive proof systems in much the same way as it was applied to
short quantum games in Corollary \ref{cor:4:sqg2opt}.
That is, we identify a set of winning yes-provers
$Y(x) = (Y_1,\dots,Y_{r_1})$ with a set of winning transcripts
$\rho_1,\dots,\rho_{r_1}$.

\begin{cor}
\label{cor:4:dqip2opt}

Let $c,s : \mathbb{N} \to [0,1]$ satisfy $1 - c(n) - s(n) > 0$ for all
$n \in \mathbb{N}$, let $L \in \cls{DQIP}(c,s)$, and let
$V(x) = (V_0,\dots,V_{r_1},W_1,\dots,W_{r_2})$
be a $(r_1 + r_2)$-round verifier witnessing this fact.
Define $\mapset{Win}{V,c}$ to be the set of all transcripts
$\rho_1,\dots,\rho_{r_1} \in
\mapset{D}{\hilb{M}_Y \otimes \hilb{V} \otimes \hilb{M}_N \otimes \hilb{N}}$
such that $\rho_1,\dots,\rho_{r_1}$ is $\hilb{M}_Y$-consistent with
$V_0,\dots,V_{r_1 - 1}$ and
$$\Inner{ D^* D, \rho_{r_1} } \leq c$$
for all $D$ of the form
\begin{equation}
\label{eqn:4:D}
D = \Pi_\mathrm{reject} W_{r_2} N_{r_2} W_{r_2 - 1} \cdots W_1 N_1 V_{r_1}
\end{equation}
for some unitary matrices
$N_1,\dots,N_{r_2} \in \mapset{U}{\hilb{M}_N \otimes \hilb{N}}$.
Then $\mapset{Win}{V,c}$ is nonempty if and only if $x \in L$.

\end{cor}

\begin{proof}

If $x \in L$ then by definition there exists
$Y_1,\dots,Y_{r_1} \in \mapset{U}{\hilb{Y} \otimes \hilb{M}_Y}$ such that
$$\Norm{D Y_{r_1} V_{r_1 - 1} \cdots V_1 Y_1 V_0 \ket{0}}^2 \leq c$$
for every $D$ of the form (\ref{eqn:4:D}).
By the Consistency Characterization (Lemma \ref{lm:4:norminner})
there exists a transcript
$\rho_1,\dots,\rho_{r_1}$
that is $\hilb{M}_Y$-consistent with $V_0,\dots,V_{r_1 - 1}$ such that
$$
\Inner{ D^* D, \rho_{r_1} }
=
\Norm{D Y_{r_1} V_{r_1 - 1} \cdots V_1 Y_1 V_0 \ket{0}}^2,
$$
from which it follows that the transcript $\rho_1,\dots,\rho_{r_1}$ is an
element of $\mapset{Win}{V,c}$.

If $x \not \in L$ then let $N(x) = (N_1,\dots,N_{r_2})$ witness this fact, let
$$D = \Pi_\mathrm{reject} W_{r_2} N_{r_2} W_{r_2 - 1} \cdots W_1 N_1 V_{r_1},$$
and let $\rho_1,\dots,\rho_{r_1}$
be any transcript that is $\hilb{M}_Y$-consistent with $V_0,\dots,V_{r_1 - 1}$.
By the Consistency Characterization (Lemma \ref{lm:4:norminner})
there exists a yes-prover
$Y(x) = (Y_1,\dots,Y_{r_1})$ such that
$$
\Inner{ D^* D, \rho_{r_1} }
=
\Norm{D Y_{r_1} V_{r_1 - 1} \cdots V_1 Y_1 V_0 \ket{0}}^2.
$$
It follows from our choice of $D$ that the quantity on the right is at least
$1 - s > c$ and hence the transcript $\rho_1,\dots,\rho_{r_1}$ is not
an element of $\mapset{Win}{V,c}$.
\end{proof}


Just like Corollary \ref{cor:4:sqg2opt}, Corollary \ref{cor:4:dqip2opt} suggests
that any language with a double quantum interactive proof system can be decided
by ascertaining the emptiness of $\mapset{Win}{V,c}$.
Our intention is to solve this problem by extending the ideas of Section
\ref{subsec:4:SQG:start} to double quantum interactive proof systems.
That is, given a transcript $\rho_1,\dots,\rho_{r_1}$ for some yes-prover
$Y(x)$, we check to see if there exists a no-prover who wins against $Y(x)$ by
solving a certain instance of \textsc{opt}.
If such a no-prover exists then we use our solution to \textsc{opt} to construct
a hyperplane separating the transcript $\rho_1,\dots,\rho_{r_1}$ from the set
$\mapset{Win}{V,c}$ of winning transcripts.
Given the ability to construct separating hyperplanes, we will see that the
emptiness of $\mapset{Win}{V,c}$ can be determined via the ellipsoid method.

\subsection{The Ellipsoid Method}
\label{subsec:4:SQG:ellipsoid}

The problem of ascertaining the emptiness of a convex set is sometimes called
\emph{convex feasibility}.
This problem is a special case of the \emph{convex programming} problem, wherein
the task is to maximize a given convex function over some convex set of feasible
solutions.
Convex programming is a generalization of both the linear and semidefinite
programming problems mentioned in Section \ref{subsec:4:opt:sdp}.
In the case of convex feasibility, no function is given and the goal is only to
determine whether or not the set of feasible solutions is empty.

The \emph{ellipsoid method} is an iterative procedure that can often be used to
solve the convex feasibility problem in polynomial time.
Most of our discussion concerning this method is based upon material found in
the book by Gr\"otschel, Lov\'asz, and Schrijver \cite{GrotschelL+88}.
The ellipsoid method has a rich history that we do not survey here.
The interested reader is referred to the aforementioned book for a concise
account of this history up until 1988.

The method can be described informally as follows.
Given a set $\mathcal{A} \subset \mathbb{R}^n$ such that
\begin{itemize}

\item
$\mathcal{A}$ is bounded and convex; and

\item
if $\mathcal{A}$ is nonempty then $\mathcal{A}$ is a full-dimensional subset of
$\mathbb{R}^n$,

\end{itemize}
we wish to decide whether $\mathcal{A}$ is empty.
One iteration of the ellipsoid method consists of the generation of a candidate
element $x \in \mathbb{R}^n$.
If $x$ is found to belong to $\mathcal{A}$ then the algorithm terminates, having
found a certification that $\mathcal{A}$ is nonempty.
Otherwise, the convexity of $\mathcal{A}$ implies that there exists a hyperplane
that separates $x$ from $\mathcal{A}$ \cite{Rockafellar70}.
This hyperplane is used by the ellipsoid method to generate a refined candidate
$x' \in \mathbb{R}^n$ during the next iteration.
If an appropriate number of iterations pass without finding a vector in
$\mathcal{A}$ then the ellipsoid method terminates with the conclusion that
$\mathcal{A}$ must be empty.

The ellipsoid method is an \emph{oracle} algorithm in the sense that it does not
produce the separating hyperplanes used to refine candidates.
Often, producing such a hyperplane is a complicated task that depends heavily
upon the definition of $\mathcal{A}$.
An algorithm that computes a separating hyperplane for a given candidate in this
manner is called a \emph{separation oracle}.
The ellipsoid method guarantees that, given a polynomial-time separation oracle
for $\mathcal{A}$, the emptiness of $\mathcal{A}$ can be deduced in polynomial
time.

The details of the ellipsoid method are many and a proper discussion of those
details would be tedious.
In lieu of such a discussion, we cite an amusing analogy found in Reference
\cite[page 73]{GrotschelL+88} that effectively conveys an intuition of how the
ellipsoid method works:
\begin{quote}
Recall the well-known method of catching a lion in the Sahara.
It works as follows.
Fence in the Sahara, and split it into two parts; check which part does not
contain the lion, fence the other part in, and continue.
After a finite number of steps we will have caught the lion (if there was any)
because the fenced-in zone will be so small that the lion cannot move anymore.
Or we realize that the fenced-in zone is so small that it cannot contain any
lion.
\end{quote}
In this analogy, the Sahara is the vector space $\mathbb{R}^n$ and the lion is
the bounded, convex, and full-dimensional set $\mathcal{A}$.
The ellipsoid method specifies how to ``fence in'' some subset of $\mathbb{R}^n$
and the separation oracle serves to split the fenced-in area and check which
side contains the lion.

Based on this analogy, the necessity of the requirement that $\mathcal{A}$ be
bounded and full-dimensional becomes clear.
If the lion were unbounded then it would be impossible to fence him in.
On the other hand, if the lion were not full-dimensional then he would have zero
length along one axis.
We could conceivably continue fencing him in along that axis \emph{ad infinitum}
and he would still have room to move within the fenced-in area.

Fortunately, the requirement that $\mathcal{A}$ be full-dimensional can often be
dropped.
For example, the case in which $\mathcal{A}$ is a full-dimensional subset of
some ``simple polyhedron'' $\mathcal{P} \subset \mathbb{R}^n$ can also be
handled by the ellipsoid method as described in Reference
\cite[Chapter 6]{GrotschelL+88}.
Moreover, any separation oracle for $\mathcal{A}$ may assume without loss of
generality that the input vector $x \in \mathbb{R}^n$ is also an element of the
polyhedron $\mathcal{P}$.

\subsection{The Set of Winning Yes-Provers Revisited}
\label{subsec:4:SQG:WinningYesProvers}

As of now, the set $\mapset{Win}{V,c}$ is defined only loosely as a set of
winning transcripts for a double quantum interactive proof system.
In this subsection, we provide a more precise definition of this set and show
that it meets the criteria set out by the ellipsoid method.

Before we proceed, we remind the reader of some of the notation used in Section
\ref{subsec:4:opt:opt2sdp}.
In particular, $\hilb{H}^{\oplus n}$ denotes the Hilbert space with dimension
$n \dim(\hilb{H})$ and $(B_1,\dots,B_n) \in \mapset{L}{\hilb{H}^{\oplus n}}$
denotes the block-diagonal matrix whose blocks are the matrices
$B_1,\dots,B_n \in \mapset{L}{\hilb{H}}$.

Let $L \subseteq \set{0,1}^*$ be a language that has a double quantum
interactive proof system with completeness error $c$ and let
$V(x) = (V_0,\dots,V_{r_1},W_1,\dots,W_{r_2})$ be a
$(r_1 + r_2)$-round verifier witnessing this fact.
We define
$$
\mapset{Win}{V,c} \subset
\Mapset{H}{(\hilb{M}_Y \otimes \hilb{V} \otimes
\hilb{M}_N \otimes \hilb{N})^{\oplus r_1 + 1}}
$$
to be the set of all block-diagonal positive semidefinite matrices
$(X_0,\dots,X_{r_1})$ such that $X_0 = \ket{0}\bra{0}$ and the list
$X_1,\dots,X_{r_1}$ is $\hilb{M}_Y$-consistent with
$V_0, \dots, V_{r_1 - 1}$ and
$$\Inner{ D^* D, X_{r_1} } \leq c$$
for all $D$ of the form (\ref{eqn:4:D}) in Corollary \ref{cor:4:dqip2opt}.

An advantage of defining $\mapset{Win}{V,c}$ in this manner is that we can
leverage the results of Section \ref{sec:4:opt} to show that
$\mapset{Win}{V,c}$ is a proper candidate for use with the ellipsoid method.
For example, the following lemma is a straightforward consequence of the work in
that section.

\begin{lm}
\label{prop:4:YesVcconvex}
$\mapset{Win}{V,c}$ is bounded and convex.
\end{lm}

\begin{proof}
Let $\lambda \in [0,1]$ and let $X,Z \in \mapset{Win}{V,c}$ with
$X=(\ket{0}\bra{0},X_1,\dots,X_{r_1})$ and
$Z=(\ket{0}\bra{0},Z_1,\dots,Z_{r_1})$.
That $\lambda X + (1-\lambda) Z$ is block-diagonal with first block equal to
$\ket{0}\bra{0}$ follows immediately.
Moreover, we know that $X$ and $Z$ satisfy the $\hilb{M}_Y$-Consistency
Constraints (Lemma \ref{lm:4:sdpconstraints}).
Letting $(A,\alpha)$ be any one of these constraints, we have
$$
\Inner{ A, \lambda X + (1-\lambda) Z }
= \lambda \Inner{ A, X } + (1-\lambda) \Inner{ A, Z }
= \lambda \alpha + (1-\lambda) \alpha = \alpha,
$$
from which it follows that $\lambda X + (1+\lambda) Z$ also satisfies the
$\hilb{M}_Y$-Consistency Constraints (Lemma \ref{lm:4:sdpconstraints}).
By similar reasoning, any matrix $D$ for which
$\inner{D^* D, X_{r_1}} \leq c$ and $\inner{D^* D, Z_{r_1}} \leq c$
also satisfies
$$\Inner{D^* D, \lambda X_{r_1} + (1-\lambda)Z_{r_1}} \leq c,$$
completing the proof that $\mapset{Win}{V,c}$ is convex.

That $\mapset{Win}{V,c}$ is bounded follows from the
$\hilb{M}_Y$-Consistency Constraint Bound (Lemma \ref{lm:4:feasiblebound}),
which tells us that any element of $\mapset{Win}{V,c}$ has spectral norm at most
$1$.
\end{proof}

According to the discussion in Section \ref{subsec:4:SQG:ellipsoid}, if
we are to use the ellipsoid method to decide the emptiness of
$\mapset{Win}{V,c}$ then it is necessary that the set be polynomial-time
isomorphic to a full-dimensional subset of some ``simple polyhedron'' in
$\mathbb{R}^n$.
We now argue that such an isomorphism exists.

For our purposes, a \emph{simple polyhedron} in $\mathbb{R}^n$ is any set
$\mathcal{P}$ defined explicitly as an intersection of at most $p(n)$ halfspaces
for some fixed $p \in \poly$.
For example, the $\hilb{M}_Y$-Consistency Constraints define a simple polyhedron
$\mathcal{C}$ in the vector space of $n \times n$ Hermitian matrices, which is
readily identified with $\mathbb{R}^{n^2}$ as mentioned in Section
\ref{subsec:QIPinSQG:separation:preamble}.
Hence, it is suffices to prove the following lemma.

\begin{lm}
\label{lm:4:fulldim}
For every $c' > c$, $\mapset{Win}{V,c'}$ is a
full-dimensional subset of $\mathcal{C}$ if it is nonempty.
\end{lm}

\begin{proof}

By definition, every element $X \in \mapset{Win}{V,c'}$:
\begin{itemize}

\item
is block-diagonal with first block equal to $\ket{0}\bra{0}$;

\item
satisfies the $\hilb{M}_Y$-Consistency Constraints;

\item
is positive semidefinite; and

\item
satisfies $\inner{D^* D, X} \leq c < c'$ for all appropriately chosen $D$.

\end{itemize}

The first two restrictions are precisely the definition of the simple polyhedron
$\mathcal{C}$, from which it follows that
$\mapset{Win}{V,c'} \subseteq \mathcal{C}$.
As the positive semidefinite matrices are a full-dimensional subset of the
Hermitian matrices, it follows that the third restriction does not decrease the
dimension of $\mapset{Win}{V,c'}$.

For the final restriction, a simple continuity argument shows that if $X$
satisfies $\inner{D^* D, X} < c'$ for some $D$ then so must all $X'$ in some
neighbourhood of $X$.
Hence, this final restriction also does not decrease the dimension of
$\mapset{Win}{V,c'}$.
\end{proof}

Lemma \ref{lm:4:fulldim} tells us that if the verifier $V(x)$ has completeness
error strictly smaller than $c$ then $\mapset{Win}{V,c}$ is a full-dimensional
subset of the simple polyhedron $\mathcal{C}$.
As $\mapset{Win}{V,c}$ is also bounded and convex, it follows that the ellipsoid
method can decide its emptiness if provided with an efficient separation oracle.

\subsection{A Separation Oracle}

We have seen that the set $\mapset{Win}{V,c}$ qualifies for use with the
ellipsoid method provided that the verifier $V$ has completeness error strictly
smaller than $c$, but it remains to show that there exists an efficient
separation oracle for that set.
Recall from Section \ref{subsec:4:SQG:ellipsoid} that a separation oracle for
$\mapset{Win}{V,c}$ is a computational problem that takes as input a candidate
Hermitian matrix $X$ and outputs either
(i) an assertion that $X \in \mapset{Win}{V,c}$; or
(ii) a hyperplane separating $X$ from $\mapset{Win}{V,c}$.
In this subsection we formalize the statement of that problem and we provide a
polynomial-time solution.

%

The separation oracle we seek for $\mapset{Win}{V,c}$ solves the problem
$\textsc{sep}(V,c)$ defined in Figure \ref{fig:4:sep}
(compare with Reference \cite[Theorem 3.2.1]{GrotschelL+88}).
\begin{figure}
\hrulefill
\begin{description}

\item[Problem.] $\textsc{sep}(V,c)$.

\item[Input.]

A block-diagonal Hermitian matrix
$X \in
\mapset{H}{(\hilb{M}_Y \otimes \hilb{V} \otimes
\hilb{M}_N \otimes \hilb{N})^{\oplus r_1 + 1}}$
satisfying the $\hilb{M}_Y$-Consistency Constraints
(Lemma \ref{lm:4:sdpconstraints})
with first block equal to $\ket{0}\bra{0}$
and an accuracy parameter $\varepsilon > 0$.

\item[Output.]

One of the following:
\begin{enumerate}

\item

A block-diagonal Hermitian matrix
$H \in
\mapset{H}{(\hilb{M}_Y \otimes \hilb{V} \otimes
\hilb{M}_N \otimes \hilb{N})^{\oplus r_1 + 1}}$
with $\norm{H} = 1$ such that
$\inner{H,Z} < \inner{H,X} + \varepsilon$ for every
$Z \in \mapset{Win}{V,c}$.

\item

An assertion that if $\mapset{Win}{V,c}$ is nonempty then there exists
$X' \in \mapset{Win}{V,c}$ with $\norm{X - X'} < \varepsilon$.

\end{enumerate}

\end{description}
\hrulefill
\caption{Definition of $\textsc{sep}(V,c)$}
\label{fig:4:sep}
\end{figure}
As with \textsc{sdp} and \textsc{opt}, we assume that the real and
imaginary parts of all input numbers
(including $c$ and the matrices belonging to $V$)
are represented in binary notation.

As per the discussion in Sections \ref{subsec:4:SQG:ellipsoid} and
\ref{subsec:4:SQG:WinningYesProvers}, we assume without loss of generality that
the input matrix $X$ to our separation oracle already satisfies the
$\hilb{M}_Y$-Consistency Constraints (Lemma \ref{lm:4:sdpconstraints}),
as these constraints define a simple polyhedron $\mathcal{C}$ of which
$\mapset{Win}{V,c}$ is a full-dimensional subset.

It is instructive to note that $\textsc{sep}(V,c)$ implements a \emph{weak}
separation oracle in the following sense.
Output case 1 in Figure \ref{fig:4:sep} allows us to reject a ``good''
yes-prover if it is close to a ``bad'' yes prover.
Conversely, output case 2 implies that we may accept a ``bad'' yes-prover so
long as it is close to a ``good'' yes prover.
We will soon see that leeway afforded to us by the bounded-error requirement for
quantum refereed games permits this convenient relaxation.
This bounded-error requirement will also permit us the necessary assumption that
the verifier's completeness error be strictly less than $c$.

The remainder of this subsection is devoted to proving the following theorem.

\begin{thm}
\label{thm:4:sep}
$\textsc{sep}(V,c)$ can be solved in time polynomial in the bit lengths
of $(V,c)$ and the input data.
\end{thm}

\begin{proof}

Figure \ref{fig:4:sepalg} describes a polynomial-time algorithm for
$\textsc{sep}(V,c)$.
\begin{figure}
\hrulefill
\begin{enumerate}

\item
If $X$ is not positive semidefinite then let $u$ be a unit vector satisfying
$u^* X u < 0$.
Halt and output case 1, returning $-uu^*$.

\item
Write $X = (\ket{0}\bra{0},X_1,\dots,X_{r_1})$ and let
$Y_1,\dots,Y_{r_1} \in \mapset{U}{\hilb{Y} \otimes \hilb{M}_Y}$
be unitary matrices satisfying
$$
\Norm{D Y_{r_1} V_{r_1-1} \cdots V_1 Y_1 V_0 \ket{0} }^2
= \Inner{D^* D, X_{r_1}}
$$
for all matrices $D$ not acting on $\hilb{Y}$.
Let $C = Y_{r_1} V_{r_1-1} \cdots V_1 Y_1 V_0$
and compute an approximation $\tilde C$ of $C$ satisfying
$\norm{C - \tilde C} < \frac{\varepsilon}{4}$.

\item
Solve \textsc{opt} with input matrices
$V_{r_1} \tilde C, W_1, \dots, W_{r_2 - 1}, \Pi_\mathrm{reject} W_{r_2}$
and accuracy parameter $\frac{\varepsilon}{2}$.
Let $\varpi$ denote the optimal value indicated by this solution.
If $\varpi \leq c$ then halt and output case 2.

\item
Otherwise, \textsc{opt} returned matrices
$Z_1,\dots,Z_{r_2} \in
\mapset{Pos}{\hilb{Y} \otimes \hilb{M}_Y \otimes \hilb{V} \otimes \hilb{M}_N}$
$\hilb{M}_N$-consistent with
$V_{r_1} \tilde C, W_1, \dots, W_{r_2 - 1}$ such that
$$
\Inner{ W_{r_2}^* \Pi_\mathrm{reject}^* \Pi_\mathrm{reject} W_{r_2}, Z_{r_2} }
= \varpi > c.
$$
Let $N_1,\dots,N_{r_2} \in \mapset{U}{\hilb{M}_N \otimes \hilb{N}}$
be unitary matrices satisfying
$$
\Norm{\Pi_\mathrm{reject} W_{r_2} N_{r_2} W_{r_2-1} \cdots W_1 N_1 V_{r_1}
\tilde C \ket{0} }^2 = \varpi.
$$
Let $D = \Pi_\mathrm{reject} W_{r_2} N_{r_2} W_{r_2-1} \cdots W_1 N_1 V_{r_1}$
and compute an approximation $\tilde D$ of $ D$ satisfying
$\norm{D - \tilde D} < \frac{\varepsilon}{8}$.
Halt and output case 1, returning the block-diagonal matrix
$(0,\dots,0,\tilde D^* \tilde D)$.

\end{enumerate}
\hrulefill
\caption{A polynomial-time algorithm for $\textsc{sep}(V,c)$}
\label{fig:4:sepalg}
\end{figure}
We now verify the correctness of that algorithm.

If step 2 is reached then $X=(\ket{0}\bra{0},X_1,\dots,X_{r_1})$ must be
positive semidefinite where the list $X_1,\dots,X_{r_1}$ is
$\hilb{M}_Y$-consistent with $V_0, \dots, V_{r_1-1}$.
Existence of the unitary matrices $Y_1,\dots,Y_{r_1}$ then follows from the
Consistency Characterization (Lemma \ref{lm:4:norminner}).
The matrices $X_1,\dots,X_{r_1}$ represent a transcript for the
yes-prover defined by $Y(x) = (Y_1,\dots,Y_{r_1})$.

Suppose first that the halting condition is reached in step 3.
We prove that every no-prover loses to $Y(x)$.
Let $N(x) = N_1,\dots,N_{r_2}$ be any no-prover and
let $p$ (respectively $\tilde p$) denote the probability with which $N(x)$
convinces $V(x)$ to reject $x$ given $Y(x)$ (respectively $\tilde Y(x)$) so that
\begin{align*}
p &=
\norm{ \Pi_\mathrm{reject} W_{r_2} N_{r_2} W_{r_2-1} \cdots W_1 N_1 V_{r_1}
C \ket{0} }^2, \\
\tilde p &=
\norm{ \Pi_\mathrm{reject} W_{r_2} N_{r_2} W_{r_2-1} \cdots W_1 N_1 V_{r_1}
\tilde C \ket{0} }^2.
\end{align*}
By our choice of accuracy parameter for \textsc{opt} we have
$$\tilde p < \varpi + \frac{\varepsilon}{2} \leq c + \frac{\varepsilon}{2}$$
and by Lemma \ref{lm:4:precision} with
$\norm{C - \tilde C} < \frac{\varepsilon}{4}$ we have
$$\abs{ \tilde p - p } < 2 \frac{\varepsilon}{4} = \frac{\varepsilon}{2},$$
from which it follows that $p < c + \varepsilon$.
As $N(x)$ was chosen arbitrarily, it follows that the transcript $X$ is at least
``close'' to $\mapset{Win}{V,c}$ as required by output case 2 in the definition
of \textsc{sep}$(V,c)$.

Next, suppose that the algorithm proceeds to step 4.
We prove that the returned matrix $(0,\dots,0,\tilde D^* \tilde D)$ indicates an
``almost'' separating hyperplane as required by output case 1 in the definition
of $\textsc{sep}(V,c)$.

We start by showing that $\inner{\tilde D^* \tilde D, X_{r_1}}$ is large.
Existence of the unitary matrices $N_1,\dots,N_{r_2}$ in step 4 follows from the
Consistency Characterization (Lemma \ref{lm:4:norminner}).
By Lemma \ref{lm:4:precision} with
$\norm{C - \tilde C} < \frac{\varepsilon}{4}$ and
$\norm{D - \tilde D} < \frac{\varepsilon}{8}$ we have
\begin{equation}
\label{eq:4:sep1}
c < \norm{D \tilde C \ket{0}}^2
< \norm{\tilde D C \ket{0}}^2 + \frac{3}{4} \varepsilon
= \inner{\tilde D^* \tilde D, X_{r_1}} + \frac{3}{4} \varepsilon.
\end{equation}

Choose any transcript $X' \in \mapset{Win}{V,c}$ and write
$X' = (\ket{0} \bra{0}, X_1',\dots,X_{r_1}')$.
We now show that $\inner{\tilde D^* \tilde D, X_{r_1}'}$ is small.
Let $Y'(x) = (Y_1',\dots,Y_{r_1}')$ be a yes-prover giving rise to the
transcript $X'$ and let $C' = Y_{r_1}' V_{r_1 - 1} \cdots V_1 Y_1' V_0$.
We have
\begin{equation}
\label{eq:4:sep2}
c \geq \inner{D^* D, X_{r_1}'} = \norm{D C' \ket{0}}^2
> \norm{\tilde D C' \ket{0}}^2 - \frac{\varepsilon}{4}
= \inner{\tilde D^* \tilde D, X_{r_1}'} - \frac{\varepsilon}{4}
\end{equation}
Combining (\ref{eq:4:sep1}) and (\ref{eq:4:sep2}) we obtain
$$
\inner{\tilde D^* \tilde D, X_{r_1}'}
< \inner{\tilde D^* \tilde D, X_{r_1}} + \varepsilon
$$
as required by output case 1 in the definition of \textsc{sep}$(V,c)$.
The algorithm is therefore correct.

It remains only to verify that this algorithm runs in polynomial time.
Using any established method for computing a unitary matrix witnessing
the Unitary Equivalence of Purifications (Fact \ref{fact:4:purification}),
we can approximate the matrices
$Y_1,\dots,Y_{r_1}$ in step 2 and $N_1,\dots,N_{r_2}$ in step 4 to $t$ bits of
precision in time polynomial in $t$ and the dimensions of those matrices.
The desired approximations $\tilde C$ and $\tilde D$ can then be computed in
polynomial time by choosing an appropriate $t$ according to Section
\ref{subsec:4:bounds:roundoff}.
As the input matrices to \textsc{opt} in step 3 can be computed in polynomial
time and have at most polynomial dimension, the desired result follows from the
fact that \textsc{opt} admits a polynomial-time solution
(Theorem \ref{thm:4:opt2sdp}).
\end{proof}

\subsection{At Long Last}

Now that $\textsc{sep}(V,c)$ has been shown to admit a polynomial-time
solution (Theorem \ref{thm:4:sep}), we are finally ready to prove the
upper bound $\cls{SQG} \subseteq \cls{EXP}$.
Indeed, we prove that the containment holds under the following relaxations of
the definition of $\cls{SQG}$:
\begin{itemize}

\item
The verifier may exchange an exponential number of messages with the
yes-prover followed by an exponential number of messages with the no-prover.

\item
The verifier's quantum circuits may contain an exponential number of gates, so
long as they still act upon at most a polynomial number of qubits.

\item
The completeness error and soundness error may be exponentially close to
$\frac{1}{2}$ in $|x|$.

\end{itemize}

Recall the definitions of a strong verifier
(Section \ref{sec:4:bounds:QIPinEXP})
and a double quantum interactive proof system
(Section \ref{subsec:4:SQG:doubleQIP}).
We prove the following theorem.

\begin{thm}[An extension of $\cls{SQG} \subseteq \cls{EXP}$]
\label{thm:4:SQGinEXP}

Let $c,s : \mathbb{N} \to [0,1]$ be any polynomial-time computable functions
satisfying
$1 - c(n) - s(n) > 0$
for all $n \in \mathbb{N}$.
Any language $L \subseteq \set{0,1}^*$ that can be decided by a double quantum
interactive proof system with a strong verifier having completeness error $c$
and soundness error $s$ is in $\cls{EXP}$.

\end{thm}

\begin{proof}

Let $\varepsilon = 1 - c - s$.
It follows from the fact that $c$ and $s$ are polynomial-time computable that
$\varepsilon \in 2^{-\poly}$.

Figure \ref{fig:4:SQGalg} describes a deterministic exponential-time
algorithm that decides $L$.
\begin{figure}
\hrulefill
\begin{enumerate}

\item
Run the exponential-time Turing machine that generates the verifier's
quantum circuits on input string $x \in \set{0,1}^*$.
Let $V(x) = (V_0,\dots,V_{r_1},W_1,\dots,W_{r_2})$ denote the unitary matrices
associated with these circuits.

\item
Compute an approximation
$\tilde V(x)=(\tilde V_0,\dots,\tilde V_{r_1},\tilde W_1,\dots,\tilde W_{r_2})$
of $V(x)$ satisfying
$$
\norm{V_i - \tilde V_i}, \norm{W_j - \tilde W_j}
< \frac{\varepsilon}{8(r_1+r_2+1)}
$$
for every $i \in \set{0,\dots,r_1}$ and every $j \in \set{1,\dots,r_2}$.

\item
Let $c' = c + \frac{\varepsilon}{2}$ and use the ellipsoid method with an
oracle for \textsc{sep}$(\tilde V, c')$ with accuracy parameter
$\frac{\varepsilon}{4}$ to decide the emptiness of
$\mapset{Win}{\tilde V, c'}$.
If it is empty then reject $x$, otherwise accept $x$.

\end{enumerate}
\hrulefill
\caption{An exponential-time algorithm for $L \in \cls{DQIP}$}
\label{fig:4:SQGalg}
\end{figure}
To see that this algorithm is correct,
let $\tilde c$ and $\tilde s$ denote the completeness error and soundness error
respectively of $\tilde V(x)$.
It follows from Lemma \ref{lm:4:precision} with the approximations of step 2
that
$\tilde c < c + \frac{\varepsilon}{4}$ and
$\tilde s < s + \frac{\varepsilon}{4}$.
As
$$1 - c' - \tilde s > 1 - c - s - \frac{3}{4}\varepsilon > 0$$
it follows from Corollary \ref{cor:4:sqg2opt} that
$\mapset{Win}{\tilde V, c'}$ is nonempty if and only if $x \in L$.
As $\tilde V(x)$ has completeness error strictly smaller than $c'$, it follows
from the remarks in Section \ref{subsec:4:SQG:WinningYesProvers} that the set
$\mapset{Win}{\tilde V, c'}$ qualifies for use with the ellipsoid method.

It remains only to verify our choice of accuracy parameter to
\textsc{sep}$(\tilde V, c')$.
Suppose that the ellipsoid method found that
$\mapset{Win}{\tilde V, c'}$ is nonempty.
Then there must exist a transcript $X$ for which the maximum probability with
which $\tilde V(x)$ rejects $x$ is smaller than
$$c' + \frac{\varepsilon}{4} = c + \frac{3}{4}\varepsilon
= 1 - s - \frac{\varepsilon}{4} < 1 - \tilde s.$$
As this transcript violates the soundness condition, it must be the case that
$x \in L$.

Conversely, suppose the ellipsoid method found that
$\mapset{Win}{\tilde V, c'}$ is empty.
Then for every transcript $X$ the maximum probability with which
$\tilde V(x)$ rejects $x$ is larger than
$$c' - \frac{\varepsilon}{4} = c + \frac{\varepsilon}{4} > \tilde c.$$
As every transcript violates the completeness condition, it must be the case
that $x \not \in L$.

That our algorithm runs in exponential time follows from the fact that
\textsc{sep}$(\tilde V, c')$ admits a polynomial-time solution
(Theorem \ref{thm:4:sep}) and from the polynomiality of the ellipsoid method.
\end{proof}

%% file: QIPtranscript.latex
\setlength{\unitlength}{2047sp}%
\begingroup\makeatletter\ifx\SetFigFont\undefined%
\gdef\SetFigFont#1#2#3#4#5{%
  \reset@font\fontsize{#1}{#2pt}%
  \fontfamily{#3}\fontseries{#4}\fontshape{#5}%
  \selectfont}%
\fi\endgroup%
\begin{picture}(8274,2862)(589,-2311)
\thinlines
{\put(601,-511){\line( 1, 0){750}}
}%
{\put(601,-661){\line( 1, 0){750}}
}%
{\put(601,-811){\line( 1, 0){750}}
}%
{\put(601,-961){\line( 1, 0){750}}
}%
{\put(601,-1411){\line( 1, 0){750}}
}%
{\put(601,-1561){\line( 1, 0){750}}
}%
{\put(601,-1711){\line( 1, 0){750}}
}%
{\put(601,-1861){\line( 1, 0){750}}
}%
{\put(2101,-511){\line( 1, 0){750}}
}%
{\put(2101,-661){\line( 1, 0){750}}
}%
{\put(2101,-811){\line( 1, 0){750}}
}%
{\put(2101,-961){\line( 1, 0){750}}
}%
{\put(2851,389){\line(-1, 0){2250}}
}%
{\put(2851,239){\line(-1, 0){2250}}
}%
{\put(2851, 89){\line(-1, 0){2250}}
}%
{\put(2851,-61){\line(-1, 0){2250}}
}%
{\put(3601,-511){\line( 1, 0){750}}
}%
{\put(3601,-661){\line( 1, 0){750}}
}%
{\put(3601,-811){\line( 1, 0){750}}
}%
{\put(3601,-961){\line( 1, 0){750}}
}%
{\put(4351,-1411){\line(-1, 0){2250}}
}%
{\put(4351,-1561){\line(-1, 0){2250}}
}%
{\put(4351,-1711){\line(-1, 0){2250}}
}%
{\put(4351,-1861){\line(-1, 0){2250}}
}%
{\put(5101,-511){\line( 1, 0){750}}
}%
{\put(5101,-661){\line( 1, 0){750}}
}%
{\put(5101,-811){\line( 1, 0){750}}
}%
{\put(5101,-961){\line( 1, 0){750}}
}%
{\put(5851,389){\line(-1, 0){2250}}
}%
{\put(5851,239){\line(-1, 0){2250}}
}%
{\put(5851, 89){\line(-1, 0){2250}}
}%
{\put(5851,-61){\line(-1, 0){2250}}
}%
{\put(6601,-511){\line( 1, 0){750}}
}%
{\put(6601,-661){\line( 1, 0){750}}
}%
{\put(6601,-811){\line( 1, 0){750}}
}%
{\put(6601,-961){\line( 1, 0){750}}
}%
{\put(7351,-1411){\line(-1, 0){2250}}
}%
{\put(7351,-1561){\line(-1, 0){2250}}
}%
{\put(7351,-1711){\line(-1, 0){2250}}
}%
{\put(7351,-1861){\line(-1, 0){2250}}
}%
{\put(8851,389){\line(-1, 0){2250}}
}%
{\put(8851,239){\line(-1, 0){2250}}
}%
{\put(8851, 89){\line(-1, 0){2250}}
}%
{\put(8851,-61){\line(-1, 0){2250}}
}%
{\put(8101,-511){\line( 1, 0){750}}
}%
{\put(8101,-661){\line( 1, 0){750}}
}%
{\put(8101,-811){\line( 1, 0){750}}
}%
{\put(8101,-961){\line( 1, 0){750}}
}%
{\put(8101,-1411){\line( 1, 0){750}}
}%
{\put(8101,-1561){\line( 1, 0){750}}
}%
{\put(8101,-1711){\line( 1, 0){750}}
}%
{\put(8101,-1861){\line( 1, 0){750}}
}%
{\put(2851,-1111){\framebox(750,1650){$P_1$}}
}%
{\put(5851,-1111){\framebox(750,1650){$P_2$}}
}%
{\put(1351,-2011){\framebox(750,1650){$V_0$}}
}%
{\put(4351,-2011){\framebox(750,1650){$V_1$}}
}%
{\put(7351,-2011){\framebox(750,1650){$V_2$}}
}%
{\put(4006,-1981){\oval(210,210)[bl]}
\put(4006,-391){\oval(210,210)[tl]}
\put(4021,-1981){\oval(210,210)[br]}
\put(4021,-391){\oval(210,210)[tr]}
\put(4006,-2086){\line( 1, 0){ 15}}
\put(4006,-286){\line( 1, 0){ 15}}
\put(3901,-1981){\line( 0, 1){1590}}
\put(4126,-1981){\line( 0, 1){1590}}
}%
{\put(1006,-1981){\oval(210,210)[bl]}
\put(1006,-391){\oval(210,210)[tl]}
\put(1021,-1981){\oval(210,210)[br]}
\put(1021,-391){\oval(210,210)[tr]}
\put(1006,-2086){\line( 1, 0){ 15}}
\put(1006,-286){\line( 1, 0){ 15}}
\put(901,-1981){\line( 0, 1){1590}}
\put(1126,-1981){\line( 0, 1){1590}}
}%
{\put(7006,-1981){\oval(210,210)[bl]}
\put(7006,-391){\oval(210,210)[tl]}
\put(7021,-1981){\oval(210,210)[br]}
\put(7021,-391){\oval(210,210)[tr]}
\put(7006,-2086){\line( 1, 0){ 15}}
\put(7006,-286){\line( 1, 0){ 15}}
\put(6901,-1981){\line( 0, 1){1590}}
\put(7126,-1981){\line( 0, 1){1590}}
}%
\put(326,-861){\makebox(0,0)[r]{\smash{{\SetFigFont{12}{14.4}{\familydefault}{\mddefault}{\updefault}{$\hilb{M}$}%
}}}}
\put(326,-1761){\makebox(0,0)[r]{\smash{{\SetFigFont{12}{14.4}{\familydefault}{\mddefault}{\updefault}{$\hilb{V}$}%
}}}}
\put(326, 39){\makebox(0,0)[r]{\smash{{\SetFigFont{12}{14.4}{\familydefault}{\mddefault}{\updefault}{$\hilb{P}$}%
}}}}
\put(9001,-1505){$\leftarrow$\hspace*{-3mm}\parbox{2cm}{
		\begin{center}\small
			output\\[-0.7mm]
			qubit
		\end{center}}}
\put(901,-2411){\makebox(0,0)[lb]{\smash{{\SetFigFont{12}{14.4}{\rmdefault}{\mddefault}{\updefault}{$\rho_0$}%
}}}}
\put(3901,-2411){\makebox(0,0)[lb]{\smash{{\SetFigFont{12}{14.4}{\rmdefault}{\mddefault}{\updefault}{$\rho_1$}%
}}}}
\put(6901,-2411){\makebox(0,0)[lb]{\smash{{\SetFigFont{12}{14.4}{\rmdefault}{\mddefault}{\updefault}{$\rho_2$}%
}}}}
\end{picture}%

%% file: QIPtranscript2.latex
\setlength{\unitlength}{2047sp}%
\begingroup\makeatletter\ifx\SetFigFont\undefined%
\gdef\SetFigFont#1#2#3#4#5{%
  \reset@font\fontsize{#1}{#2pt}%
  \fontfamily{#3}\fontseries{#4}\fontshape{#5}%
  \selectfont}%
\fi\endgroup%
\begin{picture}(8274,2862)(589,-2311)
\thinlines
{\put(601,-511){\line( 1, 0){750}}
}%
{\put(601,-661){\line( 1, 0){750}}
}%
{\put(601,-811){\line( 1, 0){750}}
}%
{\put(601,-961){\line( 1, 0){750}}
}%
{\put(601,-1411){\line( 1, 0){750}}
}%
{\put(601,-1561){\line( 1, 0){750}}
}%
{\put(601,-1711){\line( 1, 0){750}}
}%
{\put(601,-1861){\line( 1, 0){750}}
}%
{\put(2101,-511){\line( 1, 0){750}}
}%
{\put(2101,-661){\line( 1, 0){750}}
}%
{\put(2101,-811){\line( 1, 0){750}}
}%
{\put(2101,-961){\line( 1, 0){750}}
}%
{\put(2851,389){\line(-1, 0){2250}}
}%
{\put(2851,239){\line(-1, 0){2250}}
}%
{\put(2851, 89){\line(-1, 0){2250}}
}%
{\put(2851,-61){\line(-1, 0){2250}}
}%
{\put(3601,-511){\line( 1, 0){750}}
}%
{\put(3601,-661){\line( 1, 0){750}}
}%
{\put(3601,-811){\line( 1, 0){750}}
}%
{\put(3601,-961){\line( 1, 0){750}}
}%
{\put(4351,-1411){\line(-1, 0){2250}}
}%
{\put(4351,-1561){\line(-1, 0){2250}}
}%
{\put(4351,-1711){\line(-1, 0){2250}}
}%
{\put(4351,-1861){\line(-1, 0){2250}}
}%
{\put(5101,-511){\line( 1, 0){750}}
}%
{\put(5101,-661){\line( 1, 0){750}}
}%
{\put(5101,-811){\line( 1, 0){750}}
}%
{\put(5101,-961){\line( 1, 0){750}}
}%
{\put(5851,389){\line(-1, 0){2250}}
}%
{\put(5851,239){\line(-1, 0){2250}}
}%
{\put(5851, 89){\line(-1, 0){2250}}
}%
{\put(5851,-61){\line(-1, 0){2250}}
}%
{\put(6601,-511){\line( 1, 0){750}}
}%
{\put(6601,-661){\line( 1, 0){750}}
}%
{\put(6601,-811){\line( 1, 0){750}}
}%
{\put(6601,-961){\line( 1, 0){750}}
}%
{\put(7351,-1411){\line(-1, 0){2250}}
}%
{\put(7351,-1561){\line(-1, 0){2250}}
}%
{\put(7351,-1711){\line(-1, 0){2250}}
}%
{\put(7351,-1861){\line(-1, 0){2250}}
}%
{\put(8851,389){\line(-1, 0){2250}}
}%
{\put(8851,239){\line(-1, 0){2250}}
}%
{\put(8851, 89){\line(-1, 0){2250}}
}%
{\put(8851,-61){\line(-1, 0){2250}}
}%
{\put(8101,-511){\line( 1, 0){750}}
}%
{\put(8101,-661){\line( 1, 0){750}}
}%
{\put(8101,-811){\line( 1, 0){750}}
}%
{\put(8101,-961){\line( 1, 0){750}}
}%
{\put(8101,-1411){\line( 1, 0){750}}
}%
{\put(8101,-1561){\line( 1, 0){750}}
}%
{\put(8101,-1711){\line( 1, 0){750}}
}%
{\put(8101,-1861){\line( 1, 0){750}}
}%
{\put(2851,-1111){\framebox(750,1650){$P_1$}}
}%
{\put(5851,-1111){\framebox(750,1650){$P_2$}}
}%
{\put(1351,-2011){\framebox(750,1650){$V_0$}}
}%
{\put(4351,-2011){\framebox(750,1650){$V_1$}}
}%
{\put(7351,-2011){\framebox(750,1650){$V_2$}}
}%
{\put(2431,-1981){\oval(210,210)[bl]}
\put(2431,-1291){\oval(210,210)[tl]}
\put(2446,-1981){\oval(210,210)[br]}
\put(2446,-1291){\oval(210,210)[tr]}
\put(2431,-2086){\line( 1, 0){ 15}}
\put(2431,-1186){\line( 1, 0){ 15}}
\put(2326,-1981){\line( 0, 1){690}}
\put(2551,-1981){\line( 0, 1){690}}
}%
{\put(4006,-1981){\oval(210,210)[bl]}
\put(4006,-1291){\oval(210,210)[tl]}
\put(4021,-1981){\oval(210,210)[br]}
\put(4021,-1291){\oval(210,210)[tr]}
\put(4006,-2086){\line( 1, 0){ 15}}
\put(4006,-1186){\line( 1, 0){ 15}}
\put(3901,-1981){\line( 0, 1){690}}
\put(4126,-1981){\line( 0, 1){690}}
}%
{\put(7006,-1981){\oval(210,210)[bl]}
\put(7006,-1291){\oval(210,210)[tl]}
\put(7021,-1981){\oval(210,210)[br]}
\put(7021,-1291){\oval(210,210)[tr]}
\put(7006,-2086){\line( 1, 0){ 15}}
\put(7006,-1186){\line( 1, 0){ 15}}
\put(6901,-1981){\line( 0, 1){690}}
\put(7126,-1981){\line( 0, 1){690}}
}%
{\put(5431,-1981){\oval(210,210)[bl]}
\put(5431,-1291){\oval(210,210)[tl]}
\put(5446,-1981){\oval(210,210)[br]}
\put(5446,-1291){\oval(210,210)[tr]}
\put(5431,-2086){\line( 1, 0){ 15}}
\put(5431,-1186){\line( 1, 0){ 15}}
\put(5326,-1981){\line( 0, 1){690}}
\put(5551,-1981){\line( 0, 1){690}}
}%
\put(326,-861){\makebox(0,0)[r]{\smash{{\SetFigFont{12}{14.4}{\familydefault}{\mddefault}{\updefault}{$\hilb{M}$}%
}}}}
\put(326,-1761){\makebox(0,0)[r]{\smash{{\SetFigFont{12}{14.4}{\familydefault}{\mddefault}{\updefault}{$\hilb{V}$}%
}}}}
\put(326, 39){\makebox(0,0)[r]{\smash{{\SetFigFont{12}{14.4}{\familydefault}{\mddefault}{\updefault}{$\hilb{P}$}%
}}}}
\put(9001,-1505){$\leftarrow$\hspace*{-3mm}\parbox{2cm}{
		\begin{center}\small
			output\\[-0.7mm]
			qubit
		\end{center}}}
\put(3901,-2461){\makebox(0,0)[lb]{\smash{{\SetFigFont{12}{14.4}{\rmdefault}{\mddefault}{\updefault}{$\xi_1'$}%
}}}}
\put(6901,-2461){\makebox(0,0)[lb]{\smash{{\SetFigFont{12}{14.4}{\rmdefault}{\mddefault}{\updefault}{$\xi_2'$}%
}}}}
\put(2326,-2461){\makebox(0,0)[lb]{\smash{{\SetFigFont{12}{14.4}{\rmdefault}{\mddefault}{\updefault}{$\xi_1$}%
}}}}
\put(5326,-2461){\makebox(0,0)[lb]{\smash{{\SetFigFont{12}{14.4}{\rmdefault}{\mddefault}{\updefault}{$\xi_2$}%
}}}}
\end{picture}%

%% file: QRGtranscript.latex
\setlength{\unitlength}{2047sp}%
\begingroup\makeatletter\ifx\SetFigFont\undefined%
\gdef\SetFigFont#1#2#3#4#5{%
  \reset@font\fontsize{#1}{#2pt}%
  \fontfamily{#3}\fontseries{#4}\fontshape{#5}%
  \selectfont}%
\fi\endgroup%
\begin{picture}(10074,4722)(589,-3823)
\thinlines
{\put(601,-511){\line( 1, 0){750}}
}%
{\put(601,-661){\line( 1, 0){750}}
}%
{\put(601,-811){\line( 1, 0){750}}
}%
{\put(601,-961){\line( 1, 0){750}}
}%
{\put(601,-1411){\line( 1, 0){750}}
}%
{\put(601,-1561){\line( 1, 0){750}}
}%
{\put(601,-1711){\line( 1, 0){750}}
}%
{\put(601,-1861){\line( 1, 0){750}}
}%
{\put(601,-2311){\line( 1, 0){750}}
}%
{\put(601,-2461){\line( 1, 0){750}}
}%
{\put(601,-2611){\line( 1, 0){750}}
}%
{\put(601,-2761){\line( 1, 0){750}}
}%
{\put(2851,-3211){\line(-1, 0){2250}}
}%
{\put(2851,-3361){\line(-1, 0){2250}}
}%
{\put(2851,-3511){\line(-1, 0){2250}}
}%
{\put(2851,-3661){\line(-1, 0){2250}}
}%
{\put(2101,-2311){\line( 1, 0){750}}
}%
{\put(2101,-2461){\line( 1, 0){750}}
}%
{\put(2101,-2611){\line( 1, 0){750}}
}%
{\put(2101,-2761){\line( 1, 0){750}}
}%
{\put(5101,-511){\line( 1, 0){150}}
}%
{\put(5101,-661){\line( 1, 0){150}}
}%
{\put(5101,-811){\line( 1, 0){150}}
}%
{\put(5101,-961){\line( 1, 0){150}}
}%
{\put(6001,-2311){\line( 1, 0){750}}
}%
{\put(6001,-2461){\line( 1, 0){750}}
}%
{\put(6001,-2611){\line( 1, 0){750}}
}%
{\put(6001,-2761){\line( 1, 0){750}}
}%
{\put(9001,-511){\line( 1, 0){150}}
}%
{\put(9001,-661){\line( 1, 0){150}}
}%
{\put(9001,-811){\line( 1, 0){150}}
}%
{\put(9001,-961){\line( 1, 0){150}}
}%
{\put(3601,-2311){\line( 1, 0){1650}}
}%
{\put(3601,-2461){\line( 1, 0){1650}}
}%
{\put(3601,-2611){\line( 1, 0){1650}}
}%
{\put(3601,-2761){\line( 1, 0){1650}}
}%
{\put(4351,-511){\line(-1, 0){2250}}
}%
{\put(4351,-661){\line(-1, 0){2250}}
}%
{\put(4351,-811){\line(-1, 0){2250}}
}%
{\put(4351,-961){\line(-1, 0){2250}}
}%
{\put(8251,-511){\line(-1, 0){2250}}
}%
{\put(8251,-661){\line(-1, 0){2250}}
}%
{\put(8251,-811){\line(-1, 0){2250}}
}%
{\put(8251,-961){\line(-1, 0){2250}}
}%
{\put(6751,-3211){\line(-1, 0){3150}}
}%
{\put(6751,-3361){\line(-1, 0){3150}}
}%
{\put(6751,-3511){\line(-1, 0){3150}}
}%
{\put(6751,-3661){\line(-1, 0){3150}}
}%
{\put(5251,-1411){\line(-1, 0){3150}}
}%
{\put(5251,-1561){\line(-1, 0){3150}}
}%
{\put(5251,-1711){\line(-1, 0){3150}}
}%
{\put(5251,-1861){\line(-1, 0){3150}}
}%
{\put(9901,-511){\line( 1, 0){750}}
}%
{\put(9901,-661){\line( 1, 0){750}}
}%
{\put(9901,-811){\line( 1, 0){750}}
}%
{\put(9901,-961){\line( 1, 0){750}}
}%
{\put(9901,-2311){\line( 1, 0){750}}
}%
{\put(9901,-2461){\line( 1, 0){750}}
}%
{\put(9901,-2611){\line( 1, 0){750}}
}%
{\put(9901,-2761){\line( 1, 0){750}}
}%
{\put(9901,-1411){\line( 1, 0){750}}
}%
{\put(9901,-1561){\line( 1, 0){750}}
}%
{\put(9901,-1711){\line( 1, 0){750}}
}%
{\put(9901,-1861){\line( 1, 0){750}}
}%
{\put(8251,389){\line(-1, 0){3150}}
}%
{\put(8251,239){\line(-1, 0){3150}}
}%
{\put(8251, 89){\line(-1, 0){3150}}
}%
{\put(8251,-61){\line(-1, 0){3150}}
}%
{\put(4351,389){\line(-1, 0){3750}}
}%
{\put(4351,239){\line(-1, 0){3750}}
}%
{\put(4351, 89){\line(-1, 0){3750}}
}%
{\put(4351,-61){\line(-1, 0){3750}}
}%
{\put(9151,-1411){\line(-1, 0){3150}}
}%
{\put(9151,-1561){\line(-1, 0){3150}}
}%
{\put(9151,-1711){\line(-1, 0){3150}}
}%
{\put(9151,-1861){\line(-1, 0){3150}}
}%
{\put(10651,-3211){\line(-1, 0){3150}}
}%
{\put(10651,-3361){\line(-1, 0){3150}}
}%
{\put(10651,-3511){\line(-1, 0){3150}}
}%
{\put(10651,-3661){\line(-1, 0){3150}}
}%
{\put(7501,-2311){\line( 1, 0){1650}}
}%
{\put(7501,-2461){\line( 1, 0){1650}}
}%
{\put(7501,-2611){\line( 1, 0){1650}}
}%
{\put(7501,-2761){\line( 1, 0){1650}}
}%
{\put(10651,389){\line(-1, 0){1650}}
}%
{\put(10651,239){\line(-1, 0){1650}}
}%
{\put(10651, 89){\line(-1, 0){1650}}
}%
{\put(10651,-61){\line(-1, 0){1650}}
}%
{\put(1351,-2911){\framebox(750,2550){$V_0$}}
}%
{\put(2851,-3811){\framebox(750,1650){$N_1$}}
}%
{\put(4006,-2881){\oval(210,210)[bl]}
\put(4006,509){\oval(210,210)[tl]}
\put(4021,-2881){\oval(210,210)[br]}
\put(4021,509){\oval(210,210)[tr]}
\put(4006,-2986){\line( 1, 0){ 15}}
\put(4006,614){\line( 1, 0){ 15}}
\put(3901,-2881){\line( 0, 1){3390}}
\put(4126,-2881){\line( 0, 1){3390}}
}%
{\put(5251,-2911){\framebox(750,2550){$V_1$}}
}%
{\put(6751,-3811){\framebox(750,1650){$N_2$}}
}%
{\put(7906,-2881){\oval(210,210)[bl]}
\put(7906,509){\oval(210,210)[tl]}
\put(7921,-2881){\oval(210,210)[br]}
\put(7921,509){\oval(210,210)[tr]}
\put(7906,-2986){\line( 1, 0){ 15}}
\put(7906,614){\line( 1, 0){ 15}}
\put(7801,-2881){\line( 0, 1){3390}}
\put(8026,-2881){\line( 0, 1){3390}}
}%
{\put(8251,-1111){\framebox(750,1650){$Y_2$}}
}%
{\put(4351,-1111){\framebox(750,1650){$Y_1$}}
}%
{\put(9151,-2911){\framebox(750,2550){$V_2$}}
}%
{\put(1006,-2881){\oval(210,210)[bl]}
\put(1006,509){\oval(210,210)[tl]}
\put(1021,-2881){\oval(210,210)[br]}
\put(1021,509){\oval(210,210)[tr]}
\put(1006,-2986){\line( 1, 0){ 15}}
\put(1006,614){\line( 1, 0){ 15}}
\put(901,-2881){\line( 0, 1){3390}}
\put(1126,-2881){\line( 0, 1){3390}}
}%
\put(326,-861){\makebox(0,0)[r]{\smash{{\SetFigFont{12}{14.4}{\familydefault}{\mddefault}{\updefault}{$\hilb{M}_Y$}%
}}}}
\put(326,-1761){\makebox(0,0)[r]{\smash{{\SetFigFont{12}{14.4}{\familydefault}{\mddefault}{\updefault}{$\hilb{V}$}%
}}}}
\put(326,-2661){\makebox(0,0)[r]{\smash{{\SetFigFont{12}{14.4}{\familydefault}{\mddefault}{\updefault}{$\hilb{M}_N$}%
}}}}
\put(326,-3561){\makebox(0,0)[r]{\smash{{\SetFigFont{12}{14.4}{\familydefault}{\mddefault}{\updefault}{$\hilb{N}$}%
}}}}
\put(326, 39){\makebox(0,0)[r]{\smash{{\SetFigFont{12}{14.4}{\familydefault}{\mddefault}{\updefault}{$\hilb{Y}$}%
}}}}
\put(901,814){\makebox(0,0)[lb]{\smash{{\SetFigFont{12}{14.4}{\rmdefault}{\mddefault}{\updefault}{$\rho_0$}%
}}}}
\put(3901,814){\makebox(0,0)[lb]{\smash{{\SetFigFont{12}{14.4}{\rmdefault}{\mddefault}{\updefault}{$\rho_1$}%
}}}}
\put(7801,814){\makebox(0,0)[lb]{\smash{{\SetFigFont{12}{14.4}{\rmdefault}{\mddefault}{\updefault}{$\rho_2$}%
}}}}
\end{picture}%

%% file: SQGtranscript.latex
\setlength{\unitlength}{2047sp}%
\begingroup\makeatletter\ifx\SetFigFont\undefined%
\gdef\SetFigFont#1#2#3#4#5{%
  \reset@font\fontsize{#1}{#2pt}%
  \fontfamily{#3}\fontseries{#4}\fontshape{#5}%
  \selectfont}%
\fi\endgroup%
\begin{picture}(8274,4737)(589,-4186)
\thinlines
{\put(601,-511){\line( 1, 0){750}}
}%
{\put(601,-661){\line( 1, 0){750}}
}%
{\put(601,-811){\line( 1, 0){750}}
}%
{\put(601,-961){\line( 1, 0){750}}
}%
{\put(601,-1411){\line( 1, 0){750}}
}%
{\put(601,-1561){\line( 1, 0){750}}
}%
{\put(601,-1711){\line( 1, 0){750}}
}%
{\put(601,-1861){\line( 1, 0){750}}
}%
{\put(601,-2311){\line( 1, 0){750}}
}%
{\put(601,-2461){\line( 1, 0){750}}
}%
{\put(601,-2611){\line( 1, 0){750}}
}%
{\put(601,-2761){\line( 1, 0){750}}
}%
{\put(2101,-511){\line( 1, 0){750}}
}%
{\put(2101,-661){\line( 1, 0){750}}
}%
{\put(2101,-811){\line( 1, 0){750}}
}%
{\put(2101,-961){\line( 1, 0){750}}
}%
{\put(2851,389){\line(-1, 0){2250}}
}%
{\put(2851,239){\line(-1, 0){2250}}
}%
{\put(2851, 89){\line(-1, 0){2250}}
}%
{\put(2851,-61){\line(-1, 0){2250}}
}%
{\put(3601,-511){\line( 1, 0){750}}
}%
{\put(3601,-661){\line( 1, 0){750}}
}%
{\put(3601,-811){\line( 1, 0){750}}
}%
{\put(3601,-961){\line( 1, 0){750}}
}%
{\put(4351,-1411){\line(-1, 0){2250}}
}%
{\put(4351,-1561){\line(-1, 0){2250}}
}%
{\put(4351,-1711){\line(-1, 0){2250}}
}%
{\put(4351,-1861){\line(-1, 0){2250}}
}%
{\put(2851,-3211){\line(-1, 0){2250}}
}%
{\put(2851,-3361){\line(-1, 0){2250}}
}%
{\put(2851,-3511){\line(-1, 0){2250}}
}%
{\put(2851,-3661){\line(-1, 0){2250}}
}%
{\put(5101,-2311){\line( 1, 0){750}}
}%
{\put(5101,-2461){\line( 1, 0){750}}
}%
{\put(5101,-2611){\line( 1, 0){750}}
}%
{\put(5101,-2761){\line( 1, 0){750}}
}%
{\put(5851,-3211){\line(-1, 0){2250}}
}%
{\put(5851,-3361){\line(-1, 0){2250}}
}%
{\put(5851,-3511){\line(-1, 0){2250}}
}%
{\put(5851,-3661){\line(-1, 0){2250}}
}%
{\put(5851,389){\line(-1, 0){2250}}
}%
{\put(5851,239){\line(-1, 0){2250}}
}%
{\put(5851, 89){\line(-1, 0){2250}}
}%
{\put(5851,-61){\line(-1, 0){2250}}
}%
{\put(6601,-2311){\line( 1, 0){750}}
}%
{\put(6601,-2461){\line( 1, 0){750}}
}%
{\put(6601,-2611){\line( 1, 0){750}}
}%
{\put(6601,-2761){\line( 1, 0){750}}
}%
{\put(7351,-1411){\line(-1, 0){2250}}
}%
{\put(7351,-1561){\line(-1, 0){2250}}
}%
{\put(7351,-1711){\line(-1, 0){2250}}
}%
{\put(7351,-1861){\line(-1, 0){2250}}
}%
{\put(8851,389){\line(-1, 0){2250}}
}%
{\put(8851,239){\line(-1, 0){2250}}
}%
{\put(8851, 89){\line(-1, 0){2250}}
}%
{\put(8851,-61){\line(-1, 0){2250}}
}%
{\put(8101,-511){\line( 1, 0){750}}
}%
{\put(8101,-661){\line( 1, 0){750}}
}%
{\put(8101,-811){\line( 1, 0){750}}
}%
{\put(8101,-961){\line( 1, 0){750}}
}%
{\put(8101,-2311){\line( 1, 0){750}}
}%
{\put(8101,-2461){\line( 1, 0){750}}
}%
{\put(8101,-2611){\line( 1, 0){750}}
}%
{\put(8101,-2761){\line( 1, 0){750}}
}%
{\put(8851,-3211){\line(-1, 0){2250}}
}%
{\put(8851,-3361){\line(-1, 0){2250}}
}%
{\put(8851,-3511){\line(-1, 0){2250}}
}%
{\put(8851,-3661){\line(-1, 0){2250}}
}%
{\put(4351,-2311){\line(-1, 0){2250}}
}%
{\put(4351,-2461){\line(-1, 0){2250}}
}%
{\put(4351,-2611){\line(-1, 0){2250}}
}%
{\put(4351,-2761){\line(-1, 0){2250}}
}%
{\put(7351,-511){\line(-1, 0){2250}}
}%
{\put(7351,-661){\line(-1, 0){2250}}
}%
{\put(7351,-811){\line(-1, 0){2250}}
}%
{\put(7351,-961){\line(-1, 0){2250}}
}%
{\put(5851,389){\line( 1, 0){750}}
}%
{\put(5851,239){\line( 1, 0){750}}
}%
{\put(5851, 89){\line( 1, 0){750}}
}%
{\put(5851,-61){\line( 1, 0){750}}
}%
{\put(2851,-3211){\line( 1, 0){750}}
}%
{\put(2851,-3361){\line( 1, 0){750}}
}%
{\put(2851,-3511){\line( 1, 0){750}}
}%
{\put(2851,-3661){\line( 1, 0){750}}
}%
{\put(8101,-1411){\line( 1, 0){750}}
}%
{\put(8101,-1561){\line( 1, 0){750}}
}%
{\put(8101,-1711){\line( 1, 0){750}}
}%
{\put(8101,-1861){\line( 1, 0){750}}
}%
\put(9001,-1505){$\leftarrow$\hspace*{-3mm}\parbox{2cm}{
		\begin{center}\small
			output\\[-0.7mm]
			qubit
		\end{center}}}
{\put(1351,-2911){\framebox(750,2550){$V_0$}}
}%
{\put(2851,-1111){\framebox(750,1650){$Y_1$}}
}%
{\put(4351,-2911){\framebox(750,2550){$V_1$}}
}%
{\put(5851,-3811){\framebox(750,1650){$N_1$}}
}%
{\put(7351,-2911){\framebox(750,2550){$V_2$}}
}%
\put(326,-861){\makebox(0,0)[r]{\smash{{\SetFigFont{12}{14.4}{\familydefault}{\mddefault}{\updefault}{$\hilb{M}_Y$}%
}}}}
\put(326,-1761){\makebox(0,0)[r]{\smash{{\SetFigFont{12}{14.4}{\familydefault}{\mddefault}{\updefault}{$\hilb{V}$}%
}}}}
\put(326,-2661){\makebox(0,0)[r]{\smash{{\SetFigFont{12}{14.4}{\familydefault}{\mddefault}{\updefault}{$\hilb{M}_N$}%
}}}}
\put(326,-3561){\makebox(0,0)[r]{\smash{{\SetFigFont{12}{14.4}{\familydefault}{\mddefault}{\updefault}{$\hilb{N}$}%
}}}}
\put(326, 39){\makebox(0,0)[r]{\smash{{\SetFigFont{12}{14.4}{\familydefault}{\mddefault}{\updefault}{$\hilb{Y}$}%
}}}}
{\put(4006,-3781){\oval(210,210)[bl]}
\put(4006,-391){\oval(210,210)[tl]}
\put(4021,-3781){\oval(210,210)[br]}
\put(4021,-391){\oval(210,210)[tr]}
\put(4006,-3886){\line( 1, 0){ 15}}
\put(4006,-286){\line( 1, 0){ 15}}
\put(3901,-3781){\line( 0, 1){3390}}
\put(4126,-3781){\line( 0, 1){3390}}
}%
\put(3901,-4286){\makebox(0,0)[lb]{\smash{{\SetFigFont{12}{14.4}{\rmdefault}{\mddefault}{\updefault}{$\rho_1$}%
}}}}
\end{picture}%

%% file: end.tex
\chapter{Conclusion} \label{ch:end}

The work of this thesis initiates the study of quantum refereed games.
We chose to focus on short quantum games, proving the containments
$\cls{QIP} \subseteq \cls{SQG}_*$ in Chapter \ref{ch:QIPinSQG} and
$\cls{SQG} \subseteq \cls{EXP}$ in Chapter \ref{ch:SQGinEXP}.

\begin{figure}
\begin{center}
\input{classes.latex}
\end{center}
\caption{Relationships among complexity classes discussed in this thesis}
\label{fig:classes}
\end{figure}
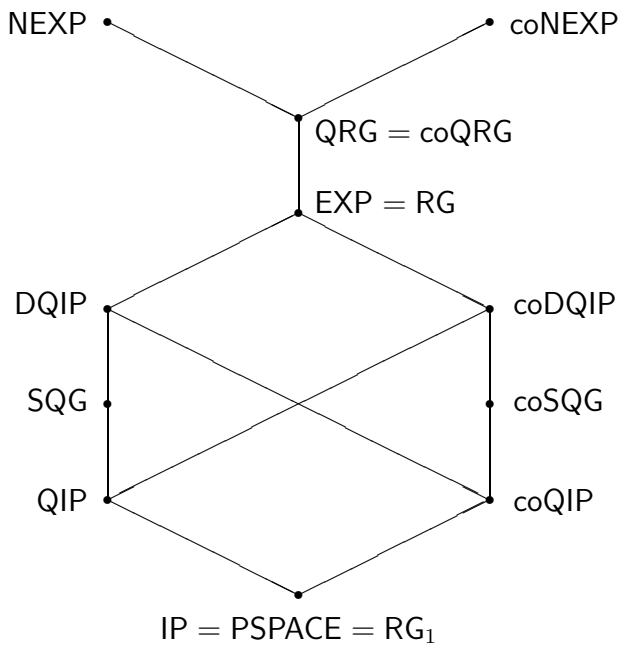

Figure \ref{fig:classes} summarizes some known relationships among the
complexity classes considered in this thesis.
In that figure, $\cls{DQIP}$ denotes the class of languages with double quantum
interactive proof systems as defined in Section
\ref{subsec:4:SQG:doubleQIP} and $\cls{coDQIP}$ its complement.
As with Figure \ref{fig:introcls}, a class $\cls{A}$ contains class $\cls{B}$ if
$\cls{A}$ can be reached from $\cls{B}$ by following a path of only upwardly
sloped edges.

\section{Open Problems}

We now discuss several open questions relating to the material covered in this
thesis.

\subsection{Parallelization}
\label{subsec:end:open:parallelization}

It is known that any $k$-message classical interactive proof system can be
simulated by a two-message interactive proof system for any constant
$k \in \mathbb{N}$ \cite{Babai85, GoldwasserS89}.
The complexity class corresponding to two-message interactive proof systems is
known as $\cls{AM}$ and is contained in $\cls{\Pi}^\cls{P}_2$, the second level
of the polynomial-time hierarchy.

As $\cls{IP} = \cls{PSPACE}$, it is widely believed that interactive proof
systems with a polynomial number of messages are strictly more powerful than
$k$-message interactive proof systems.
In contrast, we mentioned in Section \ref{subsec:defs:remarks:results} that
\emph{any} quantum interactive proof system can be simulated by a three-message
quantum interactive proof system \cite{KitaevW00}.

One can also ask whether a similar parallelization result holds for refereed
games.
In the classical case, we mentioned in Section
\ref{subsec:intro:complexity:results} that one-round refereed games characterize
$\cls{PSPACE}$ and that many-round refereed games characterize $\cls{EXP}$
\cite{FeigeK97}.
However, little is known about the power of refereed games intermediate between
these two extremes.
For example, games with a constant number of rounds may correspond to
$\cls{PSPACE}$, $\cls{EXP}$, or some complexity class between the two.

Even less is known in the quantum case.
For example, it is unclear how to solve \textsc{close-images} with a short
quantum game if the verifier is not permitted to process the yes-prover's
message before sending a message to the no-prover
(Section \ref{sec:QIPinSQG:game}).
Does this ability separate one-round quantum refereed games from short quantum
games?

\subsection{Parallel Repetition}

Suppose we wish to reduce the error of a given interactive protocol without
increasing the number of messages in that protocol.
In Section \ref{subsec:QIPinSQG:error:results} we described an approach to this
problem called \emph{parallel repetition}.
Essentially, the idea is to run many copies of the interaction in parallel and
accept or reject based upon a vote of the outcomes of the individual
repetitions.
The hope is that Chernoff bounds can be used to prove that the error of the
repeated game decreases exponentially in the number of repetitions.
Of course, we must take into account the fact that the provers need not
cooperate with the verifier by treating each repetition independently, and
therein lies the rub.

Although parallel repetition has been successfully applied to single- and
multi-prover classical interactive proof systems (see, for instance, Raz
\cite{Raz98}), this problem has not been completely solved in the quantum
setting.
It is known that parallel repetition followed by a unanimous vote of the
outcomes works to reduce the soundness error for three-message quantum
interactive proof systems with zero completeness error \cite{KitaevW00}.
We extended that result in this thesis to obtain a partial robustness result
for short quantum games (Theorem \ref{thm:QIPinSQG:error}).

However, several questions remain unanswered.
Does parallel repetition work in the quantum setting if it is followed by a
majority vote of the outcomes instead of a unanimous vote?
Is it even possible to improve the error of $k$-round quantum refereed games for
$k \geq 2$ without increasing the number of rounds?

\subsection{Insight into $\cls{QIP}$ and $\cls{SQG}$}

The cumulative results of this thesis can be viewed as wedging several
complexity classes between $\cls{QIP}$ and $\cls{EXP}$---in particular,
$$\cls{QIP} \subseteq \cls{SQG}_* \subseteq \cls{SQG} \subseteq \cls{DQIP}
\subseteq \cls{EXP}.$$
In a sense, it seems as though $\cls{QIP}$ is buried deeply inside $\cls{EXP}$.
Can we prove $\cls{QIP} = \cls{PSPACE}$?

It is clear that $\cls{DQIP}$ contains both $\cls{QIP}$ and $\cls{coQIP}$.
Does $\cls{SQG}$ also contain $\cls{coQIP}$?
Are either of $\cls{QIP}$ or $\cls{SQG}$ closed under complement?

\subsection{Do Quantum Refereed Games Characterize $\cls{EXP}$?}
\label{subsec:end:open:QRGeqEXP}

In this thesis we proved that $\cls{QRG} \subseteq \cls{NEXP}$
(Theorem \ref{thm:4:QRGinNEXP}).
It follows immediately from the fact that $\cls{QRG}$ is closed under complement
that $\cls{QRG} \subseteq \cls{coNEXP}$.
Combined with the fact that $\cls{EXP} \subseteq \cls{QRG}$, we have
$$\cls{EXP} \subseteq \cls{QRG} \subseteq \cls{NEXP} \cap \cls{coNEXP}.$$

In the realm of polynomial-time computation, problems known to be in
$\cls{NP} \cap \cls{coNP}$ yet not known to be in $\cls{P}$ are rare and often
the subject of intense study.
In many cases, a problem with this property is later discovered to lay in
$\cls{P}$ and the accompanying proof of this fact can sometimes offer new
insights in complexity theory.
Popular examples of this trend include the linear programming problem
\cite{Khachiyan79} and the primality testing problem \cite{AgrawalKS02}.
Based upon this historical precedent and upon recent unpublished work by the
author, we make the following conjecture:

\begin{con}
$\cls{QRG} = \cls{EXP}$.
\end{con}

Aside from offering a rare quantum characterization of a classical complexity
class, such a collapse would also imply that classical refereed games are
polynomially equivalent in power to quantum refereed games.
If true, this equivalence would be a powerful negative example of a case in
which the use of quantum information offers no advantage over the use of
classical information.

%% file: classes.latex
\setlength{\unitlength}{3947sp}%
\begingroup\makeatletter\ifx\SetFigFont\undefined%
\gdef\SetFigFont#1#2#3#4#5{%
  \reset@font\fontsize{#1}{#2pt}%
  \fontfamily{#3}\fontseries{#4}\fontshape{#5}%
  \selectfont}%
\fi\endgroup%
\begin{picture}(4067,4112)(151,-2290)
{\thinlines
\put(901,1739){\circle*{50}}
}%
{\put(2101,1139){\circle*{50}}
}%
{\put(3301,1739){\circle*{50}}
}%
{\put(2101,539){\circle*{50}}
}%
{\put(3301,-61){\circle*{50}}
}%
{\put(901,-61){\circle*{50}}
}%
{\put(3301,-1261){\circle*{50}}
}%
{\put(3301,-661){\circle*{50}}
}%
{\put(901,-661){\circle*{50}}
}%
{\put(901,-1261){\circle*{50}}
}%
{\put(2101,-1861){\circle*{50}}
}%
{\put(901,1739){\line( 2,-1){1200}}
}%
{\put(2101,1139){\line( 2, 1){1200}}
}%
{\put(2101,1139){\line( 0,-1){600}}
}%
{\put(901,-61){\line( 2, 1){1200}}
}%
{\put(2101,539){\line( 2,-1){1200}}
}%
{\put(901,-61){\line( 2,-1){2400}}
}%
{\put(3301,-61){\line( 0,-1){1200}}
}%
{\put(3301,-61){\line(-2,-1){2400}}
}%
{\put(2101,-1861){\line( 2, 1){1200}}
}%
{\put(901,-1261){\line( 2,-1){1200}}
}%
{\put(901,-61){\line( 0,-1){1200}}
}%
\put(776,1664){\makebox(0,0)[r]{\smash{{\SetFigFont{12}{14.4}{\familydefault}{\mddefault}{\updefault}{$\cls{NEXP}$}%
}}}}
\put(3426,1664){\makebox(0,0)[l]{\smash{{\SetFigFont{12}{14.4}{\familydefault}{\mddefault}{\updefault}{$\cls{coNEXP}$}%
}}}}
\put(776,-96){\makebox(0,0)[r]{\smash{{\SetFigFont{12}{14.4}{\rmdefault}{\mddefault}{\updefault}{$\cls{DQIP}$}%
}}}}
\put(3451,-96){\makebox(0,0)[l]{\smash{{\SetFigFont{12}{14.4}{\familydefault}{\mddefault}{\updefault}{$\cls{coDQIP}$}%
}}}}
\put(2201,989){\makebox(0,0)[l]{\smash{{\SetFigFont{12}{14.4}{\familydefault}{\mddefault}{\updefault}{$\cls{QRG} = \cls{coQRG}$}%
}}}}
\put(2201,539){\makebox(0,0)[l]{\smash{{\SetFigFont{12}{14.4}{\familydefault}{\mddefault}{\updefault}{$\cls{EXP} = \cls{RG}$}%
}}}}
\put(776,-706){\makebox(0,0)[r]{\smash{{\SetFigFont{12}{14.4}{\familydefault}{\mddefault}{\updefault}{$\cls{SQG}$}%
}}}}
\put(3451,-706){\makebox(0,0)[l]{\smash{{\SetFigFont{12}{14.4}{\familydefault}{\mddefault}{\updefault}{$\cls{coSQG}$}%
}}}}
\put(776,-1336){\makebox(0,0)[r]{\smash{{\SetFigFont{12}{14.4}{\familydefault}{\mddefault}{\updefault}{$\cls{QIP}$}%
}}}}
\put(3451,-1336){\makebox(0,0)[l]{\smash{{\SetFigFont{12}{14.4}{\familydefault}{\mddefault}{\updefault}{$\cls{coQIP}$}%
}}}}
\put(2101,-2136){\makebox(0,0){\smash{{\SetFigFont{12}{14.4}{\familydefault}{\mddefault}{\updefault}{$\cls{IP} = \cls{PSPACE} = \cls{RG}_1$}%
}}}}
\end{picture}%